\pdfoutput=1
\documentclass[11pt]{article}
\usepackage{jheppub}

\usepackage{customprelude}

\usepackage{breqn}
\usepackage{array}
\usepackage{multirow}

\allowdisplaybreaks

\let\oldnormalfont=\normalfont

\newcommand{\be}{\begin{equation}}
\newcommand{\ee}{\end{equation}}
\newcommand{\ba}{\begin{eqnarray}}
\newcommand{\ea}{\end{eqnarray}}

\newcommand\varpm{\mathbin{\vcenter{\hbox{%
  \oalign{\hfil$\scriptstyle+$\hfil\cr
          \noalign{\kern-.3ex}
          $\scriptscriptstyle({-})$\cr}%
}}}}
\newcommand\varmp{\mathbin{\vcenter{\hbox{%
  \oalign{$\scriptstyle({+})$\cr
          \noalign{\kern-.3ex}
          \hfil$\scriptscriptstyle-$\hfil\cr}%
}}}}

\title{\centering Strongly coupled phases of
 \alt{$\mathcal{N}=1$}{N=1} S-duality}

\author[a]{I\~naki Garc\'ia-Etxebarria}
\author[b]{and Ben Heidenreich}

\affiliation[a]{Max Planck Institute for Physics, F\"ohringer Ring 6, 80805 Munich, Germany}
\affiliation[b]{Department of Physics, Harvard University, Cambridge, MA
  02138, USA}

\emailAdd{inaki@mpp.mpg.de}
\emailAdd{bjheiden@physics.harvard.edu}

\abstract{We analyze S-duality of orientifolds of the Calabi-Yau cone over the first del Pezzo surface ($dP_1$). The S-duals of known phases, described by quiver gauge theories, contain intrinsically strongly-coupled sectors. These sectors are realized by a higher multiplicity intersection of NS5 branes and D5 branes atop an O5 plane, and can be thought of as stuck at the infinite coupling point between two Seiberg-dual gauge theories. We argue that such sectors appear generically in orientifolds of non-orbifold singularities, where in many examples every orientifold phase contains such a sector. Understanding such sectors is therefore key to understanding orientifolds of Calabi-Yau singularities. We construct the strongly-coupled sectors for $dP_1$ orientifolds using deconfinement, and show that they have interesting, non-trivial properties. Using this construction, we verify the predictions of S-duality for $dP_1$.}

\begin{document}

\hypersetup{pageanchor=false} 

\makeatletter
\let\old@fpheader\@fpheader
\renewcommand{\@fpheader}{\old@fpheader\hfill
MPP-2015-122}
\makeatother

\maketitle
\newpage

\hypersetup{pageanchor=true} 

\section{Introduction} 

S-duality, where two dual theories are related by an exactly marginal
deformation, is a ubiquitous phenomenon in theories with extended
supersymmetry. Perhaps the most elegant example is Montonen-Olive
duality in $\cN=4$ gauge theories, which exchanges electric and
magnetic fields and replaces the gauge group with its Langlands
dual. A subclass of these dualities admit a simple embedding in string
theory as the self-duality of the world-volume gauge theory on D3
branes induced by S-duality in type IIB string theory, as well as the
various dualities which arise when the D3 branes are coincident with
an O3 plane. The latter case was concisely described
by~\cite{Witten:1998xy}, where it was shown that the O3 plane type is
characterized in the gravity dual by a pair of discrete holonomies
(hereafter referred to as discrete
torsion) for the RR and NSNS two-form potentials, and that the
S-dualities between the corresponding gauge theories are explained by
the well-known action of type IIB S-duality on these potentials.

There are a number of ways that S-duality can be realized in
$\mathcal{N}=1$ gauge theories. It occurs in mass deformations of
S-dual theories with extended
supersymmetry~\cite{Leigh:1995ep,Argyres:1999xu} as well as in
$\mathcal{N}=1$ gaugings of Gaiotto
dualities~\cite{Benini:2009mz}. The distinct phenomenon of
universality, where two theories are related by renormalization group
flow, is ubiquitous in $\cN=1$ gauge
theories~\cite{Seiberg:1994pq,Seiberg:1995ac}, and bears some relation
to S-dualities in theories with extended
supersymmetry~\cite{Leigh:1995ep}.

Another manifestation of $\cN=1$ S-duality occurs in the chiral world-volume gauge theory on D3 branes probing a Calabi-Yau singularity~\cite{dualities1,Bianchi:2013gka,dualities2}, analogous the realization of Montonen-Olive duality on D3 branes in a flat background.\footnote{$\cN=0$ generalizations of these dualities have also been argued to exist~\cite{Uranga:1999ib,Sugimoto:2012rt,Hook:2013vza}.} As in the $\cN=4$ case, the dual theories correspond to different weakly-coupled ``cusps'' along the same fixed line, where $g_s \to 0$ in some dual description of the parent string theory. The addition of orientifold planes leads to non-zero beta functions near the cusp, hence the infrared fixed point is interacting and there is a non-trivial relationship between the dynamics at distinct dual cusps.\footnote{Even in the infrared of these orientifold gauge theories, the $g_s = 0$ cusp is a distinguished point along the fixed line, where (depending on the theory) there are enhanced global symmetries and/or free vector multiplets.}

These S-dualities are distinguished by their chiral, intrinsically $\cN=1$ nature\footnote{See also~\cite{Gaiotto:2015usa, Franco:2015jna, Hanany:2015pfa} for some recent work on chiral $\cN=1$ S-dualities arising from compactifications of 6d $(1,0)$ theories.} and their rich infrared dynamics near the cusp, with examples exhibiting confinement, chiral symmetry breaking, and a dynamically generated superpotential. Moreover, they provide an important cross-check on the physics of Calabi-Yau orientifold singularities --- where the S-dualities originate from the $\SL(2,\bZ)$ self-duality of type IIB string theory --- a role which will prove to be crucial in the present paper.

In~\cite{dualities2}, a systematic understanding of S-dualities arising in orientifolds of isolated orbifold singularities was obtained by repeating the analysis of discrete torsion in~\cite{Witten:1998xy} for these geometries. In this class of theories, the pattern of S-dualities predicted by string theory is reproduced perfectly in the dual gauge theories, passing all available checks. 

However, applying the same arguments to more general isolated
singularities immediately gives rise to a conundrum. The geometry
allows for additional discrete torsions, yet the number of known dual gauge theories is insufficient to fill out
all the possible choices of torsion, and moreover these gauge theories
do not fill out complete $\SL(2,\bZ)$ multiplets, a problem which was
recognized in~\cite{dualities1} for perhaps the simplest non-orbifold
example, the Calabi-Yau cone over the first del Pezzo surface
($dP_1$). The problem is worse still for the higher del Pezzo
singularities, where for most orientifold involutions no dual gauge
theory is known~\cite{Franco:2010jv}, despite the rapidly growing set
of discrete torsions. Other non-orbifold singularities exhibit similar
problems.

In this paper, we resolve this puzzle for the $dP_1$ singularity,
developing techniques which will generalize to other cases. We find
that there are more orientifold phases than previously known, with the
new phases precisely filling out the allowed discrete torsions. The
new phases consist of a quiver gauge theory coupled to an
intrinsically strongly-coupled sector, dual to a higher multiplicity
intersection of NS5 and D5 branes atop an O5 plane.
   We describe this strongly-coupled sector via deconfinement~\cite{Berkooz:1995km,Pouliot:1995me}, which produces a large collection of quiver gauge theories in the same universality class as it. This sector exhibits novel behavior, such baryons transforming in the spinor representation of an $\SO(2N)$ flavor symmetry which emerges accidentally in the gauge theory description.

These new sectors can be thought of as arising at the midpoint of a Seiberg duality, where the presence of an orientifold plane prevents a deformation to one or the other of the Seiberg-dual descriptions, trapping the theory at strong coupling. We argue that these sectors occur in certain orientifold phases of every toric non-orbifold singularity, including those with trivial discrete torsion. For most orientifold involutions of more complicated isolated singularities such as cones over the higher del Pezzo surfaces, every phase includes a strongly coupled sector, explaining the lack of known gauge theory duals.

Combining the strongly-coupled sector with known ingredients, we
obtain a list of theories describing the $dP_1$ orientifold which is
in complete agreement with the discrete torsion classification, and
which exhibits the S-dualities predicted by that classification,
including but generalizing the $dP_1$ duality found
in~\cite{dualities1}. We test these dualities by computing the
superconformal index for all the phases, including those involving the
strongly-coupled sector. We find that the theories predicted to be
S-dual have indices which match for a large number of low-lying states
(limited only by computational power), whereas theories not predicted
to be S-dual have distinct indices.

\medskip

Our paper is organized as follows. In~\S\ref{sec:delPezzo-review}, we
describe a class of del Pezzo orientifolds, including our main focus,
the $dP_1$ orientifold. We then summarize the known gauge theory duals
and classify the available discrete torsions, highlighting the
mismatch between the two. In~\S\ref{sec:dP1-tiling}, we determine the
discrete torsions of the known $dP_1$ orientifold gauge theories by
partially resolving to the $dP_0$ orientifold singularity studied
in~\cite{dualities1,dualities2} plus an O3 plane. 
We compare this torsion classification with a brane tiling construction of the orientifold,
suggesting that
the missing theories contain a sector dual to
a higher-multiplicity intersection of NS5 branes atop a stack of D5
branes, frozen in place by an O5 plane. We then comment on the
expected properties of these strongly-coupled ``quad CFT''
sectors. In~\S\ref{sec:deconfinement}, we obtain a gauge-theory
description of the quad-CFTs by engineering deconfinement in the brane
tiling, resolving the higher-multiplicity intersection. In the
process, extra information is uncovered (in the form of flavor branes
in the deconfined theory) which is crucial to reproducing the RR
discrete torsion in the gravity dual. We describe some novel
properties of the quad CFTs and compare their behavior with our
expectations, providing non-trivial evidence that our description is
correct. Finally, in~\S\ref{sec:dP1-S-duality} we apply the
deconfinement construction of the quad CFTs to describe the missing
phases of the $dP_1$ orientifold. We verify that the new theories
neatly fill out the missing discrete torsions, and provide highly
non-trivial evidence that the predicted S-dualities are realized in
these theories, based on comparing the superconformal indices between
the various phases. We conclude in~\S\ref{sec:conclusions}.

Appendix~\ref{app:delPezzo} contains a brief review of the geometry of del Pezzo singularities,
appendix~\ref{app:dP1-microscopics} gives a
microscopic description of the previously-known phases of $dP_1$ in terms of
exceptional collections, useful for determining the brane charges, and appendix~\ref{app:SCI} summarizes some of our explicit results for the superconformal indices of all phases of the $dP_1$ orientifold.

\section{Del Pezzo orientifolds}
\label{sec:delPezzo-review}

In this section, we construct a simple class of orientifold singularities and explore their properties.

\subsection{The orientifold geometry}
\label{subsec:delPezzo-geom}


The Calabi-Yau cone over the first del Pezzo surface ($dP_1$) is the toric variety described by the following gauged linear sigma model (GLSM, see~\cite{Witten:1993yc} for an introduction):
\begin{equation}
  \label{eq0:dP1-GLSM}
  \begin{array}{c|ccccc}
    & z_1 & z_2 & z_3 & z_4 & t\\
    \hline
    U(1)_a & 1 & 0 & 1 & 1 & -3\\
    U(1)_b & 0 & 1 & 0 & 1 & -2
  \end{array}
\end{equation}
subject to the D-term conditions
\be
  \label{eq0:dP1-GLSM-Dterms}
    |z_1|^2 + |z_3|^2 + |z_4|^2 - 3 |t|^2 = \xi_a \qquad,\qquad 
    |z_2|^2 + |z_4|^2 - 2 |t|^2 = \xi_b \; ,
\ee
where the Fayet-Iliopoulos (FI) parameters $\xi_a$ and $\xi_b$ control the resolutions of the singularity. For $\xi_a > \xi_b > 0$ the exceptional divisor $t=0$ is $dP_1$, and the O3/O7 involution $t \to -t$ describes an O7 plane wrapping this divisor.



We now generalize this construction to the other del Pezzo surfaces.
Let $K_\Sigma$ denote the canonical bundle of a smooth divisor $\Sigma$ embedded in a smooth Calabi-Yau threefold $Y$. We have~\cite{GriffithsHarris}
\be \label{eqn:canonicalembedding}
K_\Sigma = \left. K_Y\right|_{\Sigma} \otimes N_\Sigma
\ee
where $N_\Sigma$ is the normal bundle of $\Sigma \subset Y$, and $K_Y$ is the canonical bundle of $Y$. The Calabi-Yau condition implies that $K_Y$ is trivial, hence $K_\Sigma = N_\Sigma$. Thus, the local embedding of $\Sigma$ in $Y$ is isomorphic to the canonical bundle $K_\Sigma$ on $\Sigma$, and is independent of the global structure of $Y$.

Locally, we can always construct an involution $\sigma:Y\to Y$ by reflecting the fibers of $N_\Sigma$ ($\bC \to - \bC$). $\sigma$ is holomorphic by construction with an isolated fixed plane $\Sigma$, and corresponds to wrapping an O7 plane on $\Sigma$.\footnote{Unless $\Sigma$ is a K3 surface or $T^4$, $N_\Sigma$ must be non-trivial, hence tadpole cancellation requires the presence of D7 branes intersecting or wrapping $\Sigma$. To avoid ``flavor'' D7 branes, we assume that four D7 branes wrap $\Sigma$, cancelling the $C_0$ charge of the O7 plane.} As above, the local geometry is independent of the global structure of $Y$, though $\sigma$ may or may not extend to a globally defined holomophic involution, depending on $Y$.

Let $[\Sigma] \in H^{(1,1)}(Y)$ denote the divisor class of $\Sigma$. If $-[\Sigma]$ is a positive class then there is a corresponding K\"ahler modulus of $Y$ with $\Sigma$ as the exceptional divisor. In this case, the adjunction formula implies that $-K_\Sigma=-N_\Sigma$ is ample, hence $\Sigma$ is a del Pezzo surface: either $\bP^2$ blown up at $k$ generic points ($0\le k \le 8$) --- denoted $dP_k$ --- or the zeroth Hirzebruch surface, $\bF_0 \cong \bP^1 \times \bP^1$.

For each of these ten cases, we can construct an affine cone by setting the corresponding K\"ahler modulus of $Y$ to zero and reading off the local geometry $Y_p \subset Y$ near the singular point $p$. An orientifold $X_p = Y_p/\sigma$ can be constructed using the involution described above, where $p$ is an isolated fixed point of $\sigma$.
\footnote{The affine varieties $Y_p$ and $X_p$ depend only on the choice of del Pezzo surface --- and on the $2k-8$ complex structure moduli of $dP_k$ for $k\ge 5$ --- and not on the global details of $Y$. However, all ten singularities admit global embeddings, see e.g.~\cite{Malyshev:2007yb}, where $\sigma$ can be defined globally by $w\to -w$ in the notation of that paper.}
For completeness, concrete geometric realizations of all ten varieties and their orientifolds are presented in appendix~\ref{app:delPezzo}.

\subsection{Known gauge theory constructions}
\label{sec:known-constructions}

We now present the known gauge theory duals of D3 branes probing
the orientifold singularities described in the previous
subsection. These are only known for the $dP_0$, $dP_1$ and $\bF_0$
orientifold singularities, and we later argue that this list is
incomplete even in the $dP_1$ and $\bF_0$ cases.

We begin with the toric cases: $dP_k$ for $k\le 3$ and $\bF_0$. In these cases, we can use the dimer model technology developed in~\cite{Hanany:2005ve,Franco:2005rj} and the orientifold rules laid out in~\cite{Franco:2007ii} to derive the gauge theories corresponding to these orientifold singularities. We briefly review the required tools, referring the interested reader to the extensive literature (see e.g.~\cite{Kennaway:2007tq,Yamazaki:2008bt} and references therein) for further details.

\begin{figure}
  \centering
  \begin{subfigure}{0.38\textwidth}
    \centering
    \includegraphics{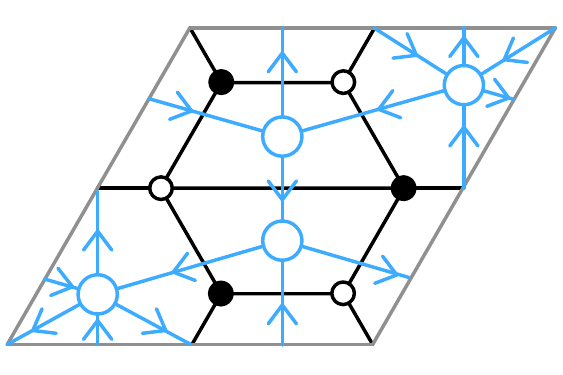}
    \caption{A dimer model and its dual quiver.}
    \label{sfig:tilingQuiver}
  \end{subfigure}
  \begin{subfigure}{0.29\textwidth}
    \centering
    \includegraphics{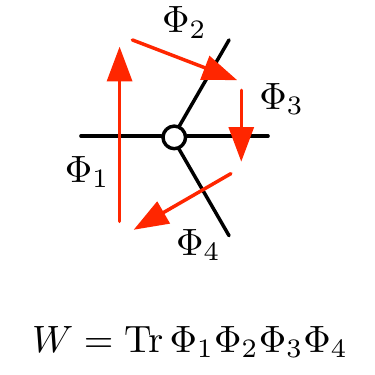}
    \caption{Reading off $W$.}
    \label{sfig:Wreadout}
  \end{subfigure}
  \begin{subfigure}{0.31\textwidth}
    \centering
    \includegraphics{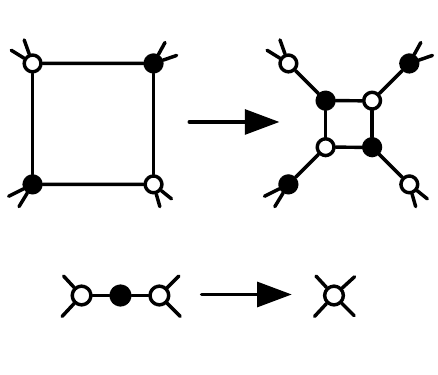}
    \caption{Seiberg duality in the dimer.}
    \label{sfig:urbanRenewal}
  \end{subfigure}
  \caption{\subref{sfig:tilingQuiver}~A dimer model (black) and its
    dual quiver (blue). This dimer corresponds to the $dP_1$
    singularity. \subref{sfig:Wreadout}~Each vertex in the dimer
    represents a
    superpotential term. \subref{sfig:urbanRenewal}~Seiberg duality (top)
    and integrating out (bottom) have a graphical interpretation in
    the dimer.\label{fig:tilings}}
\end{figure}

As illustrated in figure~\ref{fig:tilings}, a dimer model is a
bipartite graph embedded on the torus such that the surface of the
torus is divided into contractible faces by the edges of the
graph.\footnote{More technically, we are only interested in
  ``non-degenerate'' dimers, in the sense of~\cite{Ishii:2007}. Other
  ``consistency'' conditions are often imposed (see
  e.g.~\cite{Hanany:2005ss}), but we will not do so for reasons which
   become clear in~\S\ref{sec:deconfinement}.} A dimer model corresponds to a quiver
gauge theory by graph dualization: each face of the dimer model
corresponds to a node in the quiver, and each edge to an arrow in the
quiver, oriented so that the arrows circulate clockwise
(counterclockwise) around the white (black) vertices of the bipartite
graph. The vertices encode the superpotential: the loop in the
quiver surrounding each white (black) vertex corresponds to a term in
the superpotential with coefficient $+1$ ($-1$).

The mesonic moduli space of this quiver gauge theory is a toric
Calabi-Yau singularity which can be read off using the forward
algorithm~\cite{Feng:2000mi,Franco:2005rj}. Conversely, for each toric
Calabi-Yau singularity, one or more dimer models can be constructed
using the inverse algorithm~\cite{Feng:2000mi,Hanany:2005ss}. When
more than one dimer model is obtainable using the inverse algorithm,
these \emph{toric phases} are related to each other by Seiberg
duality~\cite{Beasley:2001zp}, which takes the form of ``urban
renewal'' in the dimer~\cite{Franco:2005rj}, see
figure~\ref{sfig:urbanRenewal}.

\begin{figure}
  \begin{center}
    \begin{subfigure}{0.4\textwidth}
      \centering
      \includegraphics[height=3.2cm]{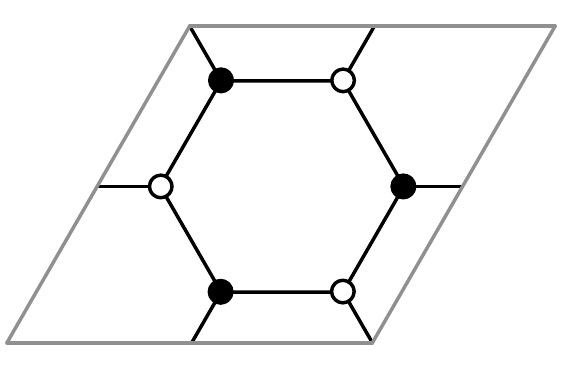}
      \caption{$dP_0$}
      \label{sfig:dP0tiling}
    \end{subfigure}
    \begin{subfigure}{0.28\textwidth}
      \centering
      \includegraphics[height=3.2cm]{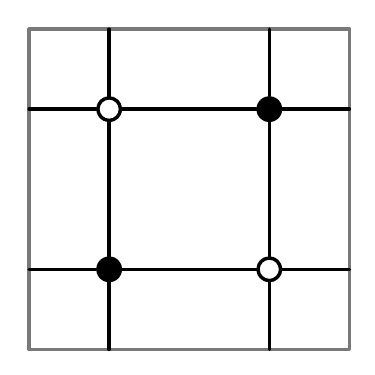}
      \caption{$\bF_0$ phase I}
      \label{sfig:F0tilingI}
    \end{subfigure}
    \begin{subfigure}{0.28\textwidth}
      \centering
      \includegraphics[height=3.2cm]{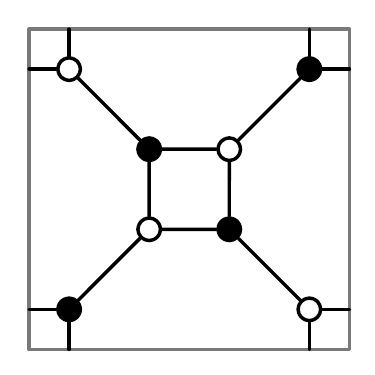}
      \caption{$\bF_0$ phase II}
      \label{sfig:F0tilingII}
    \end{subfigure}
  \end{center}
  \caption{Dimer models for \subref{sfig:dP0tiling}~$dP_0$, \subref{sfig:F0tilingI}~$\bF_0$ phase I, and \subref{sfig:F0tilingII}~$\bF_0$ phase II. The dimer for $dP_1$ is shown in figure~\ref{sfig:tilingQuiver}.}
\end{figure}

\begin{figure}
\begin{center}
  \begin{subfigure}{0.33\textwidth}
    \centering
    \includegraphics{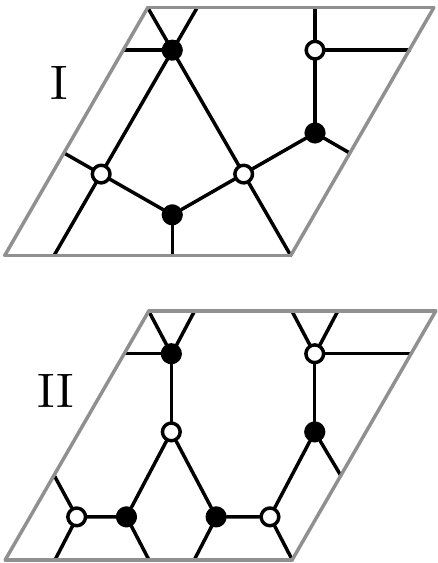}
    \caption{$dP_2$ phases}
    \label{sfig:dP2tilings}
  \end{subfigure}
  \begin{subfigure}{0.66\textwidth}
    \centering
    \includegraphics{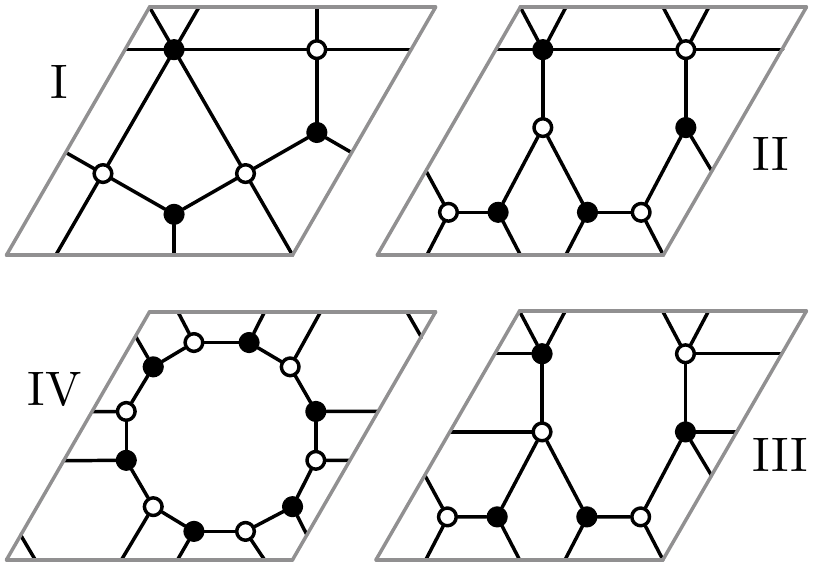}
    \caption{$dP_3$ phases}
    \label{sfig:dP3tilings}
    \end{subfigure}
  \end{center}
  \caption{Dimer models for \subref{sfig:dP2tilings}~$dP_2$ and \subref{sfig:dP3tilings}~$dP_3$.}
\end{figure}

The quiver gauge theories obtained by the inverse algorithm describe the infrared of the worldvolume gauge theory on D3 branes probing the toric singularity in question. $dP_0$ (figure~\ref{sfig:dP0tiling}) and $dP_1$ (figure~\ref{sfig:tilingQuiver}) each have a single toric phase, whereas $\bF_0$ (figures~\ref{sfig:F0tilingI}-\ref{sfig:F0tilingII}) and $dP_2$ (figure~\ref{sfig:dP2tilings}) have two phases each, and $dP_3$ (figure~\ref{sfig:dP3tilings}) has four phases.

\begin{figure}
  \begin{center}
    \begin{subfigure}[b]{0.40\textwidth}
      \centering
      \includegraphics{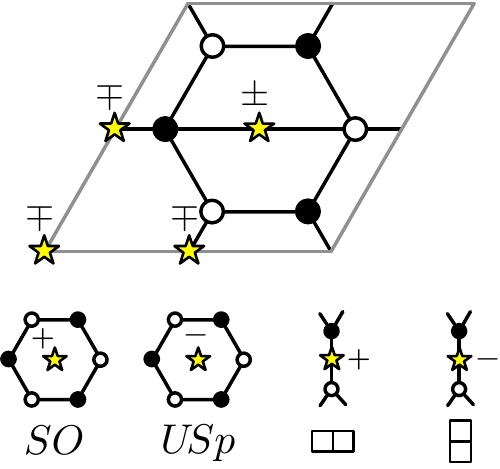}
      \caption{$dP_1$ orientifold and dictionary.}
      \label{sfig:dP1orientifoldtiling}
    \end{subfigure}
    \hfill
    \begin{subfigure}[b]{0.50\textwidth}
      \centering
      \includegraphics[width=\textwidth]{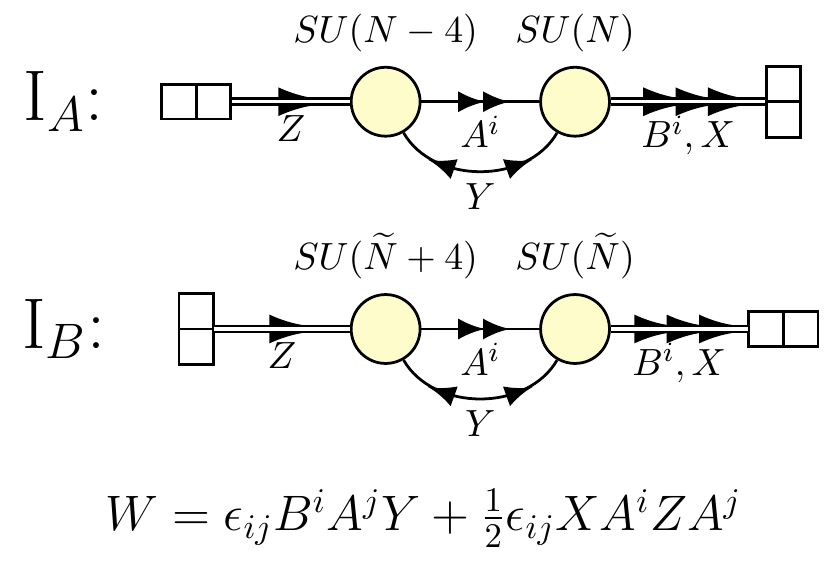}
      \caption{$dP_1$ quiver and superpotential.}
      \label{sfig:dP1orientifoldquiver}
    \end{subfigure}
  \end{center}
  \caption{\subref{sfig:dP1orientifoldtiling}~(top) Orientifolds of
    the $dP_1$ dimer model which preserve the $\SU(2)\times U(1)$
    isometry of $dP_1$. (bottom) The orientifold gauge theory can be
    read off from the dimer using this
    dictionary. \subref{sfig:dP1orientifoldquiver}~Quivers for the
    $dP_1$ orientifolds, where $A$ ($B$) corresponds to the
    upper (lower) choice of fixed point signs
    in~\subref{sfig:dP1orientifoldtiling}. Our notation for quivers is
    explained in figure~\ref{fig:quiverdictionary}.}
  \label{fig:dP1-classical-phases}
\end{figure}

Following~\cite{Franco:2007ii}, we can construct the worldvolume gauge theory on D3 branes probing an orientifold of a toric singularity by orientifolding the dimer model itself. The geometric involution considered above preserves all the isometries of the toric singularity, so that the quotient space is also toric. Geometric involutions of this type correspond to involutions of the dimer model with four isolated fixed points mapping white nodes to black nodes and vice versa. Each fixed point has an associated sign, or ``T-parity''~\cite{Imamura:2008fd}, subject to the requirement that the product of all four T-parities is positive (negative) when the number of white vertices in the dimer is even (odd). A (generalized) quiver gauge theory can be read off as before, where now each face and its image corresponds to a single $\SU(n)$ gauge group, each edge and its image to a single matter multiplet, and each vertex and its image to a single superpotential term.\footnote{The sign of the coefficient is no longer important.} When a face is mapped to itself, it corresponds to an $\SO$ ($\Sp$) gauge group if the enclosed fixed point is positive (negative), whereas an edge mapped to itself corresponds to symmetric (antisymmetric) tensor matter if the fixed point it crosses is positive (negative), see figure~\ref{sfig:dP1orientifoldtiling}.

\begin{figure}
  \centering
  \includegraphics[width=\textwidth]{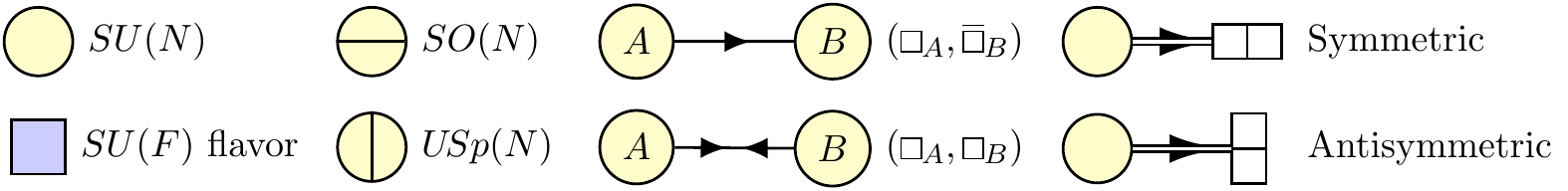}
  \caption{Summary of our quiver notation. Reversing
    the directions of the arrows corresponds to taking the charge conjugate of the
    representation, and multiple arrowheads in the same direction indicate multiple fields in the same representation.}
  \label{fig:quiverdictionary}
\end{figure}

We consider $dP_1$ first.\footnote{The $dP_1$ and $\bF_0$
  orientifold gauge theories derived below were first written down
  in~\cite{dualities1}. The $dP_0$ orientifold theories have been
  known much
  longer~\cite{Angelantonj:1996uy,Lykken:1997ub,Kakushadze:1998tr}.}
The corresponding dimer model is shown in
figure~\ref{sfig:tilingQuiver}. Up to the choice of fixed point signs,
a single fixed-point involution is possible. The relative signs of the
fixed points can be fixed by the requirement that the $SU(2)\times
U(1)$ isometry of $dP_1$ is unbroken, giving two distinct
orientifolds, which we call \IA{} and \IB. The orientifolded dimer model
 is shown in figure~\ref{sfig:dP1orientifoldtiling}, and the
resulting quivers and superpotential are displayed in
figure~\ref{sfig:dP1orientifoldquiver} using the generalized quiver
notation defined in figure~\ref{fig:quiverdictionary}. The $dP_0$ orientifold studied in~\cite{dualities1,dualities2} can be obtained from here by Higgsing the tensor $Z$.

\begin{figure}
  \begin{center}
    \begin{subfigure}[b]{0.48\textwidth}
      \centering
      \includegraphics{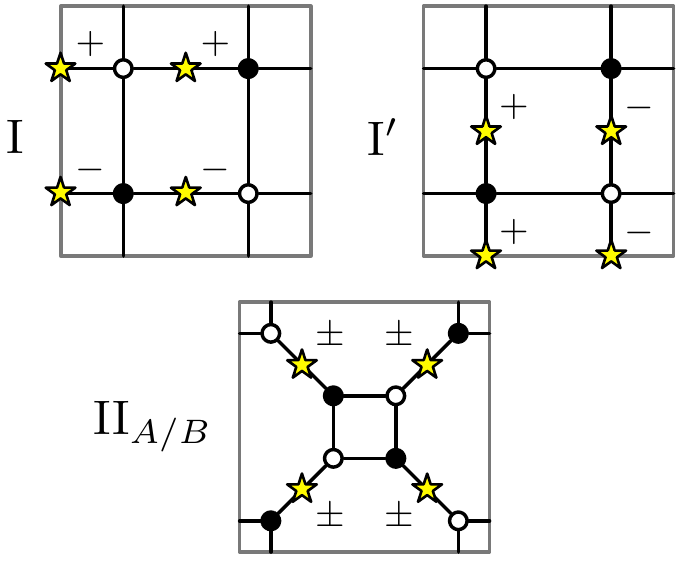}
      \vspace{0.8cm}
      \caption{$\bF_0$ orientifolds.}
      \label{sfig:F0orientifoldtilings}
    \end{subfigure}
    \hfill
    \begin{subfigure}[b]{0.48\textwidth}
      \centering
      \includegraphics[width=\textwidth]{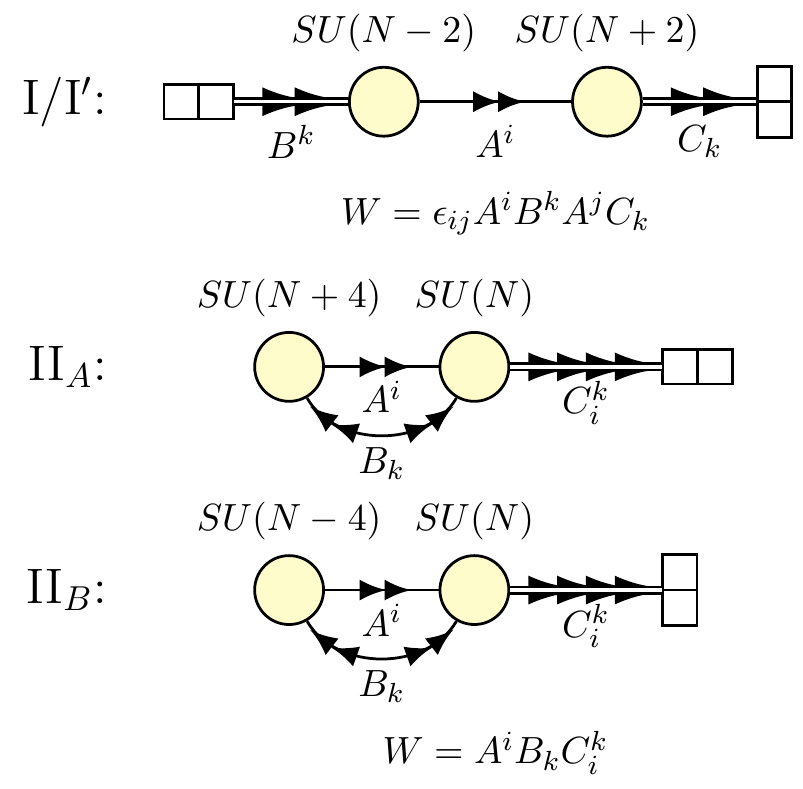}
      \caption{Orientifold quivers and superpotentials.}
      \label{sfig:F0orientifoldquivers}
    \end{subfigure}
  \end{center}
  \caption{\subref{sfig:F0orientifoldtilings}~Orientifolds of $\bF_0$
    phase I and phase II, where II$_A$ (II$_B$) corresponds to the
    upper (lower) choice of fixed point signs. The orientifolds I and
    I$'$ are related by a rotation of the dimer model, which
    generates an outer automorphism of the global symmetry group of
    the corresponding gauge
    theory. \subref{sfig:F0orientifoldquivers}~Quivers for the $\bF_0$
    orientifolds. The orientifolds I and I$'$ give identical gauge
    theories up to the labeling of the global symmetries. The
    orientifolds II$_A$ and II$_B$ are Seiberg duals.}
\end{figure}

For $\bF_0$, there are two toric phases to consider. In phase I, two different involutions are possible, but they are related to each other by rotating the dimer model by $90^\circ$, exchanging the two $\bP^1$ factors. As above, the relative signs of the fixed points can be fixed by requiring that the $\SU(2)\times\SU(2)$ isometry of $\bF_0 \cong \bP^1 \times \bP^1$ is unbroken, whereas the two remaining choices are related by translating the torus by half a period, hence they are equivalent. In phase II, only one involution is possible, but the two sign choices consistent with the $\SU(2)\times\SU(2)$ isometry are not equivalent, giving two distinct orientifold theories, II$_A$ and II$_B$. However, unlike phases I$_A$ and I$_B$ of $dP_1$ or $dP_0$, II$_A$ and II$_B$ are related by Seiberg duality, and therefore lie in the same universality class. The orientifolded dimers are shown in figure~\ref{sfig:F0orientifoldtilings} and the resulting quiverfolds and superpotentials in figure~\ref{sfig:F0orientifoldquivers}.

The higher del Pezzos $dP_2$ and $dP_3$ are strikingly different. In
these cases, no fixed-point involutions are possible in any toric
phase~\cite{Franco:2010jv}! For instance, $dP_2$ phase II and $dP_3$ phases III and IV each have a unique largest face with either $8$ or $12$ sides. The involution must map this face to itself, but this is incompatible with the rule that white vertices are mapped to black vertices, so no fixed-point involution exists for these phases. Likewise, $dP_2$ phase I and $dP_3$ phase I each have a black vertex with valence $5$ or $6$ and no white vertices with valence greater than four, once again violating the rule that white vertices are mapped to black vertices. By contrast, $dP_3$ phase II admits a unique involution that exchanges white and black vertices, but this involution --- a horizontal translation by half a period --- has no fixed points and breaks the toric isometries, hence it cannot correspond to one of the orientifolds considered above. In fact, as shown in~\S\ref{sec:brane-tiling}, the absence of fixed-point involutions for these theories generalizes to a broad class of isolated toric singularities, and is one of the central mysteries that we aim to address in this work.

\medskip

We comment briefly on the non-toric cases, $dP_k$ for $4\le k\le 8$. In these cases, the toric technology of~\cite{Franco:2005rj,Franco:2007ii} is unavailable, though quivers and superpotentials for the parent gauge theory are known~\cite{Wijnholt:2002qz}, and --- following~\cite{Wijnholt:2007vn} (see also~\cite{dualities1}) --- we can in principle orientifold these quivers to obtain the orientifold gauge theories of interest.\footnote{One can show that in the toric case the approaches of~\cite{Wijnholt:2007vn} and~\cite{Franco:2007ii} are equivalent~\cite{BJHorientifoldnote}.} However, since there are an infinite number of Seiberg-dual quivers to orientifold, a systematic approach may not be possible.\footnote{It may be that only a finite number of quivers admit an involution. It would be interesting to explore this further.} Instead, we consider a few examples to illustrate the difficulties which arise.

The $dP_k$ for $k\ge5$ have $2k-8$ complex structure moduli, none of which are projected out by the orientifold considered above, and we expect that the correct gauge theory dual will have corresponding exactly marginal deformations. (These appear as superpotential deformations in the unorientifolded quivers of~\cite{Wijnholt:2002qz}.) For $dP_5$ and $dP_6$ there are points in complex structure moduli space where the del Pezzo surface degenerates and the del Pezzo singularity becomes a non-isolated toric singularity. For instance, the cubic $X Y Z - W^3 = 0$ --- equivalent to the orbifold singularity $\bC^3/(\bZ_3\times \bZ_3)$ via the embedding $X = x^3$, $Y= y^3$, $Z=z^3$, $W=x y z$ --- is a degeneration of the $dP_6$ singularity and the complete intersection of quadrics $Z^2 = X Y = U V$ --- equivalent to a $\bZ_2\times\bZ_2$ orbifold of the conifold via the embedding $X = x^2$, $Y=y^2$, $U=u^2$, $V=v^2$ and $Z = x y u v$ --- is a degeneration of the $dP_5$ singularity. There is also a degeneration of the $dP_5$ singularity which is a partial resolution of $\bC^3/(\bZ_3\times \bZ_3)$.

\begin{figure}
  \begin{center}
    \begin{subfigure}{0.32\textwidth}
      \centering
      \includegraphics{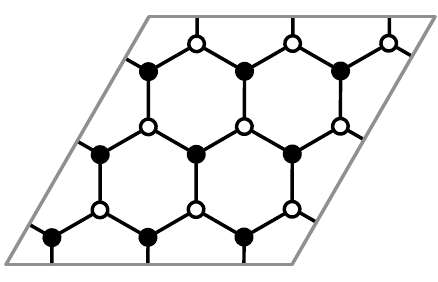}
      \caption{Dimer for $\bC^3/(\bZ_3\times \bZ_3)$.}
      \label{sfig:C3Z3Z3tiling}
    \end{subfigure}
    \begin{subfigure}{0.32\textwidth}
      \centering
      \includegraphics{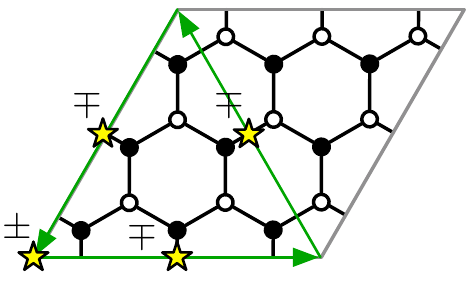}
      \caption{$\bC^3/(\bZ_3\times \bZ_3)$ orientifold.}
      \label{sfig:C3Z3Z3orientifold}
    \end{subfigure}
    \begin{subfigure}{0.32\textwidth}
      \centering
      \includegraphics{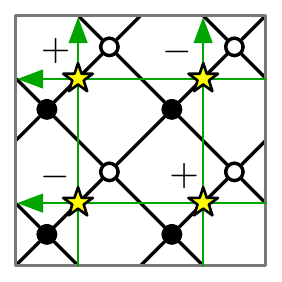}
      \caption{$\cC/(\bZ_2\times \bZ_2)$ orientifold.}
      \label{sfig:ConZ2Z2orientifold}
    \end{subfigure}
  \end{center}
  \caption{\subref{sfig:C3Z3Z3tiling}~The dimer model for
    $\bC^3/\bZ_3\times\bZ_3$, which is a toric degeneration of the
    $dP_6$ singularity. \subref{sfig:C3Z3Z3orientifold}~An orientifold
    of $\bC^3/\bZ_3\times\bZ_3$ with fixed-point signs chosen to
    correspond to the del Pezzo orientifolds considered in the
    text. The green arrows represent cubic mesons which are projected
    out in the orientifold theory, obstructing the superpotential
    deformation required for generic complex structure
    moduli. \subref{sfig:ConZ2Z2orientifold}~One toric phase of the
    $\bZ_2\times \bZ_2$ orbifold of the conifold (which is a toric
    degeneration of the $dP_5$ singularity) and its orientifold. The
    green arrows represent quartic mesons which are projected out in
    the orientifold theory, obstructing deformation of the
    superpotential as above. The other toric phases are similar.}
\end{figure}

Since there are points in moduli space where the Calabi-Yau is toric, it may be possible to construct the gauge theory dual of the orientifold at these points using dimer models, and then to generalize the result by deforming it. For instance, the dimer model for $\bC^3/(\bZ_3 \times \bZ_3)$ is shown in figure~\ref{sfig:C3Z3Z3tiling}. The entire complex structure moduli space of $dP_6$ can be reproduced in the parent theory with a general superpotential, whereas the toric superpotential encoded by the dimer corresponds to one point in that moduli space (the degeneration given above). There is a unique involution of the dimer, with two possible choices of fixed-point signs which correspond to the desired geometric involution,\footnote{Up to an overall sign, the fixed point signs are determined by the meson sign rules of~\cite{Franco:2007ii}.} see figure~\ref{sfig:C3Z3Z3orientifold}.

To deform these candidate orientifold theories, we modify the superpotential as in the parent theory. At a generic point in the moduli space of the parent theory, all 27 possible cubic superpotential terms are present~\cite{Wijnholt:2002qz}. However, not all of these terms can be switched on in the orientifold theory described above, as some mesons vanish identically, see figure~\ref{sfig:C3Z3Z3orientifold}. Thus, these gauge theories fail to describe the $dP_6$ singularity at a generic point in moduli space.\footnote{It would be interesting to understand the string-theoretic interpretation of these orientifold theories, if any.}

A similar story applies to the toric degenerations of $dP_5$, see e.g.\ figure~\ref{sfig:ConZ2Z2orientifold}. While this does not rule out the possibility of obtaining the correct gauge theory dual by orientifolding one of the many Seiberg dual quivers describing these singularities, it does imply that this cannot be done for the ``toric'' phases of $dP_5$ and $dP_6$ (phases with equal rank nodes regardless of the superpotential), and the failure of the approach of~\cite{Franco:2007ii} for $dP_2$ and $dP_3$ already suggests that the solution to these problems lies elsewhere.

\subsection{Discrete torsion in del Pezzo orientifolds}
\label{sec:dP-torsion}

So far, we have made no distinction between different supersymmetric orientifolds with the same geometric involution. This is too naive --- even in a flat background --- because orientifold planes carry RR charges, and typically come in at least two variants (O$p^+$ and O$p^-$ planes). In a curved background, orientifold planes may carry additional charges, and more than one orientifold plane can be hidden at a singularity, potentially leading to a large number of distinct orientifolds with the same geometric involution.

Fortunately, the AdS/CFT correspondence allows us to translate the difficult problem of classifying BPS orientifold planes at a singularity into the much easier one of classifying closed-string charges in the near-horizon geometry of the singularity, which is smooth if the singularity is isolated. 

The idea, as in~\cite{Witten:1998xy,dualities1,dualities2}, is that distinct BPS O3/O7 orientifolds are distinguished by their RR and NSNS three-form charges, classified by $H^3(X_5, \tbZ)$ where $X_5 = Y_5/\sigma$ is the orientifolded horizon and $\tbZ$ denotes local cofficients, twisted by $\sigma$ due to the projection $\sigma^\ast F_3 = - F_3$,  $\sigma^\ast H_3 = - H_3$.\footnote{A cohomology classification is sufficient for our present purposes~\cite{Evslin:2006cj}, and is able to reproduce the observed S-duality structure of the orientifold theories. A K-theoretic approach is difficult due in part to its unequal treatment of the RR and NSNS two-forms.}

\begin{figure}
\begin{center}
\includegraphics{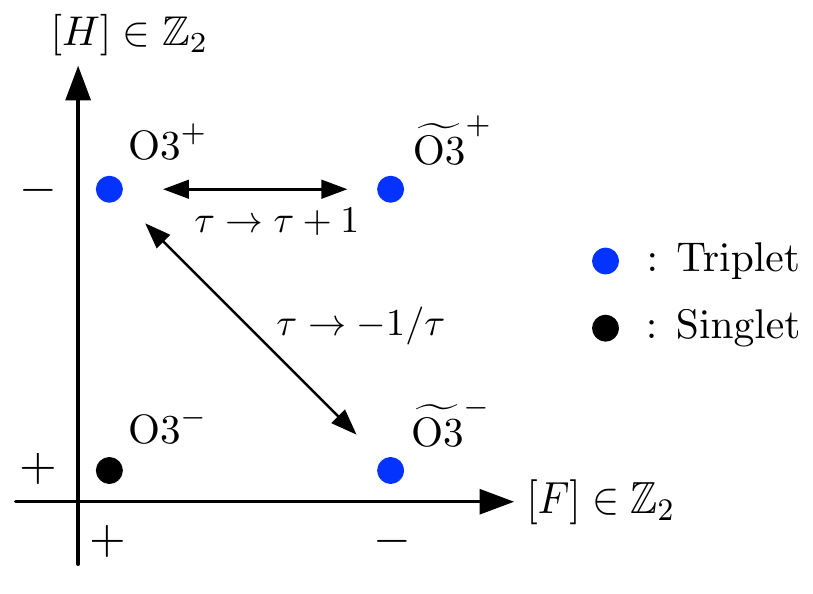}
\end{center}
\caption{O3 planes are classified by the RR and NSNS torsion classes $[F]$ and $[H]$ in $H^3(S^5/\bZ_2, \tbZ) \cong \bZ_2$. The known action of $\SL(2,\bZ)$ on the RR and NSNS three-forms $F_3$ and $H_3$ dictates its action on the O3 plane variants.\label{fig:O3planes}}
\end{figure}

For instance, O3 planes in a flat background are classified by $[F], [H] \in \bZ_2$~\cite{Witten:1998xy}, where $H^3(S^5/\bZ_2, \tbZ) \cong \bZ_2$ for the O3 involution $z^i \to - z^i$ of $\bC^3$, see figure~\ref{fig:O3planes}. In this case, the NSNS torsion $[H]$ is trivial (nontrivial) for O3$^-$ (O3$^+$) planes, whereas the RR torsion $[F]$ is trivial (nontrivial) for the O3$^-$ ($\widetilde{\mathrm O3}^- = \mathrm{O3}^-+\frac{1}{2} \mathrm{\, D3}$) variants, and likewise for the variants O3$^+$ and $\widetilde{\mathrm O3}^+$. (The latter are perturbatively equivalent, being related by $\tau \to \tau+1$ in the $\SL(2,\bZ)$ self-duality group of type IIB string theory, but they differ in their nonperturbative spectra in the presence of D3 branes.) Orientifold planes of other codimensions can be classified in an analogous manner, see e.g.~\cite{Hanany:2000fq}.

In the remainder of this section, we compute the twisted homology groups $H_i(X_5, \tbZ)$ for the del Pezzo orientifolds considered above. This determines $H^3(X_5, \tbZ)$ by Poincare duality, and will help to clarify where the gaps in the collection of known field theory duals enumerated in the previous section lie.

Our calculation is based upon the long exact sequence~\cite{AT}:
\begin{equation} \label{eqn:longexactTwisted}
\ldots \longrightarrow H_i(X,\tilde{\bZ}) \longrightarrow H_i(Y,\bZ) \;\overset{p_\ast^i}{\longrightarrow}\; H_i(X,\bZ) \longrightarrow H_{i-1}(X,\tilde{\bZ}) \longrightarrow \ldots
\end{equation}
where $X=Y/\sigma$, and $p_\ast^i$ is induced by the projection
$p: Y \to X$. In our case, it will turn out that $p_\ast^i$ is
injective for every $i$, hence the long exact sequence breaks into
short exact sequences
\begin{equation} \label{eqn:shortexactTwisted}
0 \longrightarrow H_i(Y,\bZ) \;\overset{p_\ast^i}{\longrightarrow}\; H_i(X,\bZ) \longrightarrow H_{i-1}(X,\tilde{\bZ}) \longrightarrow 0
\end{equation}
and we can compute the twisted homology groups given the homology
groups of $X$ and $Y$ and the maps $p_\ast^i$.

We begin by describing the (co)homology of $dP_k$ (see e.g.~\cite{Liu:1998dra}), which is a complex manifold with the Hodge numbers $h^{0,0}=h^{2,2}=1, h^{1,1}=k+1$ and $h^{i,j}=0$ for $i\ne j$, formed by blowing up $\bP^2$ at $k$ generic points. There are $k+1$ two cycles, which we denote by $H$ for the hyperplane section of $\bP^2$ and $E_i$, $i=1,\ldots,k$, for the exceptional divisors of the $k$ blow-ups. The canonical class is $[K]=-3 H+\sum_i E_i$. The intersection form is $H\cdot H =1$, $E_i \cdot E_j = -\delta_{i j}$ and $H\cdot E_i = 0$.

The upstairs horizon $Y_5$ is the principal $U(1)$ bundle associated to the normal bundle $N$ of the del Pezzo embedded in the Calabi-Yau threefold, where $N \cong K$ by~(\ref{eqn:canonicalembedding}). A two-cycle $A$ on the del Pezzo lifts to a two-cycle of $Y_5$ iff the $U(1)$ bundle is trivial restricted to $A$ (and hence admits a global section). This occurs iff $\int_A [N] = \int_{dP_k} A \wedge [N] = A \cdot [N] = 0$. Thus,
\be
H_2(Y_5,\bZ) \cong \{A \in H_2(dP_k,\bZ) \,|\, A\cdot [K] = 0\} \cong \bZ^k
\ee
since the resulting two-cycle is homologically trivial iff $A$ is. By contrast, any two cycle $B$ on the del Pezzo lifts to a three-cycle on $Y_5$ (the $U(1)$ bundle over $B$), but the resulting three-cycle may be trivial even for non-trivial $B$. This can only happen if the intersection form on $Y_5$ vanishes for this three-cycle, hence if $B \propto [N]$ (since the intersection form on $dP_k$ is non-degenerate). Given that $\pi_1(Y_5) = \bZ_3$ for $k=0$ (generated by a loop once around the fiber) and $\pi_1(Y_5)=0$ for $k>0$, compatibility with Poincar\'e duality and the universal coefficient theorem implies that the three cycle is trivial iff $B = n [N]$ for $n\in \bZ$, hence
\be
H_3(Y_5,\bZ) \cong H_2(dP_k,\bZ)/\{n [K] \,|\, n \in \bZ \} \cong \begin{cases} \bZ^k, & k>0\,, \\ \bZ_3, & k=0\,. \end{cases}
\ee
We conclude that
\be
H_\bullet(Y_5,\bZ) \cong \begin{cases} \{\bZ,0,\bZ^k,\bZ^k,0,\bZ \}\,, & k>0\,, \\  \{\bZ,\bZ_3,0,\bZ_3,0,\bZ \}\,, & k =0\,, \end{cases}
\ee
which is well-known (see e.g.~\cite{Liu:1998dra}).

The downstairs horizon $X_5$ is obtained by orbifolding the fiber $U(1) \to U(1)/\bZ_2$. Equivalently, we replace $N \to 2 K$. Following the same steps as before, we find:
\be
H_\bullet(X_5,\bZ) \cong \begin{cases} \{\bZ,\bZ_2,\bZ^k,\bZ^k\oplus \bZ_2,0,\bZ \}\,, & k>0\,, \\  \{\bZ,\bZ_6,0,\bZ_6,0,\bZ \}\,, & k =0\,, \end{cases}
\ee
wbere the generators of $H_2(X_5)$ are the same as before, $H_3(X_5,\bZ) \cong H_2(dP_k,\bZ)/\{2 n [K] \,|\, n \in \bZ \}$, and $H_1(X_5) \cong \pi_1(X_5)$ is generated by a loop once around the orientifolded fiber.

With an explicit description of the generators of $H_i(X_5)$ and $H_i(Y_5)$, the projection map $p_\ast^i$ is readily described. We find that $p_\ast^{2 j}$ is an isomorphism and $p_\ast^{2 j+1}$ maps $\bZ \to 2 \bZ$ and $\bZ_3 \to \bZ_3 \subset \bZ_6$. Thus, $p_\ast^i$ is injective and using~(\ref{eqn:shortexactTwisted}), we obtain:
\be \label{eqn:dPktwistedhom}
H_\bullet(X_5,\tbZ) \cong \{\bZ_2, 0, \bZ_2^{k+1}, 0, \bZ_2, 0\}\,.
\ee
 The calculation for $\bF_0 \cong \bP^1 \times \bP^1$ is similar. In this case, $h^{1,1}=2$, generated by the two hyperplane classes $H_1, H_2$ with the intersection form $H_i \cdot H_j = 1-\delta_{i j}$ and $[K] = -2 H_1 - 2 H_2$. We find
 \be
 H_\bullet(Y_5,\bZ) \cong \{\bZ, \bZ_2, \bZ, \bZ\oplus \bZ_2, 0, \bZ \} \;\;,\;\; H_\bullet(X_5,\bZ) \cong \{\bZ, \bZ_4, \bZ, \bZ\oplus \bZ_4, 0, \bZ \}\,,
 \ee
 where the twisted homology groups are the same as those of $dP_1$.

\subsection{Counting orientifolds of del Pezzo singularities} \label{subsec:countingorientifolds}

\begin{figure}
  \centering
  \includegraphics[width=0.7\textwidth]{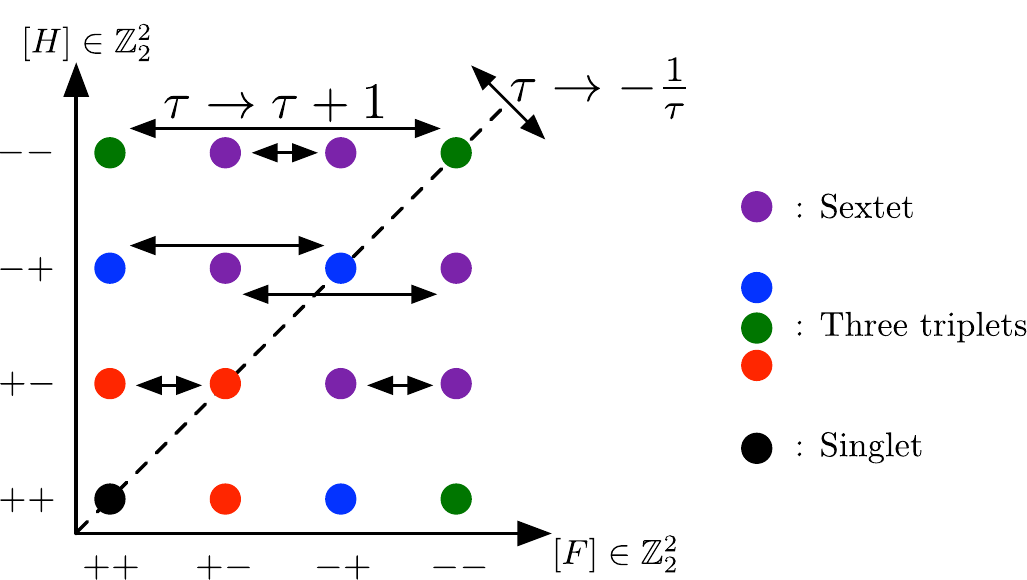}
  \caption{Choices of discrete torsion for the 
    $dP_1$ orientifold considered in the text. 
    The 16 torsions arrange themselves into
    five irreducible representations --- one singlet, three triplets, and one sextet --- under the $\SL(2,\bZ)$ symmetry of type IIB string theory. The generator $\tau\to\tau+1$
    maps the $g_s \to 0$ cusps to themselves, hence the torsions connected by horizontal arrows correspond to the same cusp at different values of axion $C_0 = \re \tau$.
    }
  \label{fig:dP1theories}
\end{figure}

With the twisted homology groups~(\ref{eqn:dPktwistedhom}) in hand, we can now classify the possible orientifolds of del Pezzo singularities with the geometric involution considered in~\S\ref{subsec:delPezzo-geom}. For $dP_k$, these are classified by $[F], [H] \in H^3(X_5, \tbZ) \cong \bZ_2^{k+1}$, for a total of $2^{2k+2}$ choices of discrete torsion. We divide these into five classes with $([F],[H]) = (1,1)$, $([F],[H]) = (\alpha,1)$, $([F],[H]) = (1,\alpha)$, $([F],[H]) = (\alpha,\alpha)$, and $([F],[H]) = (\alpha,\beta)$, where $\alpha$ and $\beta$ are distinct non-trivial elements of $\bZ_2^{k+1}$. The first class has a single member, the lone $\SL(2,\bZ)$ singlet. The second, third, and fourth classes each have $2^{k+1}-1$ members, which together form $2^{k+1}-1$ $\SL(2,\bZ)$ triplets, with orbits of the form $(\alpha,1), (1,\alpha), (\alpha,\alpha)$. The last class has $2 (2^{k+1} - 1) (2^k -1)$ members, which form $\frac{(2^{k+1} - 1) (2^k -1)}{3}$ $\SL(2,\bZ)$ sextets, with orbits of the form $(\alpha,\beta), (\beta,\alpha), (\gamma,\beta), (\beta, \gamma), (\gamma, \alpha), (\alpha, \gamma)$ where $\gamma = \alpha\beta$. The $\SL(2,\bZ)$ orbits of the various multiplets are illustrated in figure~\ref{fig:dP1theories} for the case of $dP_1$.

Not all of the $2^{2k+2}$ choices of discrete torsion will lead to
distinct $g_s \to 0$ limits of the string theory background. Torsions
related by $[F] \to [F] + [H]$ differ by a shift of the type IIB axion
$\tau \to \tau+1$, and correspond to the same ``cusp'' on the modular
curve parameterized by $\tau$.  The number of distinct cusps depends
on the $\SL(2,\bZ)$ multiplet: for singlets, there is only one,
whereas for triplets there are two, and for sextets there are
three. Adding up the results of the previous paragraph, we expect
$2^k (2^{k+1}+1)$ cusps for the $dP_k$ orientifold.

As in the well-understood $\mathcal{N}=4$ and orbifold
cases~\cite{dualities1,dualities2}, we expect each cusp to correspond
to a distinct $\mathcal{N}=1$ gauge theory, some of which are related
by S-duality. However, as shown in table~\ref{tab:torsionclass}, the
number of cusps grows rapidly with $k$, whereas the number of known
field theory duals is extremely limited, as discussed
in~\S\ref{sec:known-constructions}. For $dP_0$, the gauge theory duals
are in precise agreement with the discrete torsion
classification~\cite{dualities1,dualities2}. Conversely, for $dP_k$,
$k>1$, no CFT duals are known, despite the large and rapidly growing
number of cusps! Moreover, for $dP_1$, only four distinct theories are
known, two of which are S-dual, despite the expectation of five
different $\SL(2,\bZ)$ orbits and ten different cusps!\footnote{The
  situation for $\bF_0$ is similar, except that we need to account for
  the $\bZ_2$ discrete symmetry which exchanges the two $\bP^1$ factors
  and likewise the two generators of the twisted homology group
  $H^3(\frac{S^3 \times S^2}{\bZ_4}, \tbZ) \cong \bZ_2^2$. Up to
  $\bZ_2$-induced isomorphisms, we expect four different $\SL(2,\bZ)$
  multiplets and seven different cusps, in contrast to the six known
  gauge theories, which occupy four different universality classes,
  two of which are S-dual~\cite{dualities1}.}
\begin{table}
\begin{center}
\begin{tabular}{c|ccccc}
 & Singlets & Triplets & Sextets & $\SL(2,\bZ)$ multiplets & Cusps \\
\hline
$dP_0$ & 1 & 1 & 0 & 2 & 3\\
$dP_1, \bF_0$ & 1 & 3 & 1 & 5 & 10\\
$dP_2$ & 1 & 7 & 7 & 15 & 36\\
$dP_3$ & 1 & 15 & 35 & 51 & 136\\
$dP_4$ & 1 & 31 & 155 & 187 & 528\\
$dP_5$ & 1 & 63 & 651 & 715 & 2080\\
$dP_6$ & 1 & 127 & 2667 & 2795 & 8256\\
$dP_7$ & 1 & 255 & 10795 & 11051 & 32896 \\
$dP_8$ & 1 & 511 & 43435 & 43947 & 131328
\end{tabular}
\caption{The discrete torsion classification of orientifolds of del Pezzo singularities. The $\bF_0$ singularity and many of the $dP_k$, $k>1$ singularities have discrete symmetries which relate \emph{a priori} distinct choices of discrete torsion. The number of non-isomorphic cusps is smaller than shown here in these cases, but still very large.\label{tab:torsionclass}}
\end{center}
\end{table}

There are three ways in which this paradox might be resolved. Either
(i) the discrete torsion classification is incorrect, (ii) there are
additional gauge theory duals which have not been found yet, or (iii)
the missing cusps are not describable via perturbative gauge
theories. In the remainder of this paper, we argue through a careful
analysis of the $dP_1$ singularity that the latter possibility is the
correct one. Fortunately, using a deconfinement-like trick we will
nonetheless be able to construct perturbative gauge theories in the
same universality class as the theories at the cusps, allowing us to
describe some of their most important properties.

\section{Phases of the \alt{$dP_1$}{dP1} orientifold}
\label{sec:dP1-tiling}

\begin{figure}
  \centering
  \begin{subfigure}[b]{0.26\textwidth}
    \centering
    \includegraphics{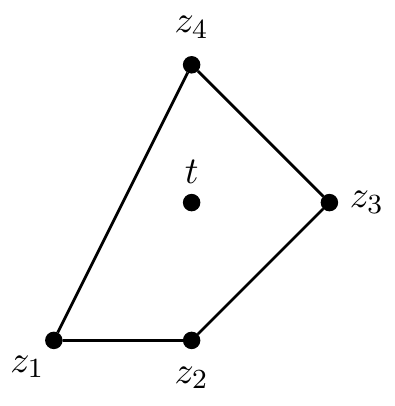}
    \vspace{0.3cm}
    \caption{Toric diagram.}
    \label{sfig:dP1-toric}
  \end{subfigure}
  \begin{subfigure}[b]{0.26\textwidth}
    \centering
    \includegraphics{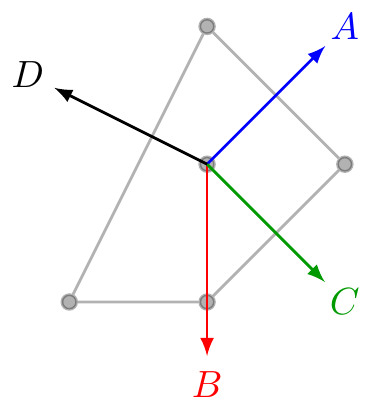}
    \vspace{0.35cm}
    \caption{Dual web diagram.}
    \label{sfig:dP1-web}
  \end{subfigure}
  \begin{subfigure}[b]{0.46\textwidth}
  \centering
  \includegraphics[width=2.5in]{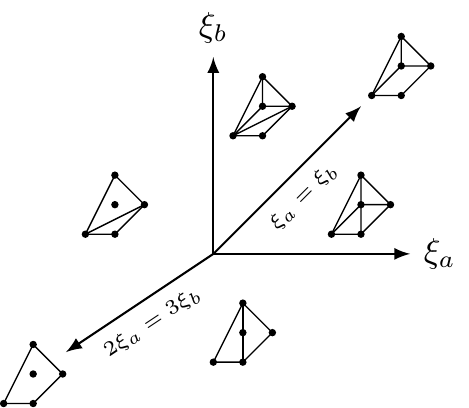}
  \caption{Partial resolutions.}
  \label{sfig:dP1-resolutions}
  \end{subfigure}
   
  \caption{\subref{sfig:dP1-toric} Toric diagram associated to the
    complex cone over $dP_1$. The labels refer to the corresponding
    fields in the GLSM~\eqref{eq:dP1-GLSM}. 
   \subref{sfig:dP1-web}
    The dual web diagram (or $(p,q)$ diagram) for the singular
    geometry. The external legs are labeled and colored for future reference.
     \subref{sfig:dP1-resolutions}~Partial resolutions of the $dP_1$ singularity are controlled by the FI parameters $\xi_a$ and $\xi_b$.}
  \label{fig:dP1-resolved-toric}
\end{figure}

We now specialize to the $dP_1$ orientifold 
described in~\S\ref{sec:delPezzo-review}. For reference, we reproduce the GLSM for the $dP_1$ singularity (\ref{eq0:dP1-GLSM}) below
\begin{equation}
  \label{eq:dP1-GLSM}
  \begin{array}{c|ccccc}
    & z_1 & z_2 & z_3 & z_4 & t\\
    \hline
    U(1)_a & 1 & 0 & 1 & 1 & -3\\
    U(1)_b & 0 & 1 & 0 & 1 & -2
  \end{array}
\end{equation}
subject to the D-term conditions (\ref{eq0:dP1-GLSM-Dterms}):
\be
  \label{eq:dP1-GLSM-Dterms}
   |z_1|^2 + |z_3|^2 + |z_4|^2 - 3 |t|^2 = \xi_a \qquad,\qquad 
    |z_2|^2 + |z_4|^2 - 2 |t|^2 = \xi_b \; ,
\ee
The toric diagram and web diagram for this singularity are shown in figure~\ref{fig:dP1-resolved-toric}. The Fayet-Iliopoulos (FI) parameters $\xi_a$ and $\xi_b$ control the resolutions of the singularity, as illustrated in figure~\ref{sfig:dP1-resolutions}, where the unresolved (affine) singularity corresponds to $\xi_a = \frac{3}{2} \xi_b \le 0$.


We consider the orientifold involution $t \to -t$, as in~\S\ref{subsec:delPezzo-geom}. 
  Blowing up in the region $\xi_a > \xi_b > 0$, the exceptional divisor $t=0$ is $dP_1$, which is wrapped by an O7 plane. Moving to the region $\xi_b > \xi_a > 0$, the exceptional divisor develops a conifold singularity (at $\xi_a = \xi_b$) which flops, changing the topology of the divisor to $dP_0$, still wrapped by an O7 plane. In the process, an O3 plane at $z_1 = z_2 = z_3 = 0$ splits off. Continuing into the region $\xi_a < \frac{3}{2} \xi_b$, $\xi_a \le 0$, the O7 plane collapses into a $\bC^3/\bZ_3$ orientifold singularity, leaving behind the O3 plane.
In this region, the $dP_1$ singularity has 
split into two orbifold singularities, 
and the $\bZ_2^2$ twisted homology group of the parent singularity maps on to the $\bZ_2$ twisted homology groups of the two components. 
Performing this partial resolution in the dual gauge theory and comparing with the known discrete torsions of orbifolds~\cite{Witten:1998xy,dualities1,dualities2},
we can read off the discrete torsion of the parent singularity. 

\subsection{Discrete torsion for the classical phases}
\label{sec:classical-phases-torsion}

\begin{figure}
  \centering
  \begin{subfigure}{0.32\textwidth}
    \centering
    \includegraphics[width=\textwidth]{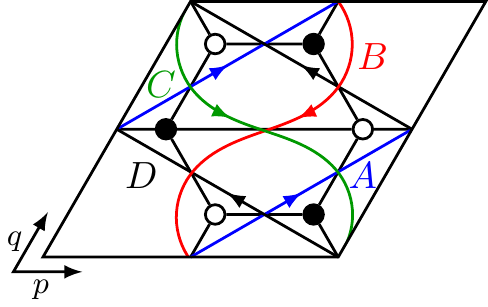}
    \caption{Zig-zag paths.}
    \label{sfig:dP1Zigzags}
  \end{subfigure}
  \begin{subfigure}{0.32\textwidth}
    \centering
    \includegraphics[width=\textwidth]{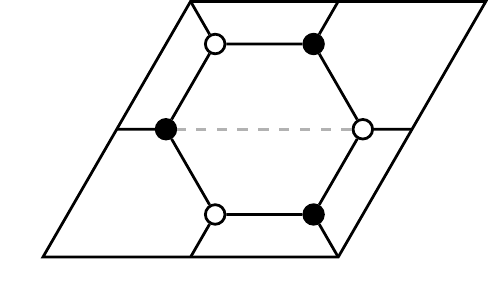}
    \caption{Higgsing to $\bC^3/\bZ_3$.}
    \label{sfig:dP1todP0}
  \end{subfigure}
  \begin{subfigure}{0.32\textwidth}
    \centering
    \includegraphics[width=\textwidth]{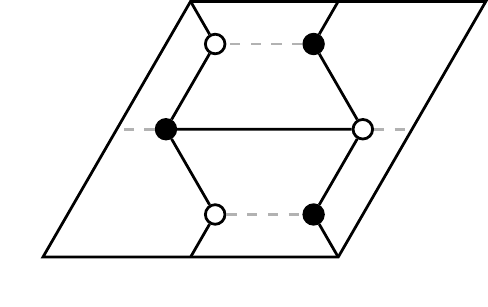}
    \caption{Higgsing to $\bC^3$.}
    \label{sfig:dP1toC3}
  \end{subfigure}
  \caption{\subref{sfig:dP1Zigzags} The zig-zag paths for the
    $dP_1$ dimer model, figure~\ref{sfig:dP1orientifoldtiling}. The winding numbers of the zig-zag paths reproduce the directions of the homonymous legs in the web diagram, figure~\ref{sfig:dP1-web}.
   \subref{sfig:dP1todP0}
    Partial resolution to $\bC^3/\bZ_3$ corresponds to a vev for the fields
    where paths $B$ and $C$ cross. \subref{sfig:dP1toC3} Partial resolution to
    $\bC^3$ corresponds to a vev for the fields where paths $A$ and $D$
    cross. An $\cN=4$ theory is obtained upon integrating out the 
    massive matter.}
  \label{fig:dP1Zigzags}
\end{figure}

We apply this approach 
to the orientifolds \IA\ and \IB\ discussed in~\S\ref{sec:known-constructions}. 
For future reference, we reproduce the charge tables for phases~\IA\ and \IB~\cite{dualities1} in tables~\ref{tab:IA-charges}, \ref{tab:IB-charges}.
To read off the effect of the partial resolution to a $\bC^3/\bZ_3$
orientifold singularity plus an O3 plane, as in the upper-left
quadrant of figure~\ref{sfig:dP1-resolutions}, we need to identify
which fields get a vev. As illustrated in figure~\ref{fig:dP1Zigzags},
this can be done systematically by analyzing the zig-zag
paths~\cite{kenyonintro,Hanany:2005ss,Feng:2005gw} of the dimer model,
see~\cite{GarciaEtxebarria:2006aq}. Giving a vev to the fields where the zig-zag paths $B$ and $C$ cross combines the corresponding legs of the web diagram, which yields the web diagram for $\bC^3/\bZ_3$, hence the baryon in question corresponds to partial resolution to the $\bC^3/\bZ_3$ orientifold singularity, as in the upper-left quadrant of figure~\ref{sfig:dP1-resolutions}, where the D3 branes remain on top of the singularity.

\begin{table}
\centering
\begin{equation*}
   \setlength{\extrarowheight}{1pt} 
  \def\arraystretch{1.15}
  \begin{array}{c|cc|ccccc}
    &\SU(N-4) & \SU(N) & \SU(2) &\U(1)_B & \U(1)_Y & \U(1)_R\\
    \hline
    A^i & \fund & \ov\fund & \fund & -\frac{1}{2N}-\frac{3}{2(N-4)} & 1+\frac{2}{N-4} & -\frac{2}{N-4} + \frac{2}{N}\\
    Y & \ov\fund & \ov\fund & {\bf 1} & -\frac{1}{2N}+\frac{3}{2(N-4)} & -1-\frac{2}{N-4} & 1 + \frac{2}{N-4} + \frac{2}{N} \\
    Z & \ov\symm & {\bf 1} & {\bf 1} & \frac{3}{N-4} & -1-\frac{4}{N-4} & 1+\frac{4}{N-4}\\
    B^i & {\bf 1} & \asymm & \fund & \frac{1}{N} & 0 & 1-\frac{4}{N}\\
    X & {\bf 1} & \asymm & {\bf 1} &\frac{1}{N} & -1 & 1-\frac{4}{N}
  \end{array}
\end{equation*}
\caption{Phase \IA}
\label{tab:IA-charges}
\end{table}

\begin{table}
\centering
\begin{equation*}
  \label{eq:IB-charges}
   \setlength{\extrarowheight}{1pt} 
  \def\arraystretch{1.15}
  \begin{array}{c|cc|ccccc}
    &\SU(\tN+4) & \SU(\tN) & \SU(2) & \U(1)_B & \U(1)_Y & \U(1)_R\\
    \hline
    \tA^i & \fund & \ov\fund & \fund & -\frac{1}{2\tN}-\frac{3}{2(\tN+4)} & 1-\frac{2}{\tN+4} & \frac{2}{\tN+4} - \frac{2}{\tN}\\
    \tY & \ov\fund & \ov\fund & {\bf 1} & -\frac{1}{2\tN}+\frac{3}{2(\tN+4)} & -1+\frac{2}{\tN+4} & 1 - \frac{2}{\tN+4} - \frac{2}{\tN} \\
    \tZ & \ov\asymm & {\bf 1} & {\bf 1} & \frac{3}{\tN+4} & -1+\frac{4}{\tN+4} & 1-\frac{4}{\tN+4}\\
    \tB^i & {\bf 1} & \symm & \fund & \frac{1}{\tN} & 0 & 1+\frac{4}{\tN}\\
    \tX & {\bf 1} & \symm & {\bf 1} &\frac{1}{\tN} & -1 & 1+\frac{4}{\tN}
  \end{array}
\end{equation*}
\caption{Phase \IB}
\label{tab:IB-charges}
\end{table}

\begin{figure}
  \centering
  \begin{subfigure}[b]{0.4\textwidth}
  \includegraphics[width=0.9\textwidth]{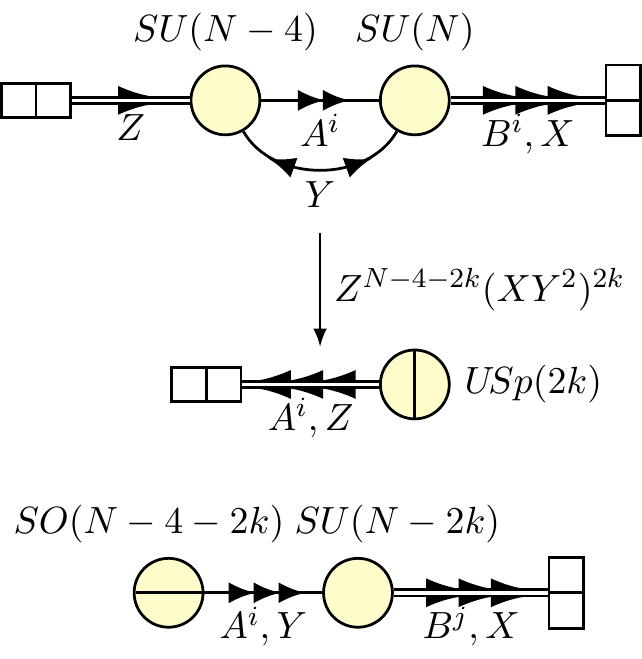}
  \caption{Partial resolution of \IA.}
  \label{subfig:quiverpartialres}
  \end{subfigure}
  \hspace{1cm}
  \begin{subfigure}[b]{0.45\textwidth}
  \centering
  \includegraphics[width=\textwidth]{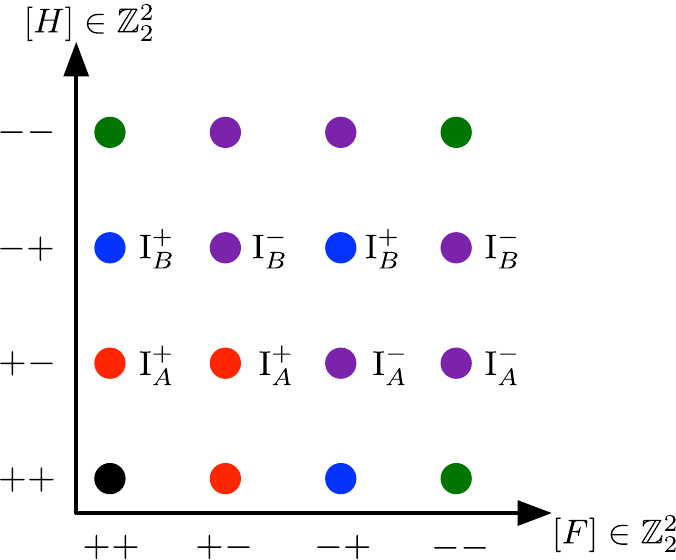}
  \caption{Discrete torsion for \IA\ and \IB.}
  \label{subfig:IABtorsions}
  \end{subfigure}
  \caption{\subref{subfig:quiverpartialres}~Partial resolution of the $dP_1$ singularity as realized in phase \IA. Phase \IB\ is analogous. \subref{subfig:IABtorsions}~Assignment of discrete torsions (in the basis associated to
    the partial resolution) to phases \IA\ and \IB.}
  \label{fig:dP1perturbativetheories}
\end{figure}

There is another baryonic branch where the D3 branes are carried away on top of the O3 plane, corresponding to a vev for the fields where the zig-zag paths $A$ and $D$ cross. More generally, some number of D3 branes can remain on the $\bC^3/\bZ_3$ orientifold singularity while the others are carried away on top of the O3 plane.\footnote{There are, of course, other directions in moduli space where the D3 branes do not lie atop either of the resulting O-planes, but these directions are not relevant to the present discussion.} The corresponding baryon -- a mixture of the two described above -- was 
 identified in~\cite{dualities1}:
\begin{equation}
\mathcal{O}_p = Z^{N-4-p} (X Y^2)^p \;\;,\;\; \widetilde{\mathcal{O}}_{\tilde{p}} = \tZ^{\tN+4-\tilde{p}} (\tX \tY^2)^{\tilde{p}} \,,
\end{equation}
for \IA\ and \IB, respectively, where $p$ and $\tN - \tilde{p}$ must be even. Turning on a vev for $\mathcal{O}_p$ or $\widetilde{\mathcal{O}}_{\tilde{p}}$ breaks
\begin{equation} \label{eqn:C3Z3partialres}
\begin{split}
\mathcal{O}_p:&\qquad \SU(N-4)\times \SU(N) \longrightarrow [\SO(N-4-p)\times \SU(N-p)] \times \Sp(p) \,, \\
\widetilde{\mathcal{O}}_{\tilde{p}}:&\qquad \SU(\tN+4)\times \SU(\tN) \longrightarrow [\Sp(\tN+4-\tilde{p})\times \SU(\tN-\tilde{p})] \times \SO(\tilde{p}) \,.
\end{split}
\end{equation}
Keeping track of the chiral superfields which are Higgsed or which acquire a mass from the superpotential, we find that the $\Sp(p)$ and $\SO(\tilde{p})$ factors decouple from the rest of the gauge group, with an $\cN=4$ spectrum and interactions, whereas the remaining fields and interactions reproduce those of the $\bC^3/\bZ_3$ orientifold studied in~\cite{dualities1}, as illustrated in figure~\ref{subfig:quiverpartialres}.

We verify that this result is consistent with D3 charge
conservation. The $\cN = 4$ theories (coming from D3 branes on top of
an O3 plane in a smooth background) have D3 brane charge
\be
Q^{\SO(p)}_{\bC^3} = p-\frac{1}{2} \;\;,\;\; Q^{\Sp(p)}_{\bC^3} =p+\frac{1}{2}\,,
\ee
whereas the $\bC^3/\bZ_3$ orientifold theories have D3 brane charge~\cite{dualities2}
\be
Q^{\SO(p-4)\times\SU(p)}_{\bC^3/\bZ_3} = p-\frac{3}{2}\;\;,\;\; Q^{\Sp(p+4)\times\SU(p)}_{\bC^3/\bZ_3} = p+\frac{3}{2}\,,
\ee
where we measure the charge in the Calabi-Yau double cover, so that a mobile D3 brane has charge $+2$.
Adding together the charges of the two components, we obtain the charge of the $dP_1$ orientifold singularity in each phase:
\be \label{eqn:dP1D3charge}
Q_{\IA}^{\SU(N-4)\times\SU(N)} = N -1 \;\;,\;\; Q_{\IB}^{\SU(\tN+4)\times\SU(\tN)} = \tN+ 1 \,.
\ee
This result can be verified using the explicit construction of the fractional branes given in appendix~\ref{app:dP1-microscopics}.

The result~(\ref{eqn:C3Z3partialres}) encodes the discrete torsion of the parent $dP_1$ orientifold singularity as follows. For the $\cN = 4$ theories as well as for the $\bC^3/\bZ_3$ orientifold theories, the gauge groups with an $\SO(p)$ factor have torsion $[F]=(-)^p$ and $[H]=+$ and those with an $\Sp(p)$ factor have torsion $[F] = \pm$ and $[H]=-$, where in the latter case the two choices of $[F]$ correspond to the same cusp at different values of $C_0$. Thus, the discrete torsion of the \IA\ theory is $[F] = ((-)^N,\pm)$ and $[H]=(+,-)$ in a basis where $(-,+)$ generates the $\bZ_2$ associated to the $\bC^3/\bZ_3$ singularity and $(+,-)$ generates the $\bZ_2$ associated to the O3 plane. Likewise, the discrete torsion for the \IB\ theory is $[F] = (\pm,(-)^\tN)$ and $[H]=(-,+)$ in the same basis.

Using this information we can start filling in some of the dots in
figure~\ref{fig:dP1theories}. We denote by $\IA^{\pm}$ the \IA\
theory with $(-1)^N=\pm 1$, and similarly by $\IB^{\pm}$ the \IB\
theory with $(-1)^\tN=\pm 1$. The result of the above 
argument is depicted in
figure~\ref{subfig:IABtorsions}. Notice that $\IA^-$ and $\IB^-$ fall in the same $\SL(2,\bZ)$ multiplet, hence they should be S-duals. Conservation of D3 charge~(\ref{eqn:dP1D3charge}) requires $\tN = N-2$. This duality was already conjectured purely on the basis of field theory evidence in~\cite{dualities1}, including a non-trivial matching between the low-$N$ dynamics. We provide further evidence for this duality in~\S\ref{subsec:SCImatching}, where we match a large number of terms in the superconformal index, limited only by the running time of an optimized program. The discrete torsion assignment in figure~\ref{fig:dP1perturbativetheories} provides a much clearer picture of why this duality exists for odd $N$ and not for even $N$. It also highlights the gaps in the first and last rows, including the S-duals of the even $N$ theories, whose absence was a mystery in~\cite{dualities1}. In fact, none of the $\SL(2,\bZ)$ multiplets are complete, including the one containing the S-dual pair $\IA^-$ and $\IB^-$.

\subsection{Brane tilings and orientifolds}
\label{sec:brane-tiling}

We now explain why these theories are absent, which is the first step along the way to constructing them.

While our main focus is $dP_1$,
  we find it convenient in the present discussion to consider a generic isolated toric singularity with $k+3$ legs in the web diagram --- for which there are $k$ two-cycles in the horizon $Y_5$ of the singularity, and for which the twisted homology group is $\bZ_2^{k+1}$~\cite{toricII} --- for example $dP_k$, $k=0,1,2,3$. 

We first review the construction of dimer models from NS5-D5 systems, as discussed in e.g.~\cite{Yamazaki:2008bt}.
 Consider type IIB string theory on $T^2 \times \bR^4$,\footnote{All of the branes we consider wrap the directions $x_0, \ldots x_3$, and none wrap the directions $x_8, x_9$.} with a flat background metric, where $x_4\sim x_4+1$ and $x_5\sim x_5+1$ parameterize the torus. We wrap $N$ D5 branes --- the double T-dual of the $N$ D3 branes we started with --- on the $T^2$. To reproduce the original Calabi-Yau geometry, we introduce NS5 branes ending on one-cycles of the torus and extending outwards along rays in the $(x_6,x_7)$ directions. Specifically, for an NS5 brane ending on the $(p,q)$ cycle of the torus, the corresponding ray --- $(x_6,x_7)=(tp, tq)$ with $t\in [0,\infty)$ --- is chosen to preserve a common $\mathcal{N}=1$ supersymmetry with the other NS5 branes and with the D5 branes. The cycles wrapped by the NS5 branes correspond to the legs of the web diagram of the toric singularity in question, see e.g.\ figure~\ref{sfig:dP1-web}.

To satisfy five-brane charge conservation, the NS5 branes cannot simply end on the torus. Instead, they divide the torus into $(N,Q)$ D5-NS5 bound states with different $Q$'s, where steps in $Q$ across the NS5 brane cycles account for the tadpole induced by the boundary of the NS5 brane. To obtain a gauge theory description, we restrict to configurations with $Q = 0, \pm1$, where the $(N,0)$ branes carry a $U(N)$ gauge group, whereas $(N,1)$ and $(N,-1)$ carry only a $U(1)$ gauge group. In the low-energy limit, the $U(1)$'s become massive or decouple, and we are left with a quiver gauge theory with the $(N,0)$ faces as nodes. Wherever two NS5 branes intersect, two of the $(N,0)$ faces touch each other, and there is bifundamental matter, which is chiral due to the presence of the NS5 branes. For each $(N,1)$ and $(N,-1)$ face, there is a corresponding disk diagram which generates a term in the superpotential out of the matter fields surrounding the face.

\begin{figure}
  \centering
  \includegraphics[width=0.75\textwidth]{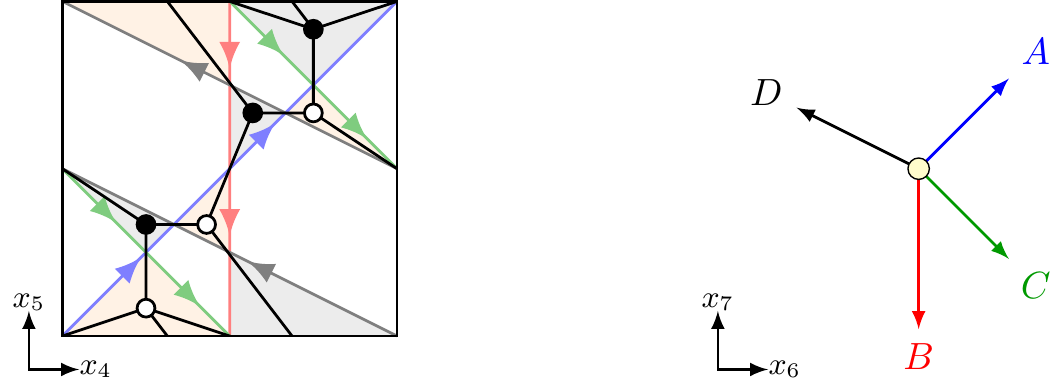}
  \caption{The brane tiling corresponding to the $dP_1$ singularity and the dual dimer model, c.f.\ figure~\ref{sfig:dP1Zigzags}.}
  \label{fig:dP1fivebranes}
\end{figure}

This construction is illustrated in figure~\ref{fig:dP1fivebranes} for $dP_1$. We refer to the configuration of five branes along the torus as a ``brane tiling''.\footnote{Our usage here is somewhat more specific than is common in the literature. We have reserved the phrase ``brane tiling'' for a physical configuration of branes, in contrast to dimer models, which represent a broader class of quiver gauge theories.} Dualizing the brane tiling, we recover the dimer model (figure~\ref{sfig:tilingQuiver}) for $dP_1$, where the $(N,0)$ branes map to faces, the $(N,1)$ and $(N,-1)$ branes to white and black nodes, respectively, and the NS5-brane intersections to edges. The NS5 branes correspond to zig-zag paths in the dimer model.

The above description of the NS5-D5 system is valid in the $g_s \to \infty$ limit, where the NS5 brane tension is much less than the D5 brane tension.\footnote{See~\cite{Yamazaki:2008bt} for a precise definition of the $g_s \to \infty$ and $g_s \to 0$ limits.} In the $g_s \to 0$ limit, the NS5 branes combine into a single NS5 brane wrapping a holomorphic curve with stacks of D5 branes (whose worldvolume is topologically a disk) ending on one-cycles on the curve. The latter description is closely related to the mirror of the original toric singularity with intersecting D6 branes and a non-trivial geometry~\cite{Feng:2005gw}. However, for the present paper we focus on the brane configuration in the $g_s \to \infty$ limit, which is easier to visualize, and will be convenient for engineering deconfinement in~\S\ref{sec:deconfinement}. 

\begin{figure}
  \centering
  \begin{subfigure}{0.2\textwidth}
    \centering
    \includegraphics[height=4.5cm]{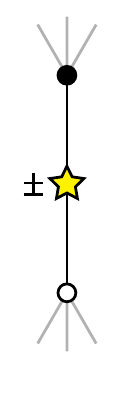}
    \caption{Dimer model.}
    \label{sfig:O5ChargesDimer}
  \end{subfigure}
  \hfill
  \begin{subfigure}{0.75\textwidth}
    \centering
    \includegraphics[height=4.5cm]{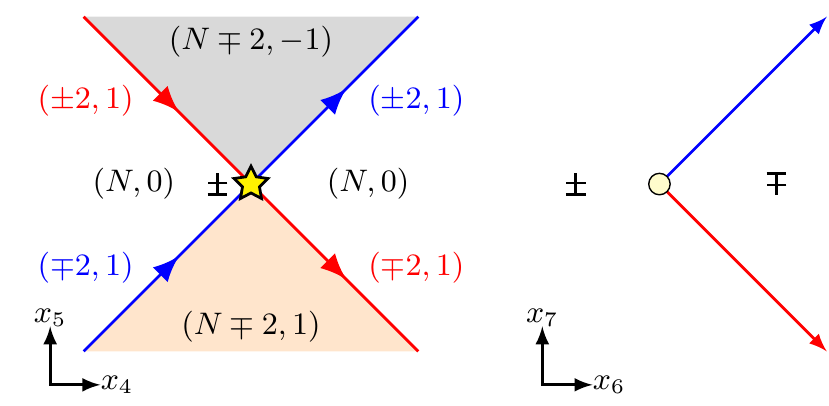}
    \caption{Brane realization.}
    \label{sfig:O5Charges}
  \end{subfigure}
  \caption{The T-parity of an orientifold fixed point in the dimer
    model (figure \subref{sfig:O5ChargesDimer}) corresponds to the RR
    charge of the O5 plane in the major angle between the NS5 branes
    in the brane tiling (figure~\subref{sfig:O5Charges}). Here the NS5
    branes carry $\pm 2$ units of D5 charge to compensate for the
    mismatch in the RR charge between the major and minor angles of
    the O5 plane.}
  \label{fig:O5Charges}
\end{figure}

Orientifolding the toric singularity corresponds to introducing O5
planes in the brane tiling~\cite{Imamura:2008fd}. In particular,
the ``fixed-point'' involutions considered
in~\S\ref{sec:known-constructions} correspond to four O5 planes
located at four points along the $T^2$ and filling the $(x_6,x_7)$
directions. The sign of the RR charge of the O5 plane changes each
time it crosses an NS5-brane in the $(x_6,x_7)$
plane~\cite{Evans:1997hk}. In the standard construction, this either
does not happen (when the O5 plane lies within an $(N,0)$ face on the
torus) or occurs twice (when the O5 plane sits at the intersection
point of two NS5 branes on the torus). In the former case, the RR
charge of the O5 plane is the same as the T-parity
associated to the fixed point in the dimer model, whereas in the
latter case the T-parity is the same as the RR charge of the O5 plane
in the major angle (greater than $180^\circ$) between the NS5 branes
in the $(x_6,x_7)$ plane, see figure~\ref{fig:O5Charges}.

Before orientifolding, there are in general moduli corresponding to the positions of the NS5-brane cycles. Accounting for a single calibration condition and translation invariance along the torus~\cite{Yamazaki:2008bt}, there are $k$ such moduli for $k+3$ legs in the web diagram. These moduli are T-dual to $B_2$ Wilson lines on the $k$ two-cycles in the horizon $Y_5$ of the toric singularity. Likewise, their superpartners --- $k$ Wilson lines on the $k+3$ NS5-brane worldvolumes (subject to one constraint and two gauge equivalences) --- are T-dual to $C_2$ Wilson lines on the same two-cycles.

Upon orientifolding, these moduli are projected out. This can be seen both in the five-brane system and in the T-dual Calabi-Yau orientifold. In the former, the five-brane positions are constrained by the orientifold. We observe that no two legs of the web diagram for $dP_1$ (figure~\ref{sfig:dP1-web}) are parallel; this is due to the fact that the $dP_1$ singularity is isolated. Since the O5 involution maps an NS5 brane ending on the cycle $(p,q)$ to another NS5 brane ending on a homologous cycle, this implies that each NS5 brane is its own image under the O5 involution. Topologically, the boundary of the NS5 brane is an $S^1$, and the O5 involution maps this boundary to itself with an orientation reversal. Thus, the involution has two fixed points along the boundary, i.e.\ each NS5 brane must pass through two of the fixed points on the torus. This completely fixes the NS5 brane moduli considered above, up to the choice of which pair of fixed points to cross (of two possible choices). 

In the T-dual Calabi-Yau geometry, the $B_2$ Wilson lines $\oint B_2$ change sign under the involution. Accounting for the shift symmetry $\oint B_2 \cong \oint B_2 + 1$, we conclude that $\oint B_2 = 0$ or $\oint B_2 = \frac{1}{2}$, and the modulus is projected out, again up to a discrete choice. But this is exactly how $[H]$ discrete torsion arises in the dual geometry!\footnote{To be precise, this accounts for a $\bZ_2^k \subset \bZ_2^{k+1}$ subgroup of the twisted homology group. The remaining $\bZ_2$ is generated by an $\bR\bP^2 \subset X_5$ which is trivial in $Y_5$, as in~\cite{Witten:1998xy}.}

Thus, we expect that the NS5 brane positions encode $[H]$ discrete torsion. To make this more precise, we count the number of choices. Since the O5 plane charge flips each time an NS5 brane is crossed in the $(x_6,x_7)$ directions, the number of NS5 branes crossing the corresponding fixed point on the torus must be even. After placing $k+2$ NS5 branes, the position of the last NS5 brane is fixed by this requirement. Moreover, translation symmetry on the torus allows us to interchange the four fixed points among each other. In particular, any two NS5 branes with $(p,q) \not\equiv (p',q') \mod 2$ intersect at exactly one fixed point,\footnote{Here we assume that the NS5 brane winding numbers are not all congruent modulo two, as happens e.g.\ for $\bF_0$. In this case, the counting is different, but the end result is the same~\cite{toricII}.} so picking some fiducial pair, we fix the translation symmetry on the torus by marking this fixed point. In placing the remaining $k$ NS5 branes, we make $k$ choices, hence there are $2^k$ possible distinct NS5 brane configurations.

To account for the final two choices of discrete torsion, we consider the RR charges of the four O5 planes at some fixed angle in the $(x_6,x_7)$ plane. Supersymmetry requires that the product of the four charges is positive~\cite{Imamura:2008fd}, so there are eight possible choices. However, changing the sign of one O5 plane relative to another will affect the meson sign rules of~\cite{Franco:2007ii}, hence changing the geometric involution. Thus, only an overall change of sign preserves the geometric involution, and there are two possible sign choices for the RR charges of the O5 planes at this point in the $(x_6,x_7)$ plane, which fixes these charges everywhere, since they change sign each time an NS5 brane is crossed.

Thus, the NS5 brane positions together with the overall sign of the RR charges of the O5 planes account for $2^{k+1}$ possibilities, exactly the same as the number of choices of $[H]$ discrete torsion. This information can be neatly encapsulated in the \emph{local charges} of one of the four O5 planes in each of the $k+3$ wedges in the $(x_6,x_7)$ plane between the legs of the web diagram. In particular, these charges specify which NS5 branes cross the corresponding fixed point on the torus (the legs of the web diagram adjoining wedges of opposite local charge), which completely fixes the NS5 brane positions. The overall RR charges of the other three O5 planes can then be fixed using the supersymmetry constraint and the meson sign rules of~\cite{Franco:2007ii}. 
There are $2^{k+3}$ choices of the local charges, but for each choice there are three equivalent ones corresponding to the local charges of the other three O5 planes, so there are $2^{k+1}$ inequivalent choices, as before.

\begin{figure}
  \centering
  \begin{subfigure}[b]{0.45\textwidth}
    \centering
    \includegraphics{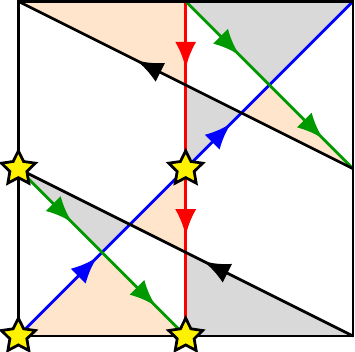}
    \caption{Double intersections.}
    \label{sfig:double-intersections}
  \end{subfigure}
  \hfill
  \begin{subfigure}[b]{0.45\textwidth}
    \centering
    \includegraphics{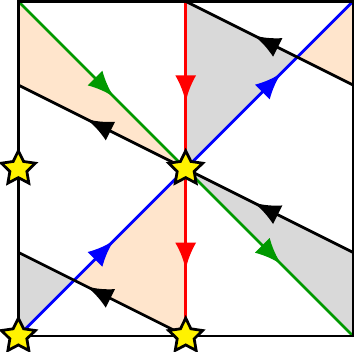}
    \caption{Quadruple intersection.}
    \label{sfig:quadruple-intersections}
  \end{subfigure}
  \caption{Possible NS5 brane positions in the brane tiling describing the $dP_1$ orientifold singularity. The white regions are $(N,0)$ branes and orange and grey regions are $(N,1)$ and $(N,-1)$ branes, respectively.}
  \label{fig:dP1-5-brane-diagrams}
\end{figure}

\begin{figure}
  \centering
  \begin{subfigure}{0.2\textwidth}
    \centering
    \includegraphics[height=3cm]{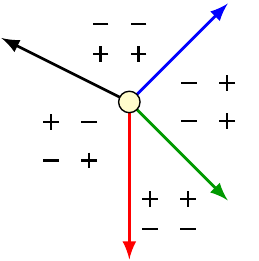}
    \caption{\IA}
    \label{sfig:IA-charges}
  \end{subfigure}
  \hfill
  \begin{subfigure}{0.2\textwidth}
    \centering
    \includegraphics[height=3cm]{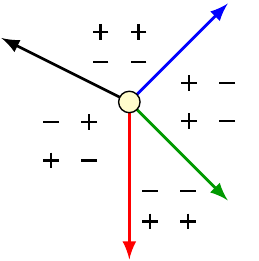}
    \caption{\IB}
    \label{sfig:IB-charges}
  \end{subfigure}
  \hfill
  \begin{subfigure}{0.28\textwidth}
    \centering
    \includegraphics[height=3cm]{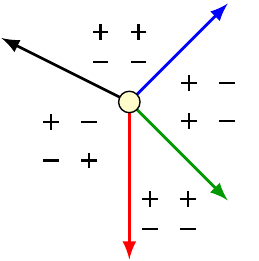}
    \caption{Unknown phase (II)}
    \label{sfig:II-charges}
  \end{subfigure}
  \hfill
  \begin{subfigure}{0.28\textwidth}
    \centering
    \includegraphics[height=3cm]{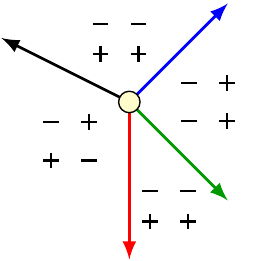}
    \caption{Unknown phase (III)}
    \label{sfig:III-charges}
  \end{subfigure}
  \caption{Local charges for the O5
    planes in the four possible configurations T-dual to branes at
    the $dP_1$ orientifold singularity with the geometric involution described in the text. Configurations~\subref{sfig:IA-charges} and
    \subref{sfig:IB-charges} correspond to the brane tiling in
    figure~\ref{sfig:double-intersections} and those in
    \subref{sfig:II-charges} and \subref{sfig:III-charges} to that
    in figure~\ref{sfig:quadruple-intersections}. The latter two are previously unknown phases.}
  \label{fig:O5-planes}
\end{figure}

Having understood the origin of $[H]$ discrete torsion in the five-brane system, we are now in a position to explain the paradoxes of~\S\ref{sec:known-constructions}--\ref{subsec:countingorientifolds}. We consider $dP_1$. From the general analysis given above, we expect that there are two distinct NS5 brane configurations, each of which admits two orientifolds related by an overall sign flip of the RR charges of the O5 planes, for a total of four choices corresponding to the four possible $[H]$ torsions. The two possible NS5 brane configurations are shown in figure~\ref{fig:dP1-5-brane-diagrams} and the four corresponding choices of local charges are shown in figure~\ref{fig:O5-planes}. Using the dimer-model / brane-tiling dictionary, it is straightforward to verify (c.f.\ figure~\ref{fig:dP1fivebranes}) that figure~\ref{sfig:double-intersections} with the local charges shown in figure~\ref{sfig:IA-charges} or~\ref{sfig:IB-charges} corresponds to the orientifolds \IA\ and \IB\ of~\S\ref{sec:known-constructions} (see figure~\ref{fig:dP1-classical-phases}), respectively.

\begin{figure}
  \centering
  \begin{subfigure}[b]{0.28\textwidth}
    \centering
    \includegraphics{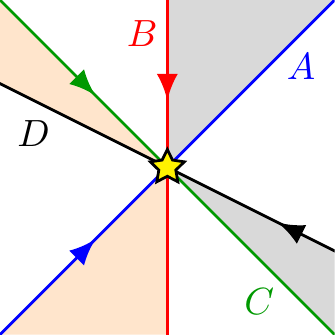}
    \caption{Four NS5 branes intersecting atop an O5 plane.}
    \label{sfig:non-Lagrangian-system-O5}
  \end{subfigure}
  \hfill
  \begin{subfigure}[b]{0.28\textwidth}
    \centering
    \includegraphics{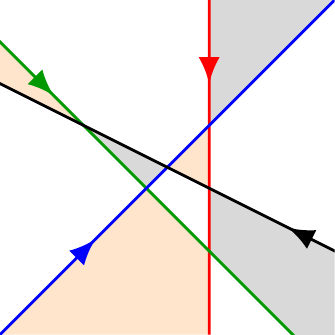}
    \caption{A deformation possible without the O5.}
    \label{sfig:non-Lagrangian-resolved}
  \end{subfigure}
  \hfill
  \begin{subfigure}[b]{0.28\textwidth}
    \centering
    \includegraphics{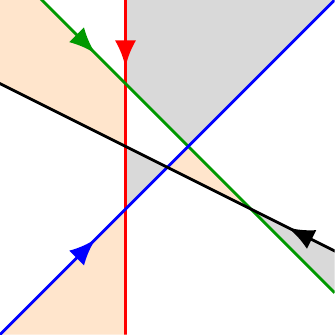}
    \caption{A different deformation possible without the O5.}
    \label{sfig:non-Lagrangian-resolved-b}
  \end{subfigure}
  \caption{\subref{sfig:non-Lagrangian-system-O5} Four NS5 branes
    intersecting over a O5 plane. 
    \subref{sfig:non-Lagrangian-resolved}--\subref{sfig:non-Lagrangian-resolved-b} Removing the
    orientifold plane, there are two deformations that allow a perturbative description, related by Seiberg duality.    
    }
  \label{fig:non-Lagrangian-system}
\end{figure}

By contrast, the configuration in figure~\ref{sfig:quadruple-intersections} is non-standard, due to the quadruple intersection of NS5 branes at one of the fixed points. We emphasize that this quadruple intersection is forced on us by the O5 involution, which projected out the NS5 brane moduli which would otherwise allow us to deform away from this special point in the moduli space, see figure~\ref{fig:non-Lagrangian-system}. Since this deformation grows a fourth $(N,0)$ face in the brane tiling, the special point lies at infinite coupling for one of the gauge groups, and we expect that there are intrinsically strongly coupled degrees of freedom localized at the quadruple intersection. Thus, the answer to the paradox of~\S\ref{sec:delPezzo-review} is that the missing phases are not gauge theories at all, but something intrinsically strongly coupled!

In fact, as shown in figures~\ref{sfig:non-Lagrangian-resolved} and \ref{sfig:non-Lagrangian-resolved-b}, there are two ways in which the quadruple crossing can be resolved, related by Seiberg duality on the shrinking $(N,0)$ face.
A configuration with formally infinite
coupling interpolating between two Seiberg dual phases is
a familiar situation in the Hanany-Witten literature~\cite{Hanany:1996ie,Giveon:1998sr}.
However, the infinite coupling makes the gauge theory description ill-defined, and the orientifolds we consider lack a modulus to deform away from the strong-coupling point, so we need a different approach to describe the low-energy physics. 

Let us take a step back and examine under what circumstances these
higher multiplicity intersections are forced to occur. Since each NS5
brane crosses two fixed points, and there are four fixed points in
total, it is evident that when there are more than four NS5 branes
(i.e.\ more than four legs in the web diagram) a higher multiplicity
intersection must occur at one or more of the fixed points, whereas in
the case of four NS5 branes a quadruple intersection will occur for
some choices of $[H]$, but not for others. This agrees exactly with
what we found in~\S\ref{sec:delPezzo-review}, where there were no
candidate duals for $dP_2$ and $dP_3$ and an incomplete list for
$dP_1$ and $\bF_0$. Moreover, in the case of three NS5 branes, as for
all isolated orbifold singularities, a higher multiplicity
intersection is obviously impossible, in perfect agreement with the
precise matching between the discrete torsion classification of
orbifolds and their dual gauge theories obtained in~\cite{dualities1,
  dualities2}.

\subsection{Quad CFTs} \label{subsec:quadCFTs}

To isolate the physics associated to the quadruple crossing and to decouple details of the rest of the brane tiling, we zoom in on the vicinity of the O5 plane, replacing $T^2$ with $\bR^2$, so that the
four $(N,0)$ faces in figure~\ref{sfig:non-Lagrangian-system-O5} become flavor branes and the superpotential terms associated to the four $(N,\pm 1)$ branes go to zero. The resulting brane configuration corresponds to a CFT in the infrared. 
Once these ``quad'' CFTs are understood, we can reintroduce gauge couplings for the non-compact branes by weakly gauging flavor symmetries of the CFT. Likewise, the missing superpotential terms can be reintroduced by deforming the CFT by a relevant operator.

\begin{figure}
  \centering
    \begin{subfigure}[b]{0.4\textwidth}
    \centering
    \includegraphics[width=\textwidth]{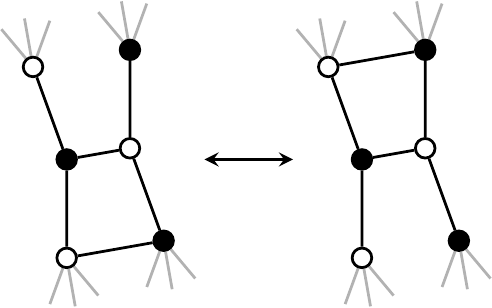}
    \vspace{0.5cm}
    \caption{Dimer models for figures~\ref{sfig:non-Lagrangian-resolved}--\subref{sfig:non-Lagrangian-resolved-b}.}
    \label{sfig:quad_dimer}
  \end{subfigure}
 \hspace{2cm}
  \begin{subfigure}[b]{0.3\textwidth}
    \centering
    \includegraphics[width=\textwidth]{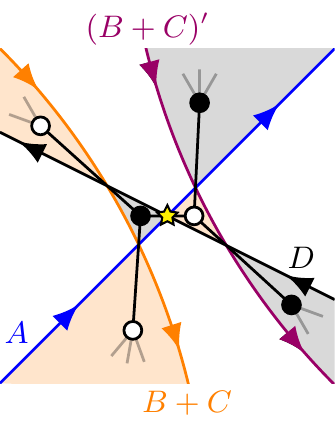}
    \caption{B-C brane recombination}
     \label{sfig:partial_resolved_quad}
  \end{subfigure}
  \caption{\subref{sfig:quad_dimer} Dimer models for the Seiberg-dual deformations of the quadruple crossing, figures~\ref{sfig:non-Lagrangian-resolved}--\subref{sfig:non-Lagrangian-resolved-b}. \subref{sfig:partial_resolved_quad} A partial resolution of the quadruple crossing in \ref{sfig:non-Lagrangian-system-O5}, which corresponds to turning on a vev where the B and C branes intersect in \ref{sfig:non-Lagrangian-resolved}--\subref{sfig:non-Lagrangian-resolved-b}.}
\end{figure}

Before orientifolding, the quadruple intersection can be resolved by a small deformation, as in figures~\ref{sfig:non-Lagrangian-resolved}--\subref{sfig:non-Lagrangian-resolved-b}, giving a pair of Seiberg dual gauge theories in the same universality class as the quad CFT, corresponding to the dimer models in figure~\ref{sfig:quad_dimer}. Heuristically, the orientifolded quad CFT corresponds to a quotient on the operator spectrum of this ``parent'' quad CFT. This is not quite correct, because the orientifold introduces a tadpole which must be cancelled by making ranks of the flavor branes unequal, but we expect it to be approximately true at large $N$, where the difference in ranks is a small effect. Thus, we can learn something about the orientifolded quad CFT by studying its parent.

We use this technique to determine the flavor symmetries associated to the quadruple crossing.
 In addition to the $\SU(N)^4$ flavor symmetry associated to the flavor branes, there is an anomaly-free $\U(1)$ flavor symmetry in the parent for each NS5 brane (zig-zag path) in the brane tiling~\cite{Imamura:2006ie}, where the sum of these $\U(1)$s decouples. Thus, the manifest non-R flavor symmetry group is $\SU(N)^4 \times \U(1)^3$. However, the superpotential generated by the finite-area $(N,\pm 1)$ faces in figures~\ref{sfig:non-Lagrangian-resolved}--\subref{sfig:non-Lagrangian-resolved-b} respects a larger $\SU(2N)\times \SU(N)^2 \times \U(1)^2$ flavor symmetry, where $\SU(2N) \supset \SU(N)^2\times \U(1)$ is broken when the flavor branes are gauged, but is conserved by the quad CFT itself.

The orientifold projection maps the NS5 branes to themselves and exchanges the $(N,0)$ flavor branes in pairs, hence the $\U(1)^3$ subgroup is invariant, whereas the $\SU(N)$ factors are exchanged in pairs, leaving the manifest flavor symmetry group $\SU(N)\times\SU(N+p)\times\U(1)^3$ after orientifolding, where $p$ is the difference in ranks required to cancel the RR tadpole associated to the O5 plane. 
Thus, the orientifold projection maps the enhanced $\SU(2N)$ symmetry of the parent theory to its charge conjugate, which suggests that $\SU(2N)$ is broken to either $\SO(2N)$ or $\Sp(2N)$ by the involution, where $\SO(2N) \supset \SU(N)\times\U(1)$ or $\Sp(2N) \supset \SU(N)\times\U(1)$ is a nonabelian enhancement of the manifest flavor symmetries. This enhancement can be seen explicitly by Higgsing the baryon associated to the B-C crossing in figures~\ref{sfig:non-Lagrangian-resolved}--\subref{sfig:non-Lagrangian-resolved-b}, recombining the NS5 branes and giving the brane tiling shown in figure~\ref{sfig:partial_resolved_quad}, which admits a perturbative, dimer model description. The superpotential terms associated to the finite-area $(N,\pm 1)$ faces preserve the enhanced symmetry $\SO(2N)$ ($\Sp(2N)$) for an O5 plane of positive (negative) T-parity.

\begin{figure}
  \centering
  \begin{subfigure}[b]{0.3\textwidth}
    \centering
    \includegraphics[width=0.9\textwidth]{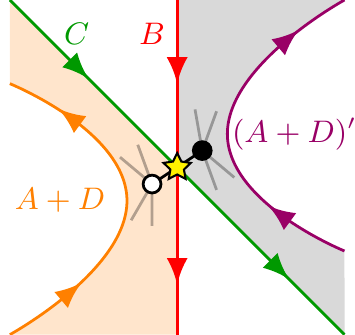}
    \caption{A-D recombination}
     \label{sfig:ADresolution}
  \end{subfigure}
  \hfill
  \begin{subfigure}[b]{0.3\textwidth}
    \centering
    \includegraphics[width=0.9\textwidth]{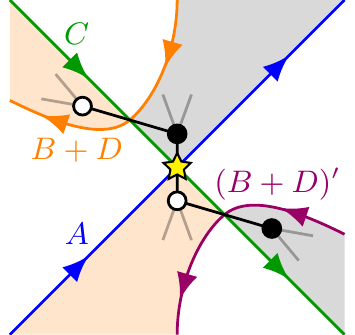}
    \caption{B-D recombination}
   \label{sfig:BDresolution}
  \end{subfigure}
   \hfill
  \begin{subfigure}[b]{0.3\textwidth}
    \centering
    \includegraphics[width=0.9\textwidth]{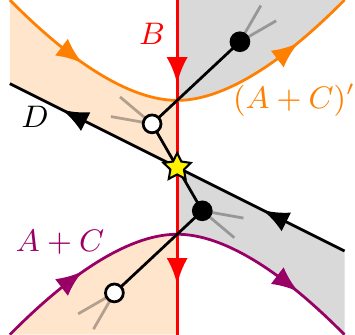}
    \caption{A-C recombination}
    \label{sfig:ACresolution}
  \end{subfigure}
  \caption{Possible brane recombinations which resolve the quadruple crossing, in addition to the B-C recombination shown in figure~\ref{sfig:partial_resolved_quad}. In each case, the corresponding dimer model is superimposed on the tiling.}
  \label{fig:quadrecombinations}
\end{figure}

We infer that there are two classes of quad CFTs, depending on the overall sign of the RR charge of the O5 plane, with non-R flavor symmetry groups $\SO(2N)\times\SU(N+p)\times \U(1)^2$ and $\Sp(2N)\times\SU(N+p')\times \U(1)^2$, respectively, for $p$ and $p'$ to be determined. We have also seen that these CFTs flow to an infrared-free theory with chiral superfields and a tree-level superpotential when we move along a baryonic branch corresponding to recombining the B and C NS5 branes. There are three other possible brane recombinations, shown in figure~\ref{fig:quadrecombinations}. In these cases, there are no finite-area $(N,\pm 1)$ faces, hence the infrared physics is described by free chiral superfields.

These brane recombinations --- which will correspond to partial resolutions of the toric singularity when the quad CFT is embedded into $T^2$ --- exhaust the baryonic directions in the moduli space. There are also mesonic directions coming from the gauge-singlet mesons in figure~\ref{sfig:quad_dimer} and their orientifold images, which are composite mesons (elementary mesons in the Seiberg dual theory). Turning on a rank $k$ vev for the gauge-singlet mesons and their orientifold images, we obtain the rank $N-k$ quad CFT plus a residual $U(k)^2$ global symmetry and various gauge-singlet chiral superfields charged under $\SU(N-k)^4\times\U(1)^3 \times \U(k)^2$. These $\U(k)^2$-charged chiral fields do not appear in the superpotential, hence they decouple from the remainder of the quad CFT. When the quad CFT is embedded into $T^2$, this direction in moduli space will correspond to removing an orientifold image pair of stacks of $k$ D3 branes from the singularity without resolving it.

Thus, even without an explicit construction of the orientifolded quad CFTs, we can infer their global symmetries and the general characteristics of their moduli space. In the next section, we will construct CFTs which match these expectations, and are therefore candidate descriptions of the quadruple crossing.

\section{Deconfinement and strong coupling} \label{sec:deconfinement}

We now turn to the task of describing five-brane systems with quadruple intersections of NS5 branes, as in figure~\ref{sfig:quadruple-intersections}. We expect that these configurations are T-dual to configurations with the missing $[H]$ torsions in figure~\ref{fig:dP1perturbativetheories}. 

Surprisingly, even though (as we have argued) these configurations are intrinsically strongly coupled, it is still possible to construct a family of UV gauge theories in the same universality class as the strongly coupled theory.
We motivate this through a brane engineering construction of the UV theories, and provide highly non-trivial evidence that our result is correct based on the quantum moduli space and superconformal index of the resulting theories, where the former matches with the five-brane description under the partial resolutions described in~\S\ref{subsec:quadCFTs}, and the latter passes all S-duality tests.
Our construction is based on the deconfinement method of~\cite{Berkooz:1995km,Pouliot:1995me}, which we now review.

\subsection{Deconfinement in the gauge theory} \label{subsec:gaugedeconfinement}

Consider an $\Sp(N-4)$ gauge theory with $N$ chiral superfields $\hat{Z}^i$ in the vector representation, transforming under an $\SU(N)$ flavor symmetry. It is well known~\cite{Intriligator:1995ne} that this theory \mbox{s-confines} with the composite $Z^{i j} = \Omega^{a b} \hat{Z}^i_a \hat{Z}^j_b$ (where $\Omega^{a b}$ is the symplectic form of $\Sp(N-4)$) and the superpotential
\be \label{eqn:Wconfined}
W = \Pf Z \,,
\ee
where $Z$ transforms in the $\asymm$ representation of $\SU(N)$. We consider deforming this theory by weakly gauging the $\SU(N)$ flavor symmetry coupled to some set of additional chiral fields to cancel the anomalies. In the limit where the $\SU(N)$ gauge coupling is small at the $\Sp(N-4)$ confinement scale, the low energy dynamics of this $\Sp(N-4)\times\SU(N)$ gauge theory are the same as those of the $\SU(N)$ gauge theory with the s-confined matter spectrum and superpotential~(\ref{eqn:Wconfined}). We can construct further duals of this theory by considering the limit where the $\SU(N)$ gauge coupling becomes large before the $\Sp(N-4)$ confinement scale. If there is no phase transition in between and the IR fixed point is isolated, this theory describes the same infrared physics as before. In the limit where the $\Sp(N-4)$ gauge coupling is small at the $\SU(N)$ dynamical scale, we can replace $\SU(N)$ with its Seiberg dual to obtain another theory with the same infrared fixed point. By dualizing nodes of the quiver in turn, we obtain a chain of dual descriptions~\cite{Berkooz:1995km}. This ``deconfinement'' technique is helpful for understanding the dynamics of the original $\SU(N)$ gauge theory with antisymmetric tensor matter.

\begin{figure}
  \centering
  \includegraphics{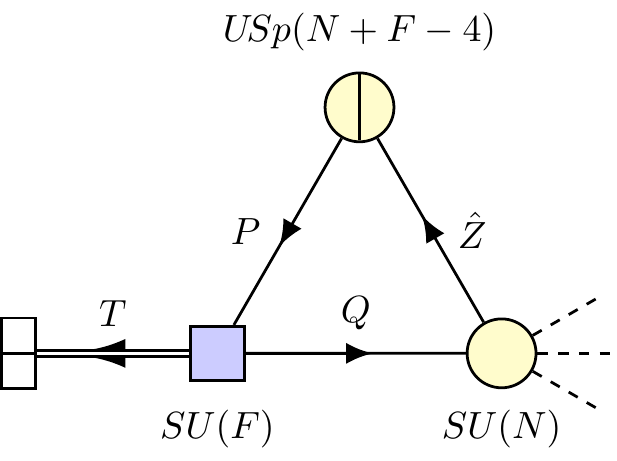}
  \caption{Deconfined description for an antisymmetric tensor.
}
  \label{fig:antisymmetric-deconfinement-quiver}
\end{figure}

So far we have assumed that the $\SU(N)$ gauge theory in question has a superpotential in the infrared of the form~(\ref{eqn:Wconfined}). To describe a theory without a superpotential, we replace $\Sp(N-4)$ with $\Sp(N+F-4)$ in the UV theory for some arbitrary $F>0$, and add $F$ additional $\Sp(N+F-4)$ vectors $P$ transforming under an $\SU(F)$ flavor symmetry, as well as $F$ antifundamentals $Q$ of $\SU(N)$ transforming under the same $\SU(F)$ and a gauge-singlet $T$ in the $\asymm$ representation of $\SU(F)$. This gives the quiver diagram shown in figure~\ref{fig:antisymmetric-deconfinement-quiver}. We introduce the $\SU(F)$-invariant superpotential
\begin{equation} \label{eqn:decW}
  W = PQ\hat{Z} + TP^2 \,.
\end{equation}
When the $\Sp(N+F-4)$ theory s-confines, it forms the composites
\be
M = \begin{pmatrix} \hat{Z}^2 & \hat{Z} P \\ -\hat{Z} P & P^2 \end{pmatrix}
\ee
with the confining superpotential $\Pf M$. However, $\hat{Z} P$ and $P^2$ get a mass with $Q$ and $T$ respectively, via the tree-level superpotential~(\ref{eqn:decW}), and can be integrated out. The $F$-term conditions for $Q$ and $T$ set $\hat{Z} P = P^2 = 0$, and the confining superpotential vanishes, leaving behind a single $\SU(N)$ antisymmetric tensor $Z = \hat{Z}^2$ with no superpotential~\cite{Pouliot:1995me}.

Note that the choice of $F>0$ is arbitrary, except that $N+F$ must be even for $\Sp(N+F-4)$ to exist. In the infrared, no fields are charged under the $\SU(F)$. Thus, despite being unbroken, the UV symmetry $\SU(F)$ is not a symmetry of the infrared fixed point, explaining how theories with different global symmetry groups in the UV and no accidental symmetries can describe the same infrared physics. We refer to such a UV global symmetry as ``trivial''.

In fact, the triviality of $\SU(F)$ imposes interesting \emph{a posteriori} constraints on the moduli space and operator spectrum. Consider turning on a vev for a gauge-invariant operator in a non-trivial $\SU(F)$ representation, spontaneously breaking the $\SU(F)$. In the UV theory, this is a D-flat direction classically, and it is straightforward to choose an operator which also satisfies the classical F-term constraints such as $T$ (for $F>2$, so that $T$ transforms under $\SU(F)$). However, triviality for $\SU(F)$ implies that the operator in question is lifted in the infrared, hence it is not a flat direction of the full theory. This is true for any operator in a non-trivial $\SU(F)$ representation. Thus, for most practical purposes, we can treat $\SU(F)$ as weakly gauged, and only consider $\SU(F)$ singlet operators. Triviality of the $\SU(F)$ implies the vanishing of the $\SU(F)^3$ and $\SU(F)^2 \U(1)$ anomalies, except that $\SU(F)^2 \U(1)_R$ receives a contribution from the $\SU(F)$ gauginos, and the R-symmetry is anomalous. Since the $\SU(F)$ is trivial, once gauged it undergoes gaugino condensation in the infrared, resulting in $F$ isolated vacua in the $\SU(F)$ sector, and an accidental R-symmetry in the $\SU(N)$ sector. While this is essentially (though not exactly) equivalent to the infrared physics of the theory before gauging the $\SU(F)$, it is usually more convenient to treat $\SU(F)$ as a global symmetry, keeping in mind that it is trivial, hence only $\SU(F)$ singlets contribute to the IR dynamics.

To illustrate the triviality of $\SU(F)$, we consider the effect of a
few $\SU(F)$-breaking vevs. Turning on a vev for $T$ gives mass to
some of the $P$'s. If $T$ has rank greater than one, then $\Sp(N+F-4)$
develops a dynamical superpotential, and supersymmetry is broken. If
$T$ has rank one and $F>2$, $\Sp(N+F-4)$ has a quantum-deformed moduli
space, $\Pf M = \Lambda^{N+F-2}$, but the F-term conditions set
$\Pf M = 0$ as before, so supersymmetry is broken. If $T$ has rank one
and $F=2$, then $T$ is a $\SU(F)$ singlet, and the result is
different. After integrating out the $P$'s in the UV theory, we are
left with a superpotential $W= Q^2 \hat{Z}^2$. As before, $\Sp(N+F-4)$
has a quantum-deformed moduli space
$\Pf (\hat{Z}^2) = \Lambda^{N+F-2}$, but now the F-term conditions do
not enforce $\Pf (\hat{Z}^2) = 0$. Instead, $\SU(N)$ is Higgsed to
$\Sp(N)$ and $Q$ acquires a mass, leaving $\SU(F)$ trivial as
before.\footnote{In fact, $T$ plays the same role in this case as
  $\hat{Z}^N = \Pf Z$ does for even $N$ and $F\ge 4$ ($\hat{Z}^N$
  vanishes identically for $F=2$).} A similar story applies to other
$\SU(F)$ breaking directions in the classical moduli space.

\subsection{Deconfinement in the brane tiling}
\label{sec:tiling-deconfinement}

In this section, we  engineer deconfinement of an antisymmetric tensor in the five-brane system discussed in~\S\ref{sec:brane-tiling}. In the process, we will inevitably encounter non-BPS --- indeed, dynamically unstable --- brane configurations. This situation is not unprecedented. For instance, many of the toric phases which arise in dimer models cannot be realized by brane tilings with straight NS5 branes. These ``geometrically inconsistent'' phases~\cite{Broomhead:2008an,Hanany:2006nm} do not correspond to any BPS configuration of branes, but in general they are Seiberg-dual to some geometrically consistent phase, hence they lie in the same universality class as some BPS configuration. With this example in mind, we engineer a configuration of branes which reproduces the gauge theory construction of~\S\ref{subsec:gaugedeconfinement}, ignoring issues of supersymmetry and stability. We then provide some post hoc justification for why this works.

We start with the brane tiling description
of a single antisymmetric tensor multiplet of $\SU(N)$, see figure~\ref{fig:O5Charges}.
As in~\S\ref{subsec:quadCFTs}, we take the worldvolume of the D5 branes to be $\bR^2$ instead of $T^2$, so that there is a single O5 plane, and the image pair of $(N,0)$ faces corresponds to an $\SU(N)$ flavor symmetry. The O5 plane is divided into an O5$^+$ plane and an O5$^-$ plane by the NS5 branes, with opposite D5 charges in the two parts, so the NS5 branes carry D5 charge $\pm2$ to cancel the tadpole~\cite{Imamura:2008fd}.

\begin{figure}
\centering
\begin{subfigure}[b]{0.50\textwidth}
  \centering
\includegraphics[width=\textwidth]{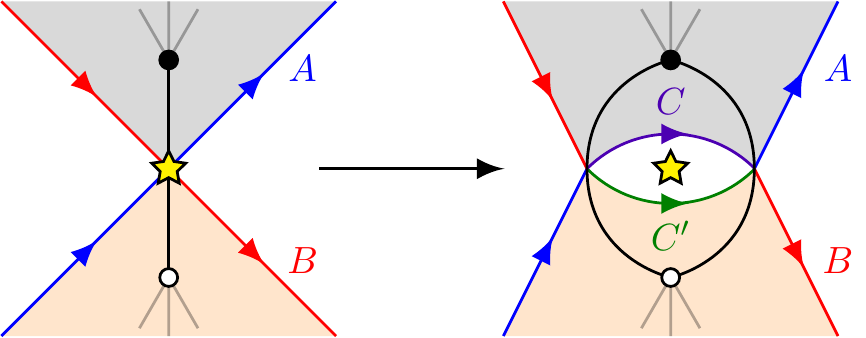}
\vspace{5mm}
\caption{Deconfinement in the dimer}
\label{subfig:deconfinementdimer}
\end{subfigure}
\hfill
\begin{subfigure}[b]{0.45\textwidth}
  \centering
\includegraphics[width=\textwidth]{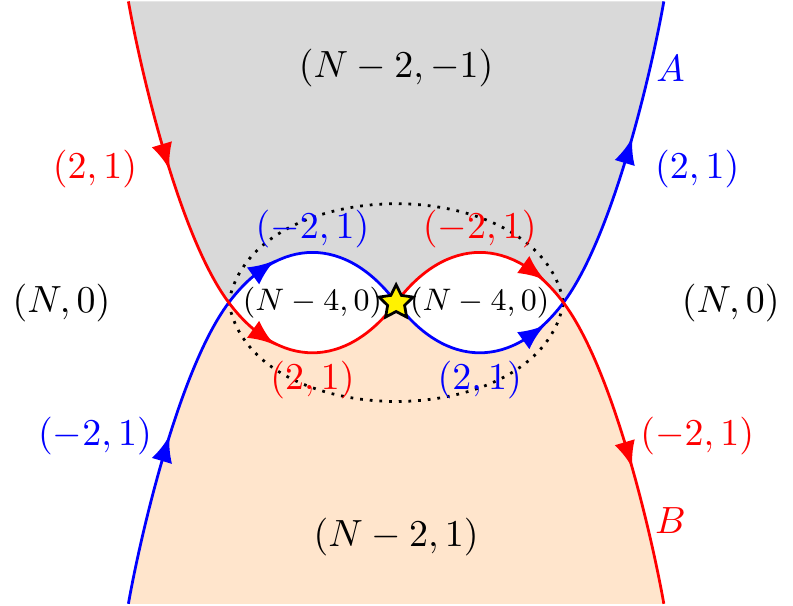}
\caption{Triple crossing of NS5 branes}
\label{subfig:triplecrossing}
\end{subfigure}
\caption{\subref{subfig:deconfinementdimer}~Deconfinement corresponds to edge doubling in the dimer model. To avoid a global change in the structure of the zig-zag paths, the original paths $A$ and $B$ are forced to recombine into new paths $C$ and $C'$ on the doubled edges. \subref{subfig:triplecrossing}~The configuration of branes prior to ``lifting'', with a triple crossing of NS5 branes. The dashed line indicates the boundary of the region to be lifted off the D5 stack, and the shaded regions within this line will become the NS5 branes $C$ and $C'$.}
\end{figure}

Leaving aside the issue of flavors and the confining superpotential temporarily, deconfinement has an obvious description in the dimer model, see figure~\ref{subfig:deconfinementdimer}. The antisymmetric tensor corresponds to an edge crossing a fixed point of negative T-parity. To deconfine this tensor, we replace the edge with an image-pair of edges surrounding a two-sided face with the fixed point in the middle. The new face is an $\Sp$ gauge group --- $\Sp(N-4)$ by anomaly cancellation --- which s-confines, reproducing the antisymmetric tensor we started with. Thus, deconfinement corresponds to ``doubling'' in the language of~\cite{Davey:2009bp,Davey:2011sw}.

In order to translate this into a five-brane system, we construct the zig-zag paths of the dimer. This immediately leads to a puzzle: before doubling the edge, there are two zig-zag paths, $A$ and $B$, which cross at the fixed point. After doubling, one end of $A$ connects to the other end of $B$, and vice versa, so the zig-zag paths get reconnected, even though we performed a local operation in the dimer. Taking the new zig-zag paths literally leads to the conclusion that dimer models of this type are ``inconsistent''~\cite{Hanany:2005ss}.

Instead, we interpret the two zig-zag paths inside the new two-sided face as new NS5 branes $C$ and $C'$ which appear after deconfinement, whereas the zig-zag paths outside the new face are the same NS5 branes $A$ and $B$ that were present before deconfinement. To understand how the two halves of $A$ and $B$ are joined together and how $C$ and $C'$ connect to $A$ and $B$, we construct the entire NS5 brane configuration in two steps, as follows. 
First, we deform the initial brane configuration by changing the
single crossing of the NS5 branes into a triple crossing, creating an
image pair of ``eyes'' surrounding the O5 plane, as in
figure~\ref{subfig:triplecrossing}. We then trace out a contour
enclosing this pair of eyes, and lift the NS5 branes within the
enclosing contour off of the D5 stack into the minor angle between the
$A$ and $B$ branes in the $(x_6,x_7)$ plane, opening up a ``bubble''
bounded by a new $(N-4,0)$ face on the D5 stack and the lifted NS5
branes, which we identify as $C$ and $C'$. The NS5 branes $A$ and $B$
now end on the upper surface of the bubble, tracing out the same
double eye as before, where they fuse with the upper boundary of the
branes $C$ and $C'$.

\begin{figure}
\centering
\begin{subfigure}[b]{0.41\textwidth}
  \centering
  \includegraphics[height=5cm]{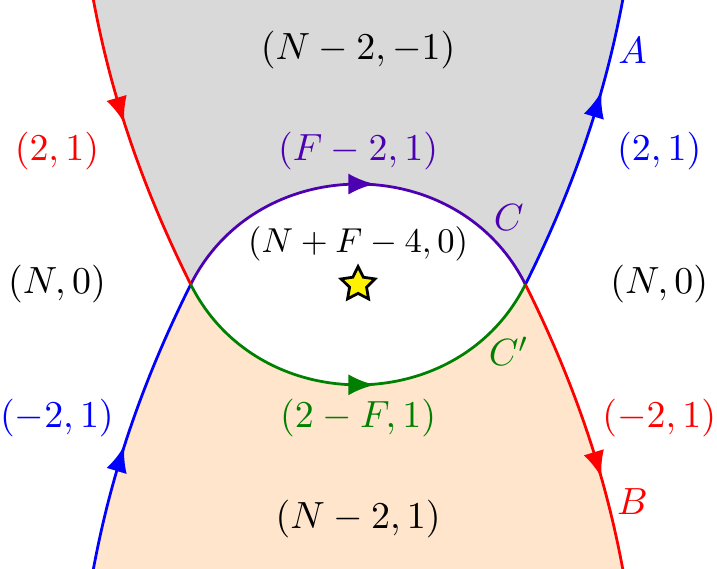}
  \caption{The D5 worldvolume}
  \label{subfig:postlift}
\end{subfigure}
\begin{subfigure}[b]{0.3\textwidth}
  \centering
  \includegraphics[height=5cm]{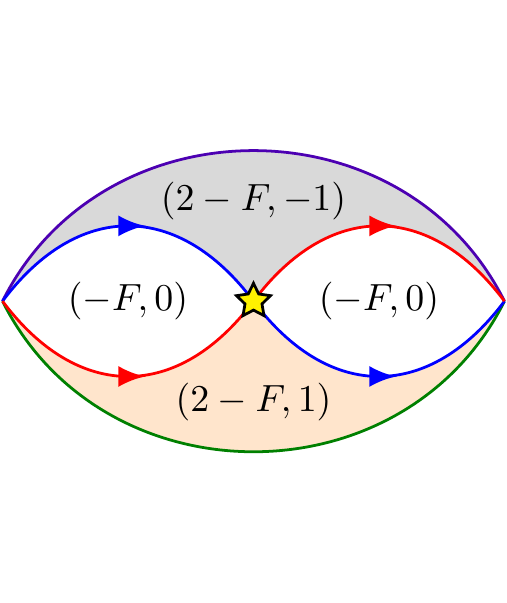}
  \caption{The top of the bubble}
  \label{subfig:liftedeye}
\end{subfigure}
\begin{subfigure}[b]{0.27\textwidth}
  \centering
\includegraphics[height=5cm]{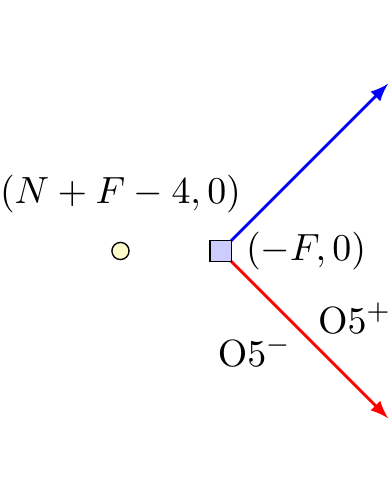}
\caption{The O5 worldvolume}
\end{subfigure}
\caption{The configuration of branes after lifting, including the pair creation of $F$ flavor branes. We use the same orientation conventions in~\subref{subfig:postlift} and~\subref{subfig:liftedeye}, so that the flavor faces on the roof of the bubble appear here as $(-F,0)$, which is the same as $(F,0)$ oppositely oriented.}
\label{fig:deconfinedbranes}
\end{figure}

We add $F$ flavors to this configuration by pair-creating $F$ pairs of
D5 branes within the larger eye and adding half of these to the
$\Sp(N-4)$ face to obtain $\Sp(N+F-4)$. Their $F$ opposite numbers are
bound to the NS5 branes $C$ and $C'$ on the roof of the bubble, where
they fill in the pair of eyes created by the boundaries of the branes
$A$ and $B$, generating $(F,0)$ faces (oriented oppositely to the
$(N+F-4,0)$ face below them) with a corresponding $\SU(F)$ symmetry.
The final brane configuration is illustrated in
figure~\ref{fig:deconfinedbranes}.

For completeness, we construct an explicit realization of the ``deconfinement bubble'' described above.
This is done merely to better illustrate our argument; we emphasize that the details of the curves have no effect on our conclusions.

We take the initial configuration of branes to be $x_5 = \pm x_4$ and $x_7 = \pm x_6$, $x_6 \ge 0$, where the upper (lower) sign applies to brane $A$ (brane $B$) and the D5 stack is located at $x_6 = x_7 = 0$.\footnote{Here we have fixed the minor angle between the branes to be $90^\circ$, but this can easily be changed.} We then deform the branes to create a triple crossing as in figure~\ref{subfig:triplecrossing}, so that their boundaries lie at $x_5 = \pm x_4 \frac{x_4^2/a^2 - 1}{x_4^2/a^2 + 1}$, where $a$ controls the size of the deformation in the $(x_4, x_5)$ plane. The deformation can be extended into the bulk as
\be
x_5 = \pm x_4 \frac{x_4^2/a^2+ x_6^2/b^2 - 1}{x_4^2/a^2+ x_6^2/b^2 + 1} \,,
\ee
where $b$ controls the size of the deformation in the $(x_6, x_7)$ plane, so that the branes return to their initial shapes asymptotically as $x_4, x_6 \to \infty$.

\begin{figure}
  \centering
  \begin{subfigure}[b]{0.52\textwidth}
    \centering
    \includegraphics[width=\textwidth]{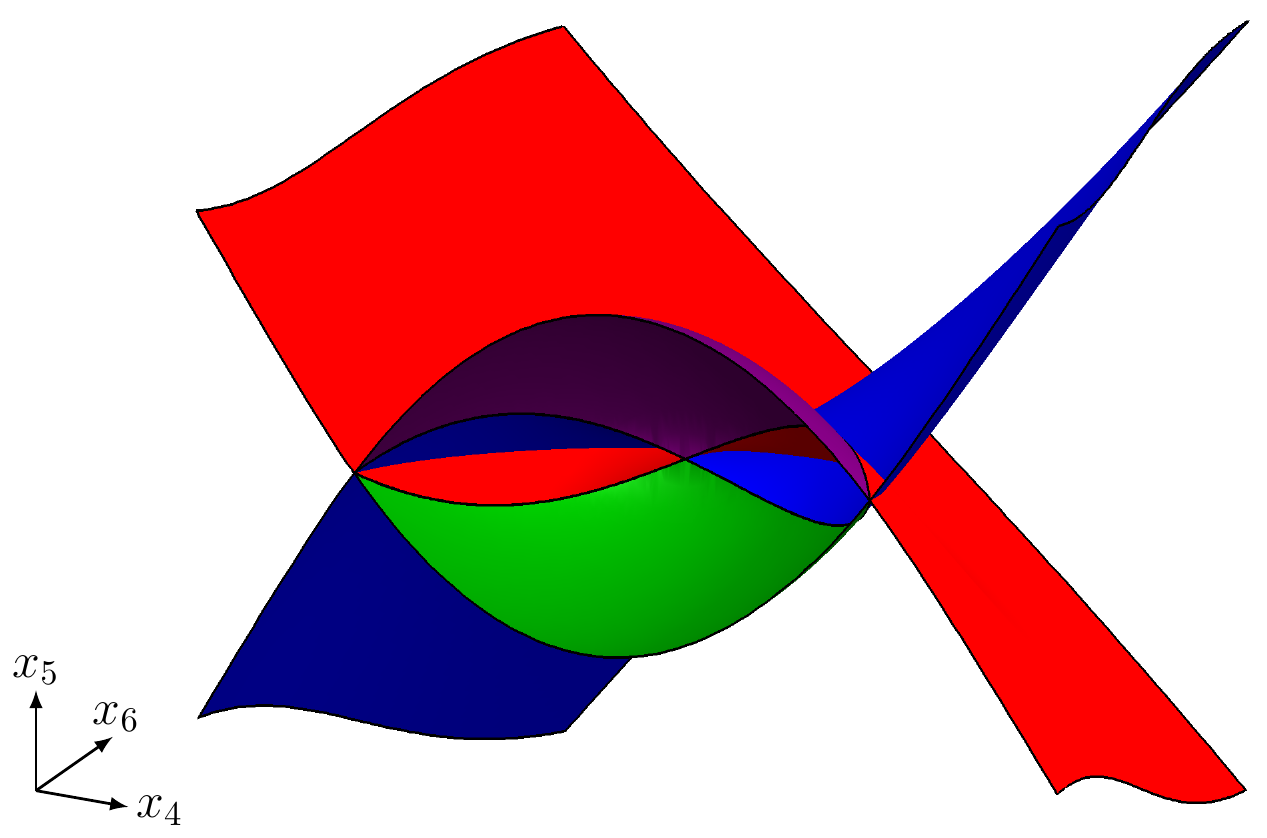}
    \caption{The $(x_4,x_5,x_6)$ directions}
    \label{sfig:decbranes456}
  \end{subfigure}
  \hfill
 \begin{subfigure}[b]{0.45\textwidth}
    \centering
    \includegraphics[width=\textwidth]{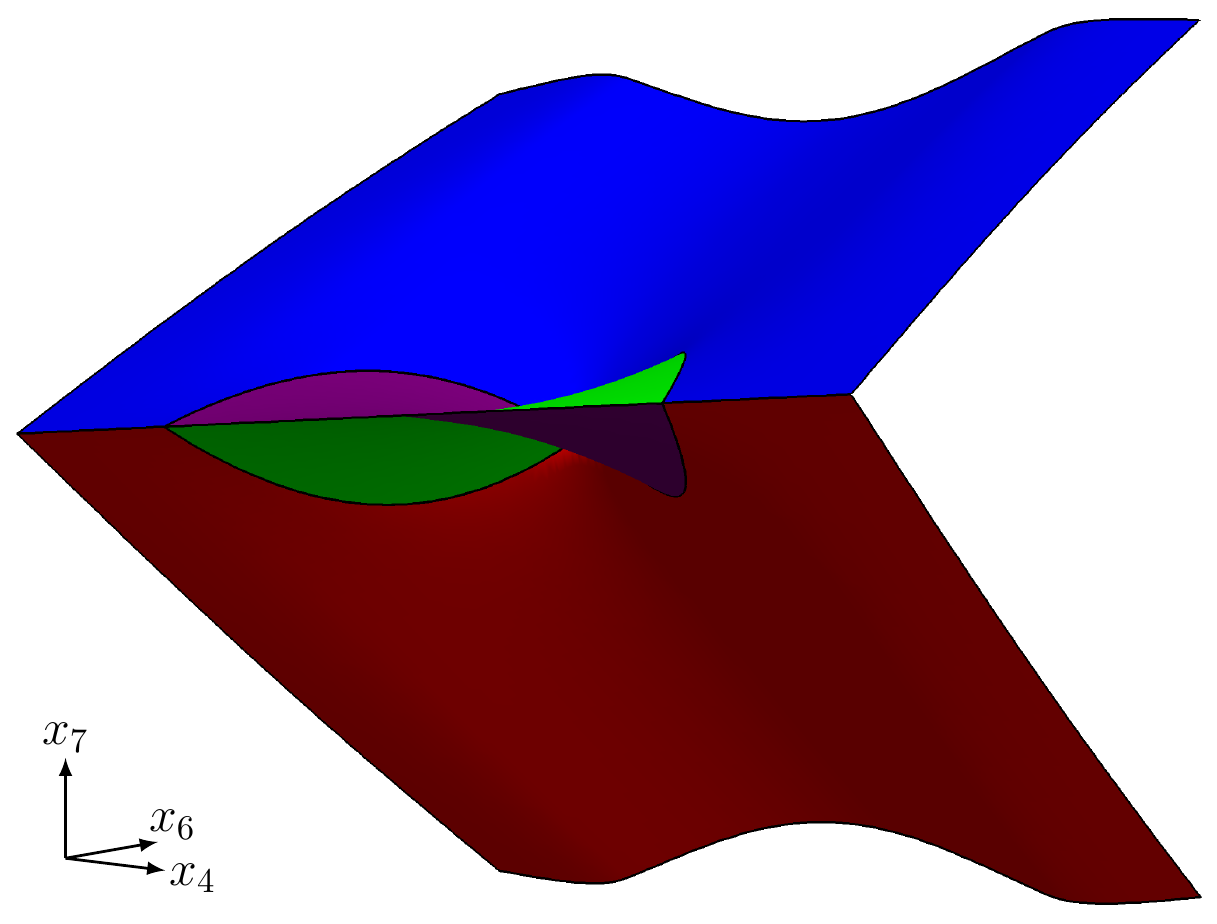}
    \caption{The $(x_4,x_6,x_7)$ directions}
    \label{sfig:decbranes467}
  \end{subfigure}
 \caption{The deconfined NS5 branes, plotted using the equations described in the text (for $c = \sqrt{2} b$). The flavor branes are not pictured: they fill in the pair of eyes where the branes $A$ and $B$ intersect the branes $C$ and $C'$.}
\end{figure}

Next, we trace out an eye surrounding the triple crossing at $x_5 = \pm \frac{a}{2} [1-x_4^2/a^2]$ for $|x_4| \le a$. After the lift, this describes boundary of the brane $C$ ($C'$) for the upper (lower) sign, and is chosen to match first derivatives with the boundaries of $A$ and $B$ at their mutual intersections. To implement the lift, we specify the displacement of $C$, $C'$ in the positive $x_6$ direction by
\be \label{eqn:Cposition}
x_5^2/a^2 + [1-x_4^2/a^2]\, x_6^2/c^2 = \frac{1}{4} [1-x_4^2/a^2]^2
\ee
for $|x_4| \le a$, where $c$ controls the size of the lift and the $x_4$-dependent prefactor of $x_6^2$ is chosen for later convenience. The branes $C$, $C'$ stretch between their boundaries on the D5 branes at $x_5 = \pm \frac{a}{2} [1-x_4^2/a^2]$ and their intersection with the branes $A$ and $B$, along the pair of curves specified by
\be \label{eqn:ABCintersection}
x_4^2/a^2\, \frac{x_4^2/a^2+ x_6^2/b^2 - 1}{x_4^2/a^2+ x_6^2/b^2 + 1} + [1-x_4^2/a^2]\, x_6^2/c^2 = \frac{1}{4} [1-x_4^2/a^2]^2 \,,
\ee
where we choose $c < 2 b$ so that the branes intersect before the pair of eyes surrounding the O5 plane close up. The final configuration of branes in the $(x_4,x_5,x_6)$ directions is plotted in figure~\ref{sfig:decbranes456}.

We now describe the positions of the branes in the $x_7$ direction. Before lifting, we had $x_7 = \pm x_6$ for branes $A$ and $B$, respectively. In order to match the slopes of the branes $A$ and $B$ with $C$ and $C'$ at their intersection point on the D5 stack, we choose $x_7 = \mp \frac{x_4}{a} x_6$ for $C$ and $C'$, respectively. To ensure that $A$ and $B$ meet the boundary of $C$ and $C'$ along the curve~(\ref{eqn:ABCintersection}), we deform the $x_7$ position of $A$ and $B$ by:
\be
x_7^2/c^2 = x_6^2/c^2 + x_5^2/a^2 -  \frac{1}{4} [1-x_4^2/a^2]^2 \,,
\ee
for $|x_4| < a$, where brane $A$ ($B$) correspond to the $x_7>0$ ($x_7 < 0$) branch of the solution. This reproduces $x_7^2 = \frac{x_4^2}{a^2} x_6^2$ on the intersection with~(\ref{eqn:Cposition}) and reduces to $x_7 = \pm x_6$ in the $x_6 \to \infty$ limit. Setting $x_4^2 = a^2$, we obtain $x_7^2/c^2 = x_6^2/c^2 + x_5^2/a^2$, which differs slightly from the initial configuration $x_7 = \pm x_6$ because $x_5 = \pm a \frac{x_6^2/b^2}{x_6^2/b^2 + 1}\ne 0$. In order to match onto this behavior for $|x_4|>a$, we take
\be
x_7^2/c^2 = x_6^2/c^2 + \frac{[3 x_4^2/a^2 - 2][2 x_4^2/a^2 + x_6^2/b^2 -2]\, x_6^2/b^2}{[x_4^2/a^2 + x_6^2/b^2 + 1]^2\, x_4^4/a^4} \,,
\ee
for $|x_4| > a$, where the second term is chosen to match first derivatives at $|x_4| = a$ and to reproduce $x_7 \to \pm x_6$ asymptotically as $x_4, x_6 \to \infty$. The final configuration of branes in the $(x_4,x_6,x_7)$ directions is plotted in figure~\ref{sfig:decbranes467}.

With a complete picture of the deconfined configuration of NS5 branes, we can read off the physics of the resulting gauge theory. From the perspective of the D5 stack near the intersection of branes $A$ and $B$, the $F$ flavor branes are ordinary minor flavor branes~\cite{Franco:2006es,Imamura:2008fd}. Thus, in addition to the bifundamental $\hat{Z}$ usually generated at the intersection point, flavored fields $P$ and $Q$ are also present, and there is a superpotential term $\hat{Z} P Q$, as in~(\ref{eqn:decW}). Similarly, from the perspective of the $F$ flavor branes near their intersection with the O5 plane, they form part of an ordinary brane tiling, hence by analogy there is an antisymmetric tensor $T$ of $\SU(F)$ at the point of intersection with the O5 plane. By the same analogy, the image pair of NS5 branes $C$ and $C'$ will generate a superpotential term $T P^2$ formed from the fields encircling their perimeter.

Thus, the configuration of branes constructed above completely reproduces the deconfined theory of~\S\ref{subsec:gaugedeconfinement}, with one exception: since the $(F,0)$ flavor faces have finite size, the $\SU(F)$ flavor symmetry is gauged. As noted in~\S\ref{subsec:gaugedeconfinement}, this describes almost the same physics as an ungauged, trivial $\SU(F)$. However, it is more convenient to work with a global $\SU(F)$, and this is easily engineered by ``puncturing'' the $(F,0)$ faces, i.e.\ by attaching long thin D5-brane tubes which end at infinity. When the $\Sp(N+F-4)$ face confines, the $\SU(F)$ flavor tubes pinch off, leaving behind a trivial $\SU(F)$ flavor symmetry, as in the gauge theory description.

We conclude that the above non-BPS, unstable configuration of branes --- which is a small deformation of the initial, BPS configuration --- produces an $\cN = 1$ gauge theory in the right universality class when we naively ignore the supersymmetry-breaking couplings. We can explain this heuristically as follows: we imagine deforming the full string theory by coupling it to external currents sourcing F-term tadpoles for the transverse scalars of the branes, arranged so that the deformed theory relaxes into the configuration of curved branes we are interested in, which is now supersymmetric due to the F-term tadpoles. In general, this deformed theory will not be UV complete, but since we are interested in the infrared physics, we ignore this and consider the effect of the deformation on the infrared fixed point. Let us suppose that the deformed theory has the same flavor symmetries as the undeformed theory, without any accidental symmetries.\footnote{Non-abelian (rank-preserving) symmetry enhancements, such as occur in~\S\ref{sec:phaseIII}, do not affect this reasoning.} In this case, the superconformal R-symmetry in the IR is a conserved symmetry of the UV theory, and the UV deformation can only induce marginal deformations in the IR. Since these deformations are forced to be flavor singlets, they are exactly marginal~\cite{Green:2010da}, and the deformed theory will flow to the same conformal manifold as the undeformed theory. In particular, in the absence of exactly marginal deformations, the two theories are in the same universality class.

Leaving aside the subtleties of this heuristic reasoning, let us see how it applies to the case at hand. Introducing the above deconfinement bubble without flavor branes ($F=0$) leads to a deformed theory with fewer flavor symmetries, due to the $\Sp(N-4)^2 \U(1)$ mixed anomalies, and we expect that the infrared physics may differ from that of the original brane configuration. Indeed, the $\Sp(N-4)$ s-confines and generates a confining superpotential~(\ref{eqn:Wconfined}) not present in the initial theory.\footnote{In the case where $\SU(N)$ is not gauged, this superpotential is irrelevant in the IR, but this is not true in general when $\SU(N)$ is gauged.} Conversely, once we add flavor branes to the deconfinement bubble in the manner described above, the symmetries of the deformed theory match those of the original theory, and indeed the infrared physics of the two descriptions match, as shown in~\S\ref{subsec:gaugedeconfinement}.

Thus, by matching the flavor symmetries of the deformed and undeformed brane configurations, we gain some control over changes to the infrared physics. With this in mind, we apply the same deconfinement construction to understand the quadruple crossing of NS5 branes encountered in~\S\ref{sec:brane-tiling}.

\subsection{Deconfining phase II} \label{sec:phaseII}

\begin{figure}
  \centering
  \begin{subfigure}[b]{0.3\textwidth}
    \centering
    \includegraphics{local-four-NS5s}
    \caption{$x_6=x_7=0$ plane}
  \end{subfigure}
  \hfill
  \begin{subfigure}[b]{0.3\textwidth}
    \centering
    \includegraphics{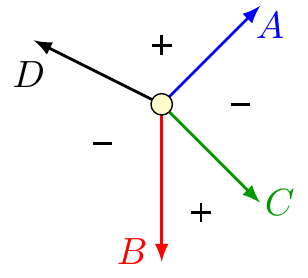}
    \caption{O5 structure in phase II}
    \label{sfig:II-O5}
  \end{subfigure}
  \hfill
  \begin{subfigure}[b]{0.3\textwidth}
    \centering
    \includegraphics{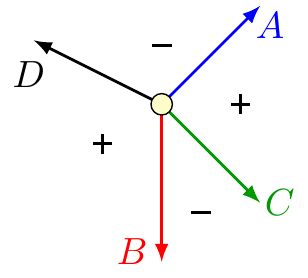}
    \caption{O5 structure in phase III}
    \label{sfig:III-O5}
  \end{subfigure}
  \caption{Neighborhood of the strongly coupled sector in phases II
    and III of the $dP_1$ orientifold.}
   \label{fig:II-III-locally}
\end{figure}

We are interested in the local physics of four NS5 branes intersecting atop an O5 plane, as in figure~\ref{fig:II-III-locally}. As above, we begin by considering the case of non-compact D5 branes with an single O5 plane in the middle, corresponding to the ``quad CFTs'' considered in~\S\ref{subsec:quadCFTs}. The deformations corresponding to deconfinement are local, so that once this configuration is understood there is no obstacle to embedding it into $T^2$.

There are two phases to consider, differing by an overall sign in the local RR charges of the O5 plane. In this section, we focus on the case shown in figure~\ref{sfig:II-O5}. Because we will only deconfine antisymmetric (rather than symmetric) tensors, the case shown in figure~\ref{sfig:III-O5} is somewhat different, and will be treated separately in~\S\ref{sec:phaseIII}.

\begin{figure}
  \centering
  \begin{subfigure}[b]{0.29\textwidth}
    \centering
    \includegraphics{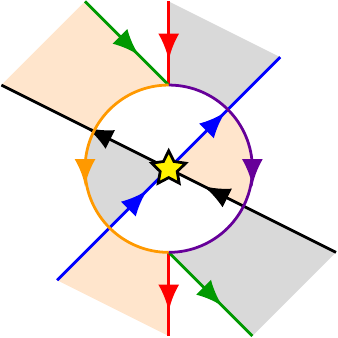}
    \vspace{0.7cm}
    \caption{The D5 worldvolume}
    \label{sfig:BC-tiling}
  \end{subfigure}
  \hfill
  \begin{subfigure}[b]{0.29\textwidth}
    \centering
    \includegraphics{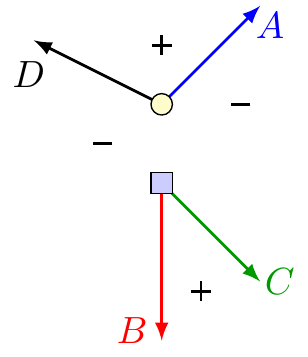}
    \vspace{0.7cm}
    \caption{The O5 worldvolume}
    \label{sfig:BC-O5}
  \end{subfigure}
  \hfill
  \begin{subfigure}[b]{0.4\textwidth}
  \includegraphics[width=\textwidth]{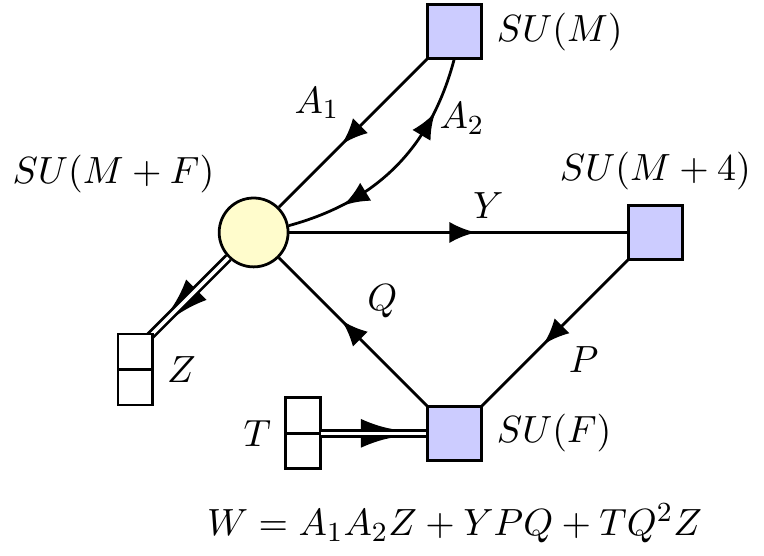}
  \caption{Quiver diagram}
  \label{sfig:BC-quiver}
  \end{subfigure}
  \caption{Local picture for the deconfinement in phase II obtained by
    deforming $B$ and $C$ together. In~\subref{sfig:BC-O5} the
    intersection of the O5 with the $x_6=x_7=0$ plane where the
    tiling lives is indicated by the yellow circle. After
    deconfinement the $B$ and $C$ branes no longer intersect over this
    point (they rather intersect over the flavor stack, indicated by
    the blue square), but $A$ and $D$ still do.}
  \label{fig:II-deconfinement}
\end{figure}

In the same spirit as~\S\ref{sec:tiling-deconfinement}, we can imagine pairing any two adjacent legs of the web diagram and blowing up a deconfinement bubble by recombining the associated NS5 branes. Analogous to deconfinement of an antisymmetric tensor, we choose to pair only legs that enclose an O5 plane with positive charge in their minor angle. Thus, for phase II we can either recombine $B$ and $C$ or recombine $A$ and $D$. We consider the former combination for now, returning to the latter shortly. After recombination, the brane configuration is the same as in~\S\ref{sec:tiling-deconfinement}, but with the branes $A$ and $D$ superimposed on top, see figure~\ref{fig:II-deconfinement}. Since the deconfinement bubble is contained within the minor angle between branes $B$ and $C$, $A$ and $D$ do not intersect it off the D5 brane stack. Moreover, the quadruple intersection has been resolved, and we can read off the resulting gauge theory exactly as in~\S\ref{sec:brane-tiling}, accounting for the flavor branes as in the same way as in~\S\ref{sec:tiling-deconfinement}. 

The resulting quiver gauge theory is shown in
figure~\ref{sfig:BC-quiver}, and the corresponding charge table is shown in table~\ref{tab:chargesII-local}, 
where the tree-level superpotential is
\be \label{eqn:WII-local}
W = A_1 A_2 Z+Y P Q+T Q^2 Z\,.
\ee
The $a$-maximized R-charge~\cite{Intriligator:2003jj} is
\be \label{eqn:UR-II-local}
U(1)_R^{\rm (sc)}=U(1)_R + \left(-\frac{4}{3}+\frac{M}{4} (\alpha_M-1)^2 + \alpha_M (\alpha_M - 2) \right) \U(1)_B + \alpha_M \U(1)_Y\,,
\ee
where $\alpha_M$ is the smallest non-negative root of
\be
9 (M+4) \alpha_M^3 - 9 M \alpha_M^2 - 3 (3M+4) \alpha_M+M=0 \,.
\ee

\begin{table}
\begin{equation*} 
  \setlength{\extrarowheight}{1pt} 
  \def\arraystretch{1.15}
  \begin{array}{c|c|ccccccc}
    & \SU(M+F) & \SU(M) & SU(M+4) & SU(F) & \U(1)_B & U(1)_X & U(1)_Y & U(1)_R\\
    \hline
    A_1 & \ov\fund & \fund & \singlet & \singlet & -\frac{1}{M+F}  & 1 & - \frac{M+4}{M+F} & 1-\frac{2}{M+F}\\
    A_2 & \ov\fund & \ov\fund & \singlet & \singlet & -\frac{1}{M+F}  & -1 & - \frac{M+4}{M+F} & 1-\frac{2}{M+F} \\
    Y & \fund & \singlet & \ov\fund & \singlet & \frac{1}{M+F} & 0 & -1+\frac{M+4}{M+F} & \frac{2}{M+F} \\
    Z & \asymm & \singlet & \singlet & \singlet & \frac{2}{M+F} & 0 & \frac{2(M+4)}{M+F} & \frac{4}{M+F} \\
    P & \singlet & \singlet & \fund & \ov\fund & \frac{1}{F} & 0 & 1 & 2 \\
    Q & \ov\fund & \singlet & \singlet & \fund & -\frac{1}{M+F} -\frac{1}{F} & 0 & - \frac{M+4}{M+F} & - \frac{2}{M+F} \\
    T & \singlet & \singlet & \singlet & \ov\asymm & \frac{2}{F} & 0 & 0 & 2
  \end{array} 
\end{equation*}
\caption{A deconfined description of the quad CFT associated to phase II}
\label{tab:chargesII-local}
\end{table}

Our task is now to argue that the infrared fixed point of this gauge
theory describes the CFT associated to the quadruple crossing of NS5
branes we started with. However, we notice immediately that this gauge
theory depends on the number of flavors, $F$, introduced during
deconfinement. As a first consistency check, we show that the infrared
fixed point depends only on the parity of $F$, and that the $\SU(F)$
global symmetry is otherwise trivial.

\begin{figure}
  \centering
  \begin{subfigure}[b]{0.3\textwidth}
    \centering
    \includegraphics{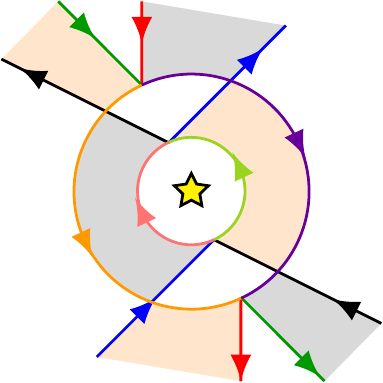}
    \vspace{0.3cm}
    \caption{A-D deconfinement}
    \label{sfig:AD-tiling-SD}
  \end{subfigure}
    \hfill
  \begin{subfigure}[b]{0.3\textwidth}
    \centering
    \includegraphics{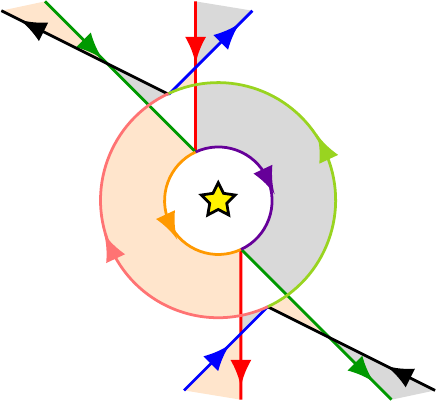}
    \vspace{0.25cm}
    \caption{Bubble size swap}
    \label{sfig:AD-tiling-deconfined}
  \end{subfigure}
    \hfill
   \begin{subfigure}[b]{0.3\textwidth}
    \centering
    \includegraphics{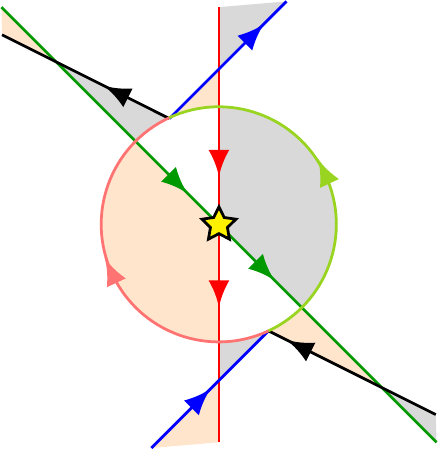}
    \caption{B-C reconfinement}
    \label{sfig:AD-tiling}
  \end{subfigure}
  \caption{Deconfinement duality in the brane tiling for the local configuration in
    phase II. \subref{sfig:AD-tiling-SD}~We deconfine the antisymmetric tensor $Z$ where NS5 branes A and D cross. \subref{sfig:AD-tiling-deconfined}~Seiberg duality on the mirror-image pair of wedge-shaped faces corresponds to passing the boundaries of the two deconfinement bubbles through each other. \subref{sfig:AD-tiling}~Reconfining the central face, we obtain a dual description.}
  \label{fig:deconfinement-transition}
\end{figure}

Our proof works by relating this resolved configuration of branes to the configuration where branes $A$ and $D$ are recombined into a bubble and branes $B$ and $C$ cross the O5 plane, illustrated in figure~\ref{sfig:AD-tiling}. To do so, we first deconfine the antisymmetric tensor $Z$ with $G$ flavors, generating a new $\Sp(M+F+G-4)$ face, where we require $F + G \equiv M \mod 2$. We then take the Seiberg dual of the $\SU(M+F)$ node, which turns into an $\SU(M+G)$ node. Keeping track of the interactions and integrating out massive matter, we find that the $\Sp(M+F+G-4)$ node is again s-confining, where now the roles of $\SU(F)$ and $\SU(G)$ have flipped. Moving to the s-confined description, all $\SU(F)$-charged states become massive, and we conclude that the $\SU(F)$ is trivial. The resulting gauge theory corresponds to the deconfined brane tiling~\ref{sfig:AD-tiling}, now with $G$ flavors. This chain of dualities, which we refer to as ``deconfinement duality'', is closely related to Seiberg duality, and does not affect the infrared physics. Notice that $G$ still encodes the parity of $F$, due to the constraint $F + G \equiv M \mod 2$, but otherwise different values of $F$ correspond to the same infrared physics. Likewise, by construction $\SU(G)$ is trivial, and the physics only depends on $(-1)^G = (-1)^{M+F}$.

\begin{figure}
  \centering
  \begin{subfigure}{0.23\textwidth}
    \centering
    \includegraphics{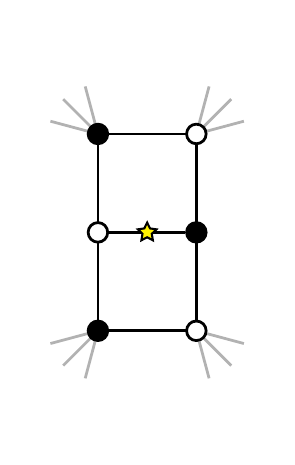}
    \caption{Dimer for fig.~\ref{sfig:BC-tiling}}
    \label{sfig:BC-dimer}
  \end{subfigure}
  \hfill
  \begin{subfigure}{0.23\textwidth}
    \centering
    \includegraphics{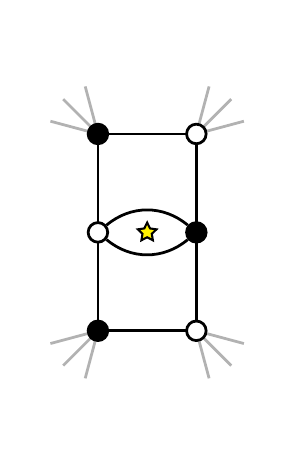}
    \caption{Dimer for fig.~\ref{sfig:AD-tiling-SD}}
    \label{sfig:AD-dimer-SD}
  \end{subfigure}
  \hfill
  \begin{subfigure}{0.23\textwidth}
    \centering
    \includegraphics{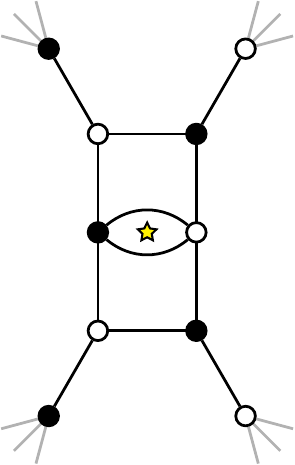}
    \caption{Dimer for fig.~\ref{sfig:AD-tiling-deconfined}}
    \label{sfig:AD-dimer-deconfined}
  \end{subfigure}
    \hfill
   \begin{subfigure}{0.23\textwidth}
    \centering
    \includegraphics{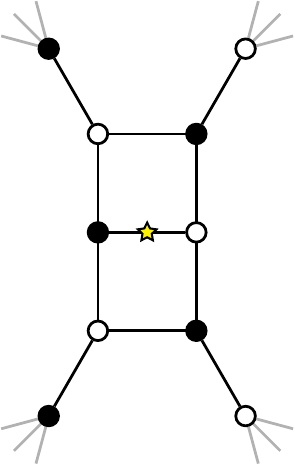}
    \caption{Dimer for fig.~\ref{sfig:AD-tiling}}
    \label{sfig:AD-dimer}
  \end{subfigure}
  \caption{Deconfinement duality from the point of view of the
    dimer model. \subref{sfig:BC-dimer}~We start with dimer model corresponding to the the B-C deconfined brane tiling
    shown in~\ref{sfig:BC-tiling}. \subref{sfig:AD-dimer-SD}~We deconfine the antisymmetric tensor. 
    \subref{sfig:AD-dimer-deconfined}~We apply Seiberg duality
    to the $\SU$ gauge group (recall figure~\ref{sfig:urbanRenewal}), and
    integrate out massive matter.
    \subref{sfig:AD-dimer}~Reconfining the central $\Sp$ node,
    we obtain the dimer model corresponding
    to the A-D deconfined brane tiling shown in
    figure~\ref{sfig:AD-tiling}.}
  \label{fig:deconfinement-transition-dimer}
\end{figure}

This chain of dualities can be described in the brane tiling using deconfinement bubbles, as in figure~\ref{fig:deconfinement-transition}, and likewise admits a simple description in the dimer model, see figure~\ref{fig:deconfinement-transition-dimer}.
However, we emphasize that the above argument is based on field theory reasoning, and does not rely on the string theory realization. 

The residual dependence on the parity of $F$ is important: there are additional data associated to the original BPS configuration that we have so far ignored. In particular, the dual Calabi-Yau singularity has moduli associated to Wilson lines of both $B_2$ and $C_2$ before orientifolding. We argued in~\S\ref{sec:brane-tiling} that the $B_2$ Wilson lines are projected to discrete values after orientifolding and contribute to the $[H]$ discrete torsion, where in the dual brane tiling these discrete Wilson lines correspond to NS5 brane positions. So far we have ignored the analogous $C_2$ discrete Wilson lines, which contribute to $[F]$ discrete torsion by the same argument. These correspond to Wilson lines on the NS5 branes in the dual brane tiling~\cite{Imamura:2007dc}, and we expect that they are likewise projected to discrete values by the O5 involution. A natural guess is that $(-1)^F$ is fixed by a combination of Wilson lines localized at the quadruple intersection, hence it contributes to RR discrete torsion. This view will be confirmed when we study the S-duality properties of these theories.

Besides the trivial $\SU(F)$ symmetry, the remaining $\SU(M)\times\SU(M+4) \times \U(1)^3 \times \U(1)_R$ flavor symmetry group displayed in table~\ref{tab:chargesII-local} matches the brane picture laid out in~\S\ref{subsec:quadCFTs}.
Moreover, the gauge theory enjoys an accidental non-abelian
enhancement to $\Sp(2 M) \supset \SU(M) \times \U(1)_X$ not manifest in the brane picture, in perfect agreement with the expectations outlined in~\S\ref{subsec:quadCFTs}.

We now describe the moduli space of these theories. The gauge-invariant chiral operators consist of mesons and baryons of $\SU(M+F)$.\footnote{Here we ignore chiral operators containing glueballs ($W_\alpha W^\alpha$) for simplicity.} However, due to the triviality of $\SU(F)$, only the $\SU(F)$-invariant subsector of these operators can correspond to flat directions in the moduli space. The other directions break $\SU(F)$, hence they are lifted by quantum effects. The $\SU(F)$-invariant mesons and baryons are
\be \label{eqn:quadCFToperators}
\Phi_i = A_i Y \;\;,\;\; \mathcal{O}_k = A_1^k A_2^{M-k} Q^F \;\;,\;\; \widetilde{\mathcal{O}}_k = Z^{\frac{F+k-4}{2}} Y^{M+4-k} \,.
\ee
In the former case, $0 \le k \le M$ is arbitrary, whereas in the latter case $0 \le k \le M+4$ must satisfy $(-1)^k = (-1)^F$.\footnote{We assume $F \ge 3$ for simplicity. For $F=1$ ($F=2$), $\widetilde{\mathcal{O}}_{1}$ ($\widetilde{\mathcal{O}}_{0}$) is not defined, and we should replace $\widetilde{\mathcal{O}}_1 \to P$ ($\widetilde{\mathcal{O}}_0 \to T$), which is an $\SU(F)$-invariant only for this value of $F$.}

Note that the due to the F-term constraint for $Z$, the baryons $\mathcal{O}_k$ combine into a single irrep of $\Sp(2M)$, the $M$-index antisymmetric tensor of dimension $\frac{(2M+2)!}{(M+1)!(M+2)!}$. This curious fact will have an interesting analog in phase III in~\S\ref{sec:phaseIII}.

\begin{figure}
\centering
\includegraphics[width=\textwidth]{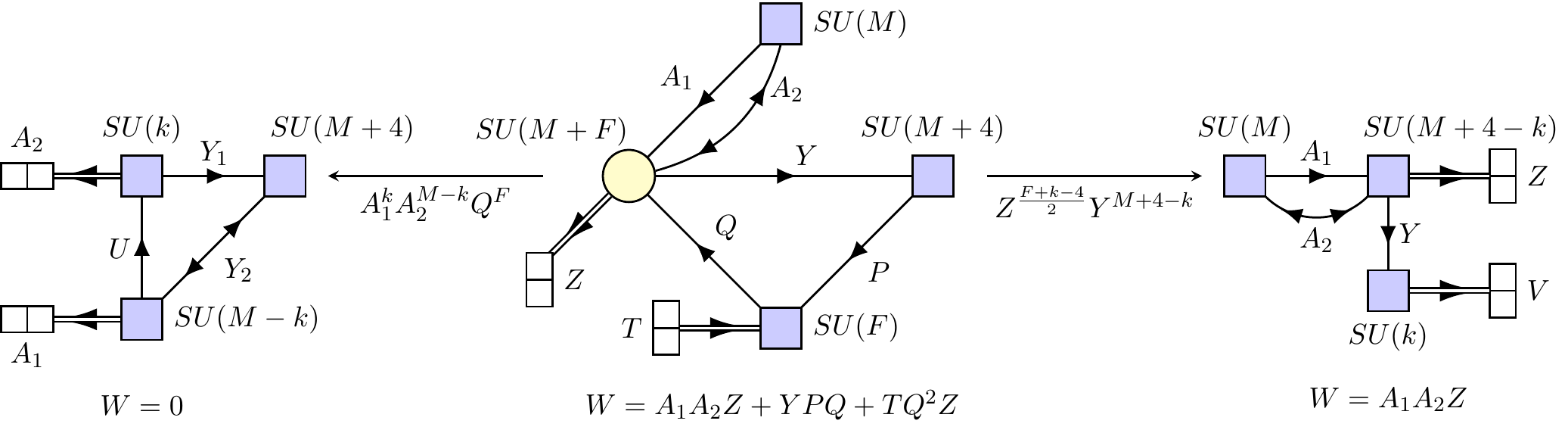}
\caption{Partial resolutions of the quad CFT for phase II. The left hand resolutions correspond to recombining B-D / A-C (cf.\ figures~\ref{sfig:BDresolution}, \ref{sfig:ACresolution}), whereas the right-hand resolutions correspond to recombining B-C / A-D (cf.\ figures~\ref{sfig:partial_resolved_quad}, \ref{sfig:ADresolution}). The fields $U$ and $Y$ are Goldstone bosons for $\SU(M) \to \SU(M-k)\times\SU(k)$ and $\SU(M+4) \to \SU(M+4-k)\times\SU(k)$ breaking, respectively.}
\label{fig:quadresolutions}
\end{figure}

We can identify the baryons with partial resolutions as
in~\S\ref{subsec:quadCFTs}. The procedure is similar to that of~\S\ref{sec:classical-phases-torsion}. A partial resolution which combines two adjacent legs of the web diagram corresponds to vevs for the baryons where the corresponding NS5 branes cross.\footnote{For this purpose, we consider the edge of the deconfinement bubble as equivalent to either of its constituent NS5 branes.} 
Thus, $\mathcal{O}_0$ describes the B-D partial resolution (figure~\ref{sfig:BDresolution}), whereas $\mathcal{O}_M$ describes the A-C partial resolution (figure~\ref{sfig:ACresolution}), and $\mathcal{O}_k$ for $0<k<M$ interpolates between the two, much as in~\S\ref{sec:classical-phases-torsion}.
Turning on a vev for $\mathcal{O}_k$, the $\SU(M+F)$ gauge group is completely Higgsed, and all $\SU(F)$-charged matter becomes massive. The remaining massless fields, shown on the left side of figure~\ref{fig:quadresolutions}, do not interact, and fall in the expected representations of the flavor brane symmetries inherited from the quad CFT, c.f.\ figures~\ref{sfig:partial_resolved_quad}, \ref{fig:quadrecombinations}, with the necessary addition of Goldstone bosons due to the breaking of global symmetries (these will be eaten when the flavor branes are gauged).

Likewise $\widetilde{\mathcal{O}}_0$ describes the B-C partial resolution (figure~\ref{sfig:partial_resolved_quad}), whereas $\widetilde{\mathcal{O}}_{M+4}$ describes the A-D partial resolution (figure~\ref{sfig:ADresolution}), and  $\widetilde{\mathcal{O}}_k$ interpolates between the two. Turning on a vev for $\widetilde{\mathcal{O}}_k$, the gauge group is Higgsed to $\Sp(F+k-4)$. Integrating out the massive matter, $\Sp(F+k-4)$ reconfines much as in~\S\ref{subsec:gaugedeconfinement}, generating a composite antisymmetric tensor $V$ without a superpotential. The entire spectrum --- including a tree-level superpotential --- is shown on the right-hand side of figure~\ref{fig:quadresolutions}. As before, the results agree with figures~\ref{sfig:ACresolution}, \ref{sfig:BDresolution}), except for the inevitable appearance of Goldstone bosons.

Finally, turning on a rank $k$ vev for $\Phi_1$ (equivalently $\Phi_2$), we obtain the rank-$(M-k)$ theory together with a residual $U(k)$ flavor symmetry and non-interacting chiral superfields charged under it and under the global symmetries of the CFT, in agreement with the behavior of the parent theory discussed in~\S\ref{subsec:quadCFTs}.

Thus, the candidate CFT constructed above has the same global symmetries and moduli space as anticipated for the quad CFT in~\S\ref{subsec:quadCFTs}.
  Moreover, one can show using the $a$-maximized R-charge~(\ref{eqn:UR-II-local}) that this CFT has no exactly marginal deformations. Therefore, the deconfinement trick of~\S\ref{sec:tiling-deconfinement} appears to generate a gauge theory \emph{in the same universality class} as the theory on four intersecting NS5 branes. Since the UV gauge theory (in contrast to the IR fixed point) depends on $F$, this is the strongest claim that can be made, and is consistent with the intuition that the higher multiplicity intersection of NS5 branes is intrinsically strongly coupled. We obtain further non-trivial evidence for this claim in~\S\ref{sec:dP1-S-duality}, where we consider S-duality of the $dP_1$ orientifold phases constructed from this CFT.

\subsection{Deconfining phase III}
\label{sec:phaseIII}

\begin{figure}
  \centering
  \begin{subfigure}[b]{0.3\textwidth}
    \centering
    \includegraphics[width=\textwidth]{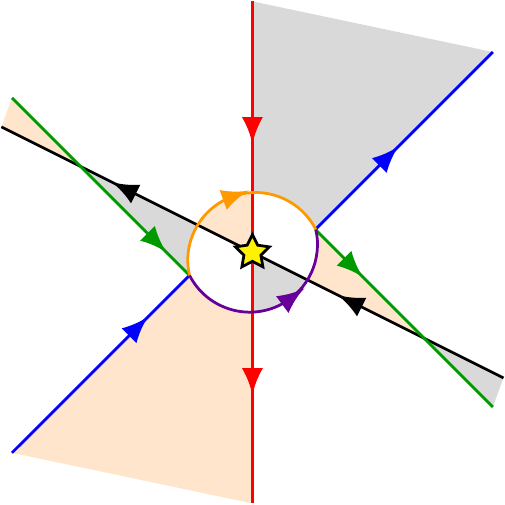}
    \caption{$x_6=x_7=0$ slice}
    \label{sfig:AC-tiling}
  \end{subfigure}
  \hfill
  \begin{subfigure}[b]{0.25\textwidth}
    \centering
    \includegraphics[width=\textwidth]{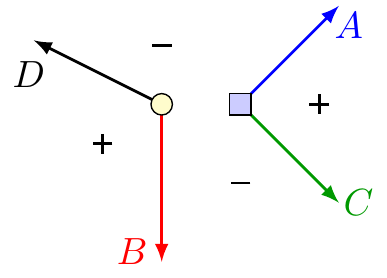}
    \vspace{0.1cm}
    \caption{$x_4=x_5=0$ slice}
    \label{sfig:AC-O5}
  \end{subfigure}
  \hfill
  \begin{subfigure}[b]{0.4\textwidth}
  \centering
  \includegraphics[width=\textwidth]{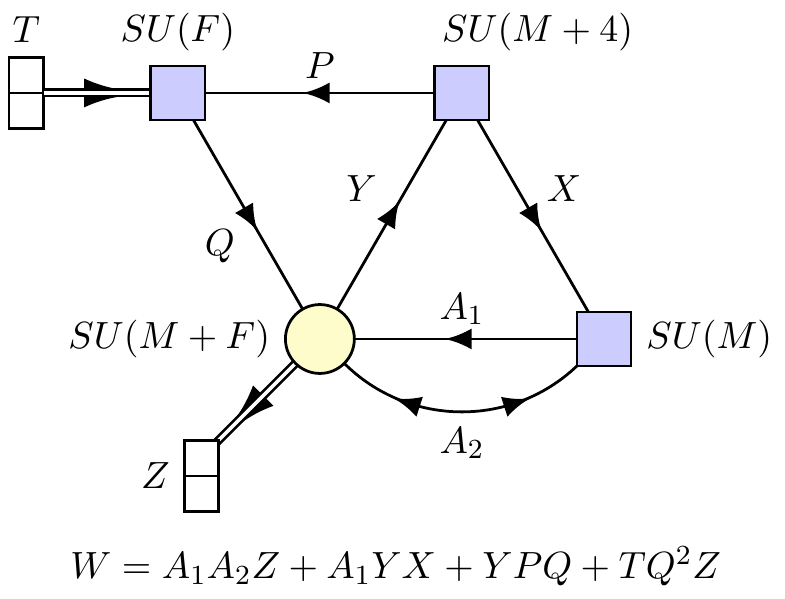}
  \caption{Quiver diagram}
  \label{sfig:localquiverIII}
  \end{subfigure}
  \caption{Local picture for the deconfinement in phase III obtained
    by deforming $A$ and $C$ together. After deconfinement the $A$ and
    $C$ branes no longer intersect the intersection of the tiling with
    the orientifold plane (depicted as the yellow circle in
    \subref{sfig:AC-O5}) but $B$ and $D$ still do.}
  \label{fig:III-deconfinement}
\end{figure}

We now construct a deconfined description of the quad CFT for phase III. Much of the discussion is analogous to phase II, described above, so we will be brief. The worldvolume of the O5 plane coincident with the quadruple intersection of NS5 branes is shown in figure~\ref{sfig:III-O5}. As before, we blow up a deconfinement bubble by pairing two adjacent NS5 branes enclosing an O5$^+$ in their minor angle. In this case, we can form the bubble using either $B$ and $D$ or $A$ and $C$. The two choices will be related via deconfinement duality, hence their infrared physics is the same. The configuration with $A$ and $C$ combined into a bubble is shown in figure~\ref{sfig:AC-tiling}--\subref{sfig:AC-O5}, and the resulting quiver gauge theory is shown in figure~\ref{sfig:localquiverIII}, with the charge table shown in table~\ref{tab:chargesIII-local} 
and the tree-level superpotential
\be
W = A_1 A_2 Z + A_1 Y X + Y P Q + T Q^2 Z \,.
\ee
Note that this is a superpotential deformation of table~\ref{tab:chargesII-local} plus the gauge singlet meson $X$, where we have chosen to redefine
\be
\U(1)_Y^{\rm (III)} = \U(1)_Y^{\rm (II)}-\frac{M+4}{2}\, \U(1)_B
\ee
for later convenience. The $a$-maximized R-charge is now
\be \label{eqn:UR-III-local}
U(1)_R^{\rm (sc)}=U(1)_R + \left(-\frac{4}{3}+\frac{M}{4} (1-\widetilde\alpha_M^2) \right) \U(1)_B + \widetilde\alpha_M \U(1)_X\,,
\ee
where $\widetilde\alpha_M$ is the smallest non-negative root of
\be
9 M \widetilde\alpha_M^3 - 9 (M+4) \widetilde\alpha_M^2 - 3 (3M+8) \widetilde\alpha_M+(M+4)=0 \,.
\ee

\begin{table}
\begin{equation*} 
  \setlength{\extrarowheight}{1pt} 
  \def\arraystretch{1.15}
  \begin{array}{c|c|ccccccc}
    & \SU(M+F) & \SU(M) & SU(M+4) & SU(F) & \U(1)_B & U(1)_X & U(1)_Y & U(1)_R\\
    \hline
    A_1 & \ov\fund & \fund & \singlet & \singlet & -\frac{1}{M+F}  & 1 & - \frac{M+4}{2(M+F)} & 1-\frac{2}{M+F}\\
    A_2 & \ov\fund & \ov\fund & \singlet & \singlet & -\frac{1}{M+F}  & -1 & - \frac{M+4}{2(M+F)} & 1-\frac{2}{M+F} \\
    X & \singlet & \ov\fund & \fund & \singlet & 0 & -1 & 1 & 1 \\
    Y & \fund & \singlet & \ov\fund & \singlet & \frac{1}{M+F} & 0 & -1+\frac{M+4}{2(M+F)} & \frac{2}{M+F} \\
    Z & \asymm & \singlet & \singlet & \singlet & \frac{2}{M+F} & 0 & \frac{M+4}{M+F} & \frac{4}{M+F} \\
    P & \singlet & \singlet & \fund & \ov\fund & \frac{1}{F} & 0 & 1-\frac{M+4}{2 F} & 2 \\
    Q & \ov\fund & \singlet & \singlet & \fund & -\frac{1}{M+F} -\frac{1}{F} & 0 & \frac{M+4}{2 F} - \frac{M+4}{2(M+F)} & - \frac{2}{M+F} \\
    T & \singlet & \singlet & \singlet & \ov\asymm & \frac{2}{F} & 0 & -\frac{M+4}{F} & 2
  \end{array} 
\end{equation*}
\caption{A deconfined description of the quad CFT associated to phase III}
\label{tab:chargesIII-local}
\end{table}

Because of the close relationship between tables~\ref{tab:chargesIII-local} and \ref{tab:chargesII-local}, many of the consistency checks performed in the previous section go through analogously for this theory. Consequently, we only discuss the novel aspects of this theory, leaving the basic checks as an exercise.

Applying deconfinement duality, we obtain a dual theory with $G \equiv M + F \mod 2$ flavors which is isomorphic to the original theory.
Thus, for even $M$ the theory is self-dual (with a non-trivial mapping of the operator spectrum to itself), whereas for odd $M$ the two choices of flavor parity are dual to each other.

 As in phase II, the manifest non-R symmetries are $\SU(M+4)\times\SU(M)\times\U(1)^3$. However, contrary to our expectations from~\S\ref{subsec:quadCFTs},
 there is no accidental non-abelian enhancement visible in the gauge theory. 
 Naively, we expect an enhancement $\SU(M+4) \times \U(1) \to \SO(2(M+4))$ for some $\U(1) \subset \U(1)^3$, but the gauge theory has no such symmetry.

The resolution is that the enhanced symmetry emerges accidentally in
the infrared. In fact, the deconfinement duality discussed above
already provides some hint of this. The even-$M$ self-duality maps
$\SU(M+4)\times \U(1)_Y$ to its charge conjugate, leaving the other
global symmetries invariant, hence the infrared CFT can have only
self-conjugate (real or pseudo-real) representations of $\SU(M+4)\times \U(1)_Y$. This is
consistent with an accidental enhancement to $\SO(2(M+4))$ because the
latter has only self-conjugate representations for even
$M$. Conversely, for odd $M$ there is no self-duality, in agreement
with the fact that $\SO(2(M+4))$ has complex spinor representations
for odd $M$.

We obtain further evidence for such an enhancement by examining the gauge-invariant operators parameterizing the moduli space. Similar to phase II, we find the following $\SU(F)$-invariant mesons and baryons:
\be \label{eqn:moduli-III}
\Phi_1 = X \;\;,\;\; \Phi_2 = A_2 Y \;\;,\;\; \mathcal{O}_k = A_1^k A_2^{M-k} Q^F \;\;,\;\; \widetilde{\mathcal{O}}_k = Z^{\frac{F+k-4}{2}} Y^{M+4-k} \,.
\ee
Here $\Phi_1$ and $\Phi_2$ transform as $\fund_1$ and $\ov\fund_{-1}$ under $\SU(M+4)\times\U(1)_Y$, respectively, hence they fill out a vector of $\SO(2(M+4))$. Likewise, $\mathcal{O}_k$ is not charged under $\SU(M+4)\times\U(1)_Y$, hence it is an $\SO(2(M+4))$ singlet. Finally, $\widetilde{\mathcal{O}}_k$ transforms as 
\be
[k]_{k-\frac{M+4}{2}} \;\;,\;\; (-1)^k = (-1)^F
\ee
under $\SU(M+4)\times\U(1)_Y$, where $[k]$ denotes the $k$-index antisymmetric tensor representation. Collecting all permissible values of $k$, we recognize the spinor representation $S$ ($\bar{S}$) of $\SO(2 (M+4))$ for even $F$ (odd $F$). Thus, the family of operators $\widetilde{\mathcal{O}}_k$ combine to form a single irrep of $\SO(2 (M+4))$, a spinor of dimension $2^{M+3}$! 

\begin{figure}
\centering
\includegraphics[width=0.85\textwidth]{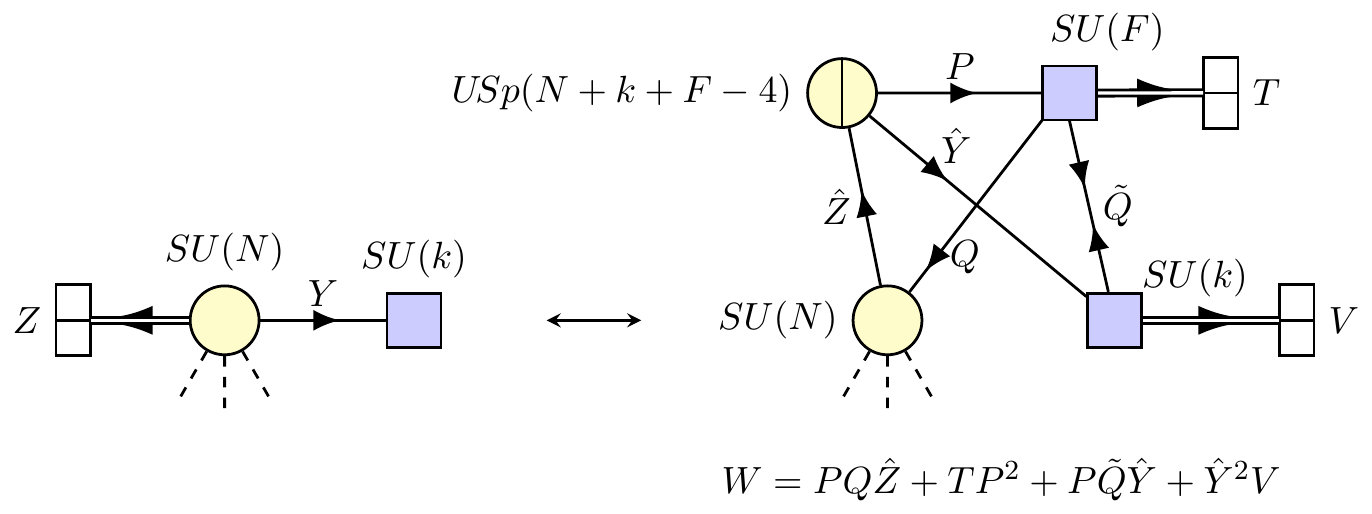}
\caption{A variant of deconfinement where an antisymmetric tensor is deconfined together with $k$ flavors. Figure~\ref{fig:antisymmetric-deconfinement-quiver} is the special case $k=0$.}
\label{fig:deconfinementvariant}
\end{figure}

Note that the spinor representations $S$ and $\bar{S}$ are related by
the $\bZ_2$ outer automorphism of $SO(2 (M+4))$. This suggests that
the two $F$-parities describe equivalent quad CFTs related by a
non-trivial mapping on the operator spectrum (the $\bZ_2$ outer
automorphism). We have already seen that this is the case for odd
$M$. To prove it for even $M$ we employ the deconfinement variant
shown in figure~\ref{fig:deconfinementvariant}, where an antisymmetric
tensor is deconfined at the same time as $k$ flavors. We apply this
variant to deconfine $Z$ together with $M+4-k$ out of the $M+4$ $Y$'s,
which explicitly breaks
$\SU(M+4) \to \SU(M+4-k)_1 \times \SU(k)_2\times U(1)_A$, so that
$\SU(M+4)$ is an accidental non-abelian enhancement in the confined
theory. It is convenient to take linear combinations $\U(1)_{1,2}$ of
$\U(1)_Y$ and $\U(1)_A$ such that $\Phi_1$ decomposes as
$(\fund_1,\singlet_0) \oplus (\singlet_0,\fund_1)$ under
$\U(M+4-k)_1 \times \U(k)_2 \subset \U(M+4)$ and likewise $\Phi_2$
decomposes as
$(\ov\fund_{-1},\singlet_0) \oplus (\singlet_0,\ov\fund_{-1})$.

The gauge group of the deconfined theory is $\SU(M+F) \times \Sp(2 M+F+G-k)$, where $G$ is the number of flavors introduced during deconfinement, constrained by $(-1)^G = (-1)^{F+k}$. The Seiberg-dual of the $\SU(M+F)$ gauge group factor is $\SU(M+G)$, where the $\Sp(2 M+F+G-k)$ factor retains a confining spectrum in the dual. Reconfining, we obtain a description with $G$ flavors and an accidental enhancement $\U(M+4-k)_1 \times \ov{\U(k)}_2 \to \U(M+4)$. This dual theory is isomorphic to the original theory up to charge conjugation of the $\U(k)$ factor. Choosing any odd value of $k$, we obtain a duality between even and odd flavor parities, hence the two are equivalent.

In fact, this variant of deconfinement duality (the usual case being $k=M+4$) is powerful enough to prove --- under mild assumptions --- that the infrared theory has the full $\SO(2 (M+4))$ symmetry, since the dual descriptions  have different $U(M+4)$ subgroups as manifest symmetries, and the smallest group which contains all of these subgroups is $\SO(2(M+4))$.\footnote{Since deconfinement and Seiberg duality are integral identities at the level of the superconformal index~\cite{Dolan:2008qi}, this is sufficient to prove that the index respects the full $\SO(2(M+4))$ symmetry, assuming that it can be reliably computed in the UV theory. We have computed low lying states in the index for small values of $M$, verifying that they fill out complete $\SO(2(M+4))$ representations.} Moreover, we can verify that the mapping induced by the deconfinement duality lies within $\SO(2(M+4))$ for even $k$ --- when the flavor parity is unchanged --- whereas it is a non-trivial $\SO(2(M+4))$ outer automorphism for odd $k$ --- when the flavor parity flips. To show this, we consider the standard embedding of $U(n)$ inside $\SO(2n)$ via the complexification
\be
z_i = x_{2i-1} + i x_{2i} \,.
\ee
Charge conjugating $k$ of the $z_i$ coordinates corresponds to flipping the sign of $k$ of the $x_{2j}$'s, hence for even $k$ the transformation lies within $\SO(2n)$, whereas for odd $k$ it lies in the disconnected component of $O(2k)$, i.e.\ it is a parity flip which exchanges the two Weyl-spinor irreps of $\SO(2n)$, $S$ and $\bar{S}$.

It is interesting to compare the physics we have just described with the quad CFT for phase II, where we previously argued that the two choices of flavor parity give distinct CFTs. The difference arises because the quad CFT for phase II is not self-dual under deconfinement duality. Applying the above variant of the duality leads to a distinct dual description with no $\U(M+4-k)_1 \times \U(k)_2 \to \U(M+4)$ enhancement of the manifest flavor symmetries, hence we cannot conclude that the two choices of flavor parity are equivalent. In fact, the two flavor parities give distinct spectra of the $\widetilde{\mathcal{O}}_k$ baryons under the manifest global symmetries, whereas a further non-abelian enhancement is not consistent with the operator spectrum,
 therefore --- in the absence of rank-enhancing accidental symmetries --- we conclude that the two flavor parities of phase II give rise to inequivalent CFTs.

Finally, we note that deformations and gaugings of this CFT may break $\SO(2(M+4))$, leading to distinct results for the different flavor parities. For instance, phase II can be reached by giving $X$ a mass with an additional gauge singlet $\bar{X}$. Since this breaks $\SO(2(M+4)) \to \U(M+4)$, the result will depend on the parity of $F$. In particular, a deconfinement duality that flips the $F$-parity will also change the deformation to a different deformation in which parts of $\bar{X}$ are coupled to components of the composite meson $\Phi_2$. Gauging the flavor faces in the brane tiling also breaks $\SO(2(M+4))$, and will have a similar effect.

\subsection{On symmetric tensor deconfinement}
\label{sec:symm-deconfinement}

We note in passing that there is a symmetric tensor analog of deconfinement~\cite{Luty:1996cg,Sakai:1997xs}, which is based on the
observation that $\SO(N+4)$ with $N$ vectors $\hat{Z}$ in the
fundamental representation of $\SU(N)$ confines with the composite
$Z = \hat{Z}^2$ in the $\symm$ representation of $\SU(N)$. There are
four branches of moduli space, two equivalent branches with a runaway
superpotential and two equivalent branches with no
superpotential~\cite{Seiberg:1994pq,Intriligator:1995id}. Thus,
treating the $\SO(N+4)$ theory with $N$ vectors on one of the latter
branches analogously to the \mbox{s-confining} $\Sp(N-4)$ theory with
$N$ vectors, the procedure of~\S\ref{subsec:gaugedeconfinement} can be repeated to obtain a
deconfined description of the symmetric tensor.

However, there are conceptual differences between the two cases. Since
there is no dynamically generated superpotential, $F>0$ is no longer strictly necessary. Instead, the case $F=0$ is distinguished by reduced flavor symmetries in the UV theory, which only conserves $\U(1)$ symmetries under which the symmetric tensor is uncharged, with the remainder emerging as accidental symmetries in the infrared.
Moreover, since the vector representation is not a faithful representation of $\SO(N)$, it cannot screen all Wilson loops, hence there are gauge-invariant order-parameters which distinguish between confining and Higgs phases and the moduli space (in particular the K\"ahler potential) will not smoothly interpolate between the two~\cite{Csaki:1996zb}.

Nonetheless, the success of the approach
in~\S\ref{sec:tiling-deconfinement}--\ref{sec:phaseIII} in describing
the quad CFTs suggests that a similar construction could be done using
symmetric tensor deconfinement. In particular, none of the details
of~\S\ref{sec:tiling-deconfinement} depended strongly on the
distinction between symmetric and antisymmetric tensors, and one can
check that theories with the correct flavor symmetries and moduli
space can be built along the lines
of~\S\ref{sec:phaseII}--\ref{sec:phaseIII}.

The primary issue with these theories is their lack of an apparent analog of flavor parity, since an $\SU(N)$ symmetric tensor can be deconfined with any number of flavors of either parity. As noted above, flavor parity will play an important role in determining the RR torsion when the quadruple crossing is incorporated into the brane tiling. Consistency demands that there is an analogous quantity in the symmetric tensor deconfined description.

A natural resolution to this puzzle is that --- like the deconfined description of a single symmetric tensor --- the theory has two supersymmetric branches of moduli space, which are no longer equivalent, but are distinguished by different spectra of light operators. For instance, consider the symmetric tensor description of phase III, whose quiver is similar to figure~\ref{sfig:BC-quiver} with symmetric tensors replacing the antisymmetric tensors and slightly different ranks for the nodes. The $\SO(2(M+4))$ flavor symmetry is now manifest but the baryons do not form a spinor of $\SO(2(M+4))$. Instead, they occupy the $(M+4)$-index antisymmetric tensor representation, which splits into self-dual (imaginary self-dual) and anti-self-dual (imaginary anti-self-dual) irreps for even $M$ (odd $M$).

One can show that --- just as $\widetilde{\mathcal{O}}_k$
in~(\ref{eqn:moduli-III}) fills out a spinor of $\SO(2(M+4))$ ---
products $\widetilde{\mathcal{O}}_{k_1} \widetilde{\mathcal{O}}_{k_2}$
fill out an $(M+4)$-index antisymmetric tensor, which is either
(imaginary) self-dual or (imaginary) anti-self-dual, depending on the
flavor parity. Thus, the symmetric-tensor deconfined theory exhibits
aspects of both flavor parities in its spectrum, consistent with our
hypothesis that the flavor parities correspond to distinct branches of
the quantum moduli space. While it would be interesting to probe this
idea further, antisymmetric tensor deconfinement presents a clearer
picture which is sufficient for our purposes. We leave further
development of the symmetric tensor deconfinement viewpoint on quad
CFTs as an interesting open problem.

\section{S-duality for all phases of \alt{$dP_1$}{dP1}}
\label{sec:dP1-S-duality}

With a description of the quad CFTs in hand, it is now straightforward to construct the missing phases of the $dP_1$ orientifold by embedding these CFTs into the brane tiling for phases II and III, figure~\ref{sfig:quadruple-intersections}.

Notionally, this will take the form of a deformation / gauging of the quad CFT. In general, we might want to distinguish between deforming the deconfined gauge theory in the UV versus deforming the quad CFT itself. In the present discussion, however, we will not distinguish between these two viewpoints,
 leaving a more careful treatment to future work. Our main result is an index, which is insensitive to such distinctions.

\begin{figure}
  \centering
  \begin{subfigure}[b]{0.3\textwidth}
    \centering
    \includegraphics[height=4cm]{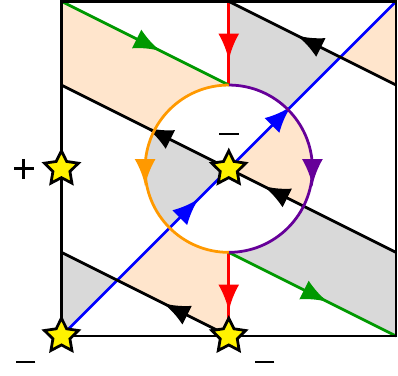}
    \caption{$x_6=x_7=0$ slice}
    \label{sfig:II-deconfined-tiling}
  \end{subfigure}
  \hspace{1cm}
  \begin{subfigure}[b]{0.5\textwidth}
    \centering
    \includegraphics[height=4cm]{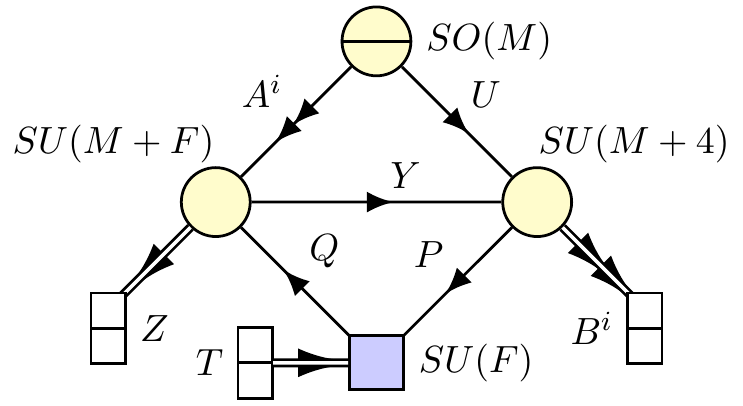}
    \caption{Quiver diagram}
    \label{sfig:II-deconfined-quiver}
  \end{subfigure}
  \caption{Deconfinement of the configuration in
    figure~\ref{sfig:quadruple-intersections} for phase
    II. \subref{sfig:II-deconfined-tiling} The deconfined brane tiling, where we hide the flavor branes for simplicity.
    \subref{sfig:II-deconfined-quiver} The corresponding quiver
    gauge theory.}
  \label{fig:II-deconfined}
\end{figure}

Embedding the deconfined description of the phase II quad-CFT into the brane tiling
leads to the deconfined tiling shown in figure~\ref{sfig:II-deconfined-tiling}, corresponding to the quiver gauge theory in figure~\ref{sfig:II-deconfined-quiver} with the superpotential
\begin{equation}
  W = \frac{1}{2} \varepsilon_{ij} A^i A^j Z + Y P Q + T Q^2 Z + \varepsilon_{ij} A^i Y B^i U\, .
\end{equation}
The corresponding charge table is shown in table~\ref{tab:II-charges}, where we have chosen yet another basis for the $\U(1)$ symmetries for future convenience. Here the gauging has broken $\Sp(2M) \to \SO(M) \times \SU(2)$. For even $M$, there is an additional $\bZ_2$ discrete symmetry of the form $\mathcal{P} (-1)^{B_1 + B_2}$, where $\mathcal{P}$ denotes the $\bZ_2$ outer automorphism of $\SO(M)$, under which $\SO(M)$ baryons are charged, and $B_{1,2}$ denote the baryon numbers associated to $\SU(M+F)$ and $\SU(M+4)$, respectively.

\begin{table}
{\small
\begin{equation*}
  \setlength{\extrarowheight}{1pt} 
  \def\arraystretch{1.2}
  \begin{array}{c|ccc|ccccc}
    & \SU(M\!+\!F) & \SO(M) & \SU(M\!+\!4) & \SU(F) & \SU(2)\!\!\!\! & \U(1)_B & \U(1)_Y & \U(1)_R\\
    \hline
    A^i & \ov\fund & \fund & \singlet & \singlet & \fund\!\!\!\!  & -\frac{1}{2(M+F)} & \frac{M+3}{2(M+F)} & 1 - \frac{M-1}{2 (M+F)}\\
    B^i & \singlet & \singlet & \asymm & \singlet & \fund\!\!\!\! &  -\frac{2}{M+4} & 1-\frac{2}{M+4} & -\frac{2}{M+4}\\
    U & \singlet & \fund & \ov\fund & \singlet & \singlet\!\!\!\! & \frac{1}{M+4} & -1+\frac{1}{M+4} & 1+\frac{1}{M+4}\\
    Y & \fund & \singlet & \ov\fund & \singlet & \singlet\!\!\!\! & \frac{1}{2(M+F)}+\frac{1}{M+4} & - \frac{M+3}{2(M+F)}+\frac{1}{M+4} & \frac{M-1}{2 (M+F)}+\frac{1}{M+4} \\
    Z & \asymm & \singlet & \singlet & \singlet & \singlet\!\!\!\! & \frac{1}{M+F} & -\frac{M+3}{M+F} & \frac{M-1}{M+F}\\
    P & \singlet & \singlet & \fund & \ov\fund & \singlet\!\!\!\! & -\frac{1}{M+4}+\frac{3}{2F} &  - \frac{1}{M+4}-\frac{M+1}{2 F} & 2  - \frac{1}{M+4}+ \frac{M-3}{2 F}\\
    Q & \ov\fund & \singlet & \singlet & \fund & \singlet\!\!\!\! &-\frac{1}{2(M+F)}-\frac{3}{2F} & \frac{M+3}{2(M+F)}+\frac{M+1}{2F}  &  - \frac{M-1}{2(M+F)}-\frac{M-3}{2F}\\
    T & \singlet & \singlet & \singlet & \ov\asymm & \singlet\!\!\!\! & \frac{3}{F} & -\frac{M+1}{F} & 2 + \frac{M-3}{F}
  \end{array}
\end{equation*}}
\caption{A deconfined description of phase II}
\label{tab:II-charges}
\end{table}

These theories fall into four classes according to $(-1)^F$ and $(-1)^M$, where the latter ``color parity'' is preserved along the mesonic moduli space due to the necessity of removing D3 branes in orientifold-image pairs in the dual Calabi-Yau geometry. We use the shorthand II$_{f}^m$ with $f=(-1)^F$ and $m=(-1)^M$ to distinguish the four classes.

To relate these theories to the brane tilings discussed in~\S\ref{sec:dP1-tiling}, we use partial resolution to read off the discrete torsion, as in~\S\ref{sec:classical-phases-torsion}. First, we need to identify the relevant baryons. Fortunately, we can rely on the similar calculation shown in figure~\ref{fig:quadresolutions} to find the correct baryon in the quad CFT, dressing it with baryons from the remainder of the brane tiling according to the standard procedure~\cite{GarciaEtxebarria:2006aq}, as reviewed in~\S\ref{sec:classical-phases-torsion}. We find (c.f.\ (\ref{eqn:quadCFToperators}))
\be
\mathcal{O}_k = Y^{2 (M+4-k)} Z^{F+k-4} U^{2 k} \,
\ee
where $k$ is constrained by $(-1)^k = (-1)^F$, and
controls the distribution of D3 branes between the residual $\bC^3/\bZ_3$ orientifold singularity and the O3 plane.

Turning on a vev for this baryon will Higgs
\be
\SU(M+F)\times\SO(M)\times\SU(M+4) \longrightarrow \Sp(F+k-4) \times \SO(M-k) \times \SU(M+4-k)\times \SO(k) \,.
\ee
Just as in figure~\ref{fig:quadresolutions}, the $\Sp(F+k-4)$ is confining, and we obtain the $\SO(M-k)\times \SU(M+4-k)$ $\bC^3/\bZ_3$ orientifold theory together with the $\SO(k)$ $\mathcal{N}=4$ theory.

Recall that for these orbifold components, the $\SO(p)\times(\ldots)$ gauge groups have torsion $[F]=(-)^p$ and $[H]=+$ and the $\Sp(2p)\times(\ldots)$ gauge groups have torsion $[F] = \pm$ and $[H]=-$, where in the latter case $[F]$ combines with $C_0$ to determine the theta angle. Thus, II$_+^{\pm}$ has $[F]$ torsion $(\pm,+)$ whereas II$_-^{\pm}$ has $[F]$ torsion $(\mp,-)$, and all four cases have trivial $[H]$ torsion.
We can also read off the D3 charge by combining the charge of the components. We obtain
\be
Q_{\rm II} = (M+4-k-3/2) + (k-1/2)= M + 2 \,.
\ee

\begin{figure}
  \centering
  \begin{subfigure}[b]{0.4\textwidth}
    \centering
    \includegraphics{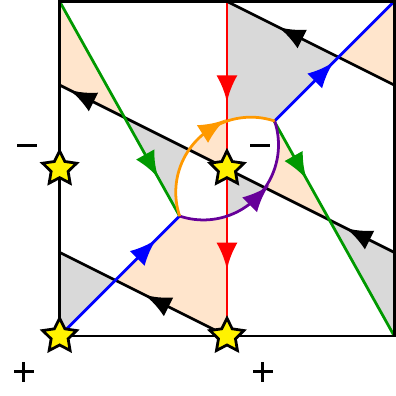}
    \caption{$x_6=x_7=0$ slice}
    \label{sfig:III-deconfined-tiling}
  \end{subfigure}
  \hfill
  \begin{subfigure}[b]{0.55\textwidth}
    \centering
    \includegraphics{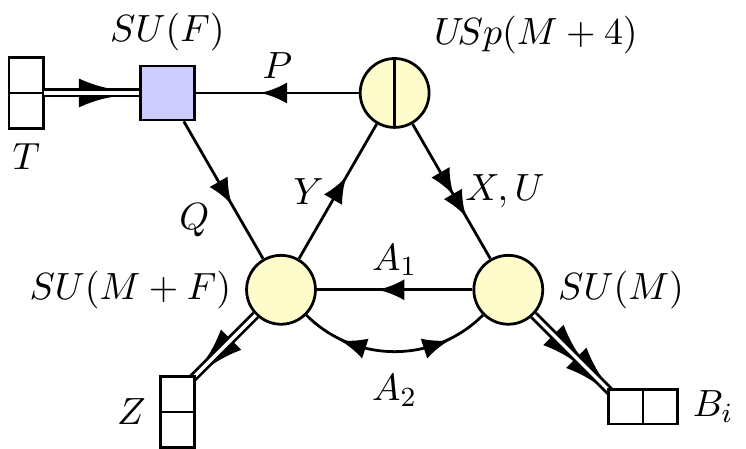}
    \caption{Quiver diagram}
    \label{sfig:III-deconfined-quiver}
  \end{subfigure}
  \caption{Deconfined phase III. \subref{sfig:II-deconfined-tiling}
    The deconfined brane tiling, where the flavor branes are hidden as before. \subref{sfig:III-deconfined-quiver}
    The associated quiver gauge theory.
    }
  \label{fig:III-deconfined}
\end{figure}

We now turn to phase III. Embedding the deconfined description of the quad CFT from~\S\ref{sec:phaseIII} into the brane tiling shown in figure~\ref{sfig:quadruple-intersections} leads to the deconfined brane tiling shown in figure~\ref{sfig:III-deconfined-tiling}. The corresponding quiver diagram, figure~\ref{sfig:III-deconfined-quiver}, has
the tree-level superpotential
\begin{equation}
  W =   Z A_1 A_2 + A_1 Y X + Y P Q + T Q^2 Z + B_1 X U + B_2 A_2 Y U \, ,
\end{equation}
and the charge table shown in table~\ref{tab:III-charges}.

For even $F$, there is an additional $\bZ_2$ discrete symmetry of the form $(-1)^{B_1 + B_3}$, where $B_{1,3}$ denotes the baryon number associated to $\SU(M+F)$ and $\SU(F)$, respectively.\footnote{Since $\SU(F)$ is trivial, we only consider $\SU(F)$ invariants, for which $B_3$ is integrally quantized.}

\begin{table}
{\small
\begin{equation*}
 \setlength{\extrarowheight}{1pt} 
  \def\arraystretch{1.2}
  \begin{array}{c|ccc|ccccc}
    & \SU(M\!+\!F) & \SU(M) & \Sp(M\!+\!4) & \SU(F)\!\!\!\! & \U(1)_B & \U(1)_X & \U(1)_Y & \U(1)_R\\
    \hline
     \!\!A_1  & \ov\fund & \fund & \singlet & \singlet\!\!\!\! & \frac{1}{M+F}-\frac{1}{M} & -\frac{M+4}{2(M+F)} & \frac{1}{M} - \frac{M+2}{2 (M+F)} & \frac{M+1}{M} + \frac{1}{M+F}\\
     \!\!A_2 & \ov\fund & \ov\fund & \singlet & \singlet\!\!\!\! & \frac{1}{M} + \frac{1}{M+F} & -\frac{M+4}{2(M+F)} & -\frac{1}{M}-\frac{M+2}{2 (M+F)} & \frac{M-1}{M} + \frac{1}{M+F}\\
    \!\!B_1 & \singlet & \symm & \singlet & \singlet\!\!\!\! & -\frac{2}{M} & -1 & 1+\frac{2}{M} & \frac{2}{M}\\
    \!\!B_2 & \singlet & \symm & \singlet  & \singlet\!\!\!\! & -\frac{2}{M} & 1 & 1+ \frac{2}{M} & \frac{2}{M}\\
    \!\!U  & \singlet & \ov\fund & \fund & \singlet\!\!\!\! & \frac{1}{M} & 0 & -1-\frac{1}{M} & \frac{M-1}{M}\\
    \!\!X  & \singlet & \ov\fund & \fund & \singlet\!\!\!\! & \frac{1}{M} & 1 & -\frac{1}{M} & \frac{M-1}{M} \\
    \!\!Y & \fund & \singlet & \fund & \singlet\!\!\!\! & -\frac{1}{M+F} & -1+\frac{M+4}{2(M+F)} & \frac{M+2}{2(M+F)} & -\frac{1}{M+F}\\
    \!\!Z & \asymm & \singlet & \singlet & \singlet\!\!\!\! & -\frac{2}{M+F} & \frac{M+4}{M+F} & \frac{M+2}{M+F} & -\frac{2}{M+F}\\
    \!\!P & \singlet & \singlet & \fund & \ov\fund\!\!\!\! & -\frac{1}{F} & 1 - \frac{M+4}{2F} & \frac{M+2}{2F} & 2 - \frac{3}{F}\\
    \!\!Q & \ov\fund & \singlet & \singlet & \fund\!\!\!\! & \frac{1}{F} + \frac{1}{M+F} & \frac{M+4}{2F}-\frac{M+4}{2 (M+F)} & -\frac{M+2}{2 F} - \frac{M+2}{2 (M+F)} & \frac{3}{F} +\frac{1}{M+F}\\
    \!\!T & \singlet & \singlet & \singlet & \ov\asymm\!\!\!\! & -\frac{2}{F} & -\frac{M+4}{F} & \frac{M+2}{F} & 2 - \frac{6}{F}
  \end{array}
\end{equation*}}
\caption{A deconfined description of phase III}
\label{tab:III-charges}
\end{table}

Here the gauging has broken $\SO(2(M+4)) \to \Sp(M+4)\times\SU(2)$,
but the $\SU(2)$ is not manifest because the original deconfined
description only conserved $\U(M+4) \subset \SO(2(M+4))$. Instead,
only $\U(1)_X \in \SU(2)$ is visible. Deconfinement duality
corresponds to charge conjugation on $\U(1)_X$, which is half-integrally quantized for gauge invariant operators, hence only (pseudo) real representations of $\U(1)_X$ with integer or half-integer charges can appear in the operator spectrum, consistent with an $\SU(2)$ enhancement.\footnote{In particular, this ensures that the superconformal index fills out complete $\SU(2)$ multiplets.}

Since $\Sp(M+4)$ is only defined for even $M$, these theories fall into two classes, which we denote III$_f$ with $f=(-1)^F$. A similar procedure can be used to identify the baryons corresponding to the partial resolution to $\bC^3/\bZ_3$. We find (c.f.\ \ref{eqn:moduli-III}):
\be
\mathcal{O}_k = A_1^k A_2^{M-k} Q^F U^{2k} \,,
\ee
where $k$ is even. Turning on a vev for this baryon will Higgs
\be
 \SU(M+F) \times \SU(M) \times \Sp(M+4) \longrightarrow \left[\Sp(M+4-k) \times \SU(M-k)\right]\times \Sp(k)
\ee
where all $\SU(F)$-charged matter becomes massive, and the remaining light fields are those of the $\Sp(M+4-k) \times \SU(M-k)$ $\bC^3/\bZ_3$ orientifold theory together with the $\Sp(k)$ $\mathcal{N}=4$ theory. Thus, the phase III theories have $[H]$ torsion $(-,-)$. We can read off the D3 charge by adding up the charge of the components, giving
\be
Q_{\rm III} = (M-k +3/2) + (k+1/2) = M+2 \,,
\ee
as in phase II.

To determine the $[F]$ torsion from the above computation, we would
need to compare theta angle of the $\mathcal{N}=4$ theory with the
phase of the exactly marginal coupling in the $\bC^3/\bZ_3$ theory,
determined in~\cite{dualities1}. Such a calculation is beyond the
scope of this work. 
Instead, we note that there are two inequivalent choices, $(\pm,\pm)$ and
$(\pm, \mp)$, where in either case the two sign choices combine with
$C_0$ to determine the phase of an exactly marginal coupling in the
gauge theory. The S-dual of the theory with trivial $[F]$ torsion is
II$^+_-$, which has a $\bZ_2$ discrete symmetry, therefore III$_+$
must correspond to $[F]$ torsion $(\pm,\pm)$ and III$_-$ to $[F]$
torsion $(\pm,\mp)$. This is a natural guess, since then even $F$ and
$M$ corresponds to trivial $[F]$ torsion, as in phase II.

\begin{figure}
  \centering
  \includegraphics[width=0.5\textwidth]{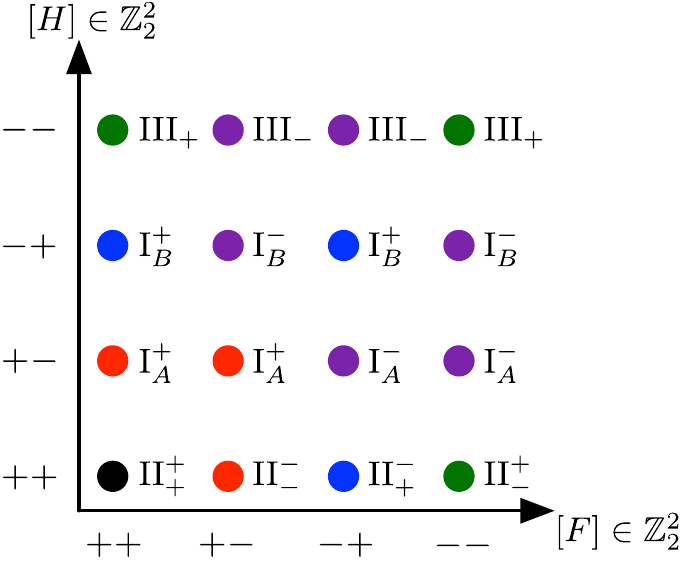}
  \caption{Discrete torsion for all phases of the
    $dP_1$ orientifold. Except for the relative position of III$_+$ and III$_-$, all torsions are fixed by the partial resolution to the $\bC^3/\bZ_3$ orientifold singularity plus an O3 plane.
    The $[F]$ torsion assignment for phase III is obtained by matching discrete symmetries between prospective S-duals.}
  \label{fig:dP1phaseIII}
\end{figure}

Thus, as illustrated in figure~\ref{fig:dP1phaseIII}, the phase II and phase III theories obtained via deconfinement neatly fill in the gaps in figure~\ref{subfig:IABtorsions}. Besides the $\mathrm{I}_A^- \longleftrightarrow \mathrm{I}_B^-$ duality already discovered in~\cite{dualities1}, this figure predicts several new S-dualities: 
\be \label{eqn:dP1dualities}
\mathrm{I}_A^+ \longleftrightarrow \mathrm{II}^-_- \hspace{1.5em},\hspace{1.5em} \mathrm{I}_B^+ \longleftrightarrow \mathrm{II}^-_+ \hspace{1.5em},\hspace{1.5em} \mathrm{II}^+_-  \longleftrightarrow \mathrm{III}_+ \hspace{1.5em},\hspace{1.5em} \mathrm{I}_A^- \longleftrightarrow \mathrm{I}_B^- \longleftrightarrow  \mathrm{III}_- \;\;,
\ee
where the previously known duality is now a triality and II$^+_+$ is expected to be self-dual. In particular, both $\mathrm{I}_A^+$ and $\mathrm{I}_B^+$ have distinct S-duals, answering a puzzle from~\cite{dualities1}. The relative ranks of the duals are predicted by matching the D3 charges,
\be \label{eqn1:dP1D3charge}
Q_{\rm D3} = N-1 = \tN+1 = M+2 \,.
\ee
The same result is obtained by anomaly matching, where the $\U(1)$ bases in tables~\ref{tab:II-charges}, \ref{tab:III-charges} were chosen to match those in tables~\ref{tab:IA-charges}, \ref{tab:IB-charges}.

\subsection{Matching the superconformal index} \label{subsec:SCImatching}

With an extensive set of predicted S-dualities in hand, we need a way to test these dualities. Anomaly matching is not sensitive enough, because all the theories with the same D3 charge (\ref{eqn1:dP1D3charge}) have matching anomalies regardless of $[F]$ and $[H]$ torsion, apart from the mismatch of the $\SU(2)^3$ Witten anomalies between $\mathrm{I}_A^+$ and $\mathrm{I}_B^+$ noted in~\cite{dualities1}, which persists in their duals.\footnote{We have not computed the Witten anomaly or the $\SU(2)^2 \bZ_2$ anomaly in phase III, where only $\U(1)_X \subset \SU(2)$ is manifest.} Based on the $\bZ_2$ discrete symmetry discussed above and on the quantization condition for $\U(1)_B$ (which is integral for I$_{A/B}^-$, III$_{\pm}$, and II$^+_-$, and otherwise half-integral), we can rule out any further dualities between the $dP_1$ theories besides those in~(\ref{eqn:dP1dualities}).

To do better, we could match the operator spectra of the dual
theories. This can be subtle due to the large number of gauge
invariant operators in a quiver gauge theory, quantum constraints on
the moduli space, etc., and we leave a more extensive treatment to
future work. Instead, we rely on a conceptually simple but very
sensitive test: we match the superconformal indices between the dual
theories.

The superconformal index~\cite{Romelsberger:2005eg,Kinney:2005ej} counts shortened superconformal multiplets of the theory quantized on $S^3 \times \bR$ up to recombination:
\be
\mathcal{I} = \Tr_{\mathcal{H} = 0} \left[(-1)^F t^{R + 2 J_L} x^{2 J_R} f \right] \,,
\ee
where $J_{L,R}$ generate the Cartan of $\SU(2)_{L,R}$, $R$ denotes the R-charge, and $f$ is an arbitrary fugacity of the flavor symmetry group. The trace is taken over states annihilated by
\be
\mathcal{H} \equiv H-2 J_L - \frac{3}{2} R \,,
\ee
with $H$ the Hamiltonian, which is the anticommutator of a supercharge $Q$ in the superconformal algebra with its conjugate (hence the index is counts $Q$ cohomology classes.)

The superconformal index can be shown to agree between Seiberg dual theories~\cite{Romelsberger:2007ec}, where it reduces to an integral identity for elliptic hypergeometric functions~\cite{Dolan:2008qi}. In the case of other dualities, such as S-dualities, the index should likewise agree, but the appropriate integral identities are not always known, see e.g.~\cite{Spiridonov:2009za, Spiridonov:2010qv}.

The index is invariant under continuous deformations~\cite{Romelsberger:2007ec}, so in particular when a weakly-coupled UV description is available --- and in the absence of rank-enhancing accidental symmetries along the flow --- the index can be computed in the UV. We have
\be \label{eqn:SCI-formula}
\mathcal{I} = \int dg\, P\!E\!\left[\sum_a i_a(t,x,g,f)\right] \,,
\ee
where $P\!E[f(x)] \equiv \exp\left(\sum_{k=1}^\infty \frac{1}{k} f(x^k)\right)$ is the plethystic exponential, the integral is taken over the gauge group with the Haar measure, and $i_a$ is the ``letter'' for the $a$th field, of the form
\be
i_V = \frac{2 t^2 - t (x+x^{-1})}{(1- t x)(1-t x^{-1})} \chi_V \;\;,\;\; i_\Phi = \frac{t^R \chi_\Phi - t^{2-R} \chi_{\Phi^\dag}}{(1-t x)(1-t x^{-1})} \,,
\ee
for vector and chiral superfields, respectively. Here $R$ is the $a$-maximized R-charge of the chiral field and $\chi_V$ and $\chi_{\Phi}$ denote the group characters of the representation under the gauge and global symmetries. In effect, (\ref{eqn:SCI-formula}) counts gauge-invariants in the symmetric product of the states in the letter, where negative contributions to the letter anticommute.

In general, the integral over the gauge group is very difficult to
perform, except at large $N$, where it simplifies
\cite{Kinney:2005ej}. Instead, we perform a series expansion in $t$ to
count low-lying states in the index. Using
\be \int dg\, \chi_r(g) =
\delta_{r,1} \;\;,\;\; \chi_{r}(g) \chi_{r'}(g) = \chi_{r \times
  r'}(g) \,,
\ee
for representations $r, r'$, where $\delta_{r,1}$
denotes the number of singlet irreps in $r$, it becomes
straightforward in principle to compute the index to any fixed order
in $t$, provided that all the chiral fields satisfy $0<R<2$.

Unfortunately, the $a$-maximized R-charges of the fundamental fields in the UV theory need not satisfy this inequality, even in the absence of accidental symmetries which mix with the R-symmetry in the infrared. For instance, the deconfined description of an $\SU(N)$ antisymmetric tensor, figure~\ref{fig:antisymmetric-deconfinement-quiver}, has the $a$-maximized R-charges:
\be
R(\hat{Z}) = \frac{1}{3} \;\;, \;\; R(P) = -\frac{N-6}{3 F} \;\;,\;\; R(Q) = \frac{5}{3} + \frac{N-6}{3F} \;\;,\;\; R(T) = 2 + 2 \frac{N-6}{3 F} \,,
\ee
where $F$ is the number of flavors in the deconfined description, with $(-1)^F = (-1)^N$. These charges satisfy the bound $0 < R < 2$ for $N<6$ and violate it otherwise, whereas in either case the infrared fixed point consists of free chiral multiplets with R-charge $2/3$, by construction.

The resolution is that the states with negative powers of $t$ completely cancel out of the index. To see how this work, consider the marginal case $N=6$ in the above deconfined theory, so that $R(P) = R(\bar{T}) = 0$, where $\bar{T}$ denotes the negative contribution to the letter for $T$. Naively, there are an infinite number of contributions to the index at $\mathcal{O}(t^0)$. However, it is straightforward to check that integrating over the symplectic group gives:
\be
\int dg\, P\!E[y \chi_{\fund}(g) \chi_{\ov\fund}(f)] = P\!E\!\left[y^2 \chi_{\ov\asymm}(f)\right] \,,
\ee
i.e.\ the gauge invariant operator $P^{2k}$ transforms in $k$th symmetric power of the conjugate antisymmetric representation of $\SU(F)$, so that the contributions of $P^{2k}$ and $\bar{T}^k$ to the $\mathcal{O}(t^0)$ term in the index cancel.

In fact, this cancellation is guaranteed by the integral identities of~\cite{Dolan:2008qi}, since the reconfinement of the deconfined description is a special case of Seiberg duality for symplectic groups, but it serves to illustrate the difficulties which arise in computing the index in more general theories whose fundamental fields do not satisfy $0 < R < 2$.

The $a$-maximized R-charge for tables~\ref{tab:IA-charges}, \ref{tab:IB-charges}, \ref{tab:II-charges}, \ref{tab:III-charges} is
\be
\U(1)_R^{\rm (sc)} = \U(1)_R + (4 - \beta)\, \U(1)_Y + Q_{\rm D3} \left(\frac{3}{\beta} - \frac{\beta}{4}\right) \U(1)_B
\ee
where $Q_{\rm D3}$ is given by (\ref{eqn1:dP1D3charge}), and $\beta$ is the unique root of the quartic equation
\begin{equation}
  (\beta-2)(\beta-6)(3 \beta^2-8 \beta-12) Q_{\rm D3}^2+16 \beta^2 (9-2
\beta)=0
\end{equation}
satisfying $\frac{2}{3} \left(2+\sqrt{13}\right)< \beta \le 4$ for
$Q_{\rm D3} \ge4$, with $\frac{2}{3}
\left(2+\sqrt{13}\right)\approx3.737$. From this, one can check that the $a$-maximized R-charges in tables~\ref{tab:IA-charges}, \ref{tab:IB-charges} satisfy $0<R<2$ for $Q_{\rm D3} > 4$,\footnote{$Q_{\rm D3} = 4$ is a special case where there is confinement with chiral symmetry breaking in the infrared~\cite{dualities1}.} whereas those in table~\ref{tab:II-charges} satisfy the constraint for $4 < Q_{\rm D3} \le 13$ with $F=1$ and for $4 < Q_{\rm D3} \le 12$ with $F=2$, and those in table~\ref{tab:III-charges} satisfy it for $4 < Q_{\rm D3} \le 10$ with $F=1,2$.

For the present paper, we confine out attention to $4 < Q_{\rm D3} \le 10$, for which the index can be computed as an expansion in $t$ in all the phases of $dP_1$.\footnote{Besides finding a systematic way to deal with the large degree of cancellation in the index, another way to proceed for $Q_{\rm D3} > 10$ would be to expand in a different fugacity. One can show that there is always a \emph{conserved} R-symmetry (not necessarily the $a$-maximized one) satisfying $0 < R < 2$ for any $Q_{\rm D3}>4$ and $F>0$, hence there is a corresponding fugacity with only positive powers in the letter, allowing a systematic expansion which avoids the above issues.} This already allows for very non-trivial tests of the duality.

Consider for example $Q_{\rm D3}=9$. The superconformal index in phase I$_A^+$ is:
\begin{multline} \label{eqn:IAplusIndex}
1+b^{{\frac{1}{2}}}y^{{-5}}t_{\text{(1.294)}}^{{-17+\frac{27}{2\beta}+\frac{31\beta}{8}}}+b^{{\frac{1}{2}}}y^{{-4}}X_1 t_{\text{(1.488)}}^{{-13+\frac{27}{2\beta}+\frac{23\beta}{8}}}+b^{{\frac{1}{2}}}y^{{-3}} X_2 t_{\text{(1.683)}}^{{-9+\frac{27}{2\beta}+\frac{15\beta}{8}}} +b^{{\frac{1}{2}}}y^{{-2}}X_3 t_{\text{(1.877)}}^{{-5+\frac{27}{2\beta}+\frac{7\beta}{8}}}  \\ -\left(1+X_2\right)t^{2}+\ldots + \left[J_1\left(3+7X_2+4X_4\right)-2J_3\left(1+X_2\right)\right]t^{5}+\ldots
\end{multline}
up to order $t^5$, where $y$ and $b$ are the flavor fugacities for $\U(1)_Y$ and $\U(1)_B$, respectively, and $X_i$ and $J_i$ denote the characters for the spin $i/2$ representation of the $\SU(2)$ flavor symmetry and of $\SU(2)_R$, respectively. In cases where $t$ is raised to an irrational exponent, we indicate a decimal approximant in the subscript. Note that we have shown only the first and last few terms in the expansion (\ref{eqn:IAplusIndex}). The full expression, shown in~\S\ref{subsec:IA-plus-index}, has 180 terms, of which only 12 have been given above!

The index for I$_B^+$ and $Q_{\rm D3}=9$ is
\begin{multline} \label{eqn:IBplusIndex}
1+b^{{\frac{3}{2}}}y^{{-4}}t_{\text{(1.021)}}^{{-12+\frac{81}{2\beta}+\frac{5\beta}{8}}}-\left(1+X_2\right) t^{2}+b^{{\frac{3}{2}}}y^{{-4}}J_1t_{\text{(2.021)}}^{{-11+\frac{81}{2\beta}+\frac{5\beta}{8}}}+b^{3}y^{{-8}}t_{\text{(2.041)}}^{{-24+\frac{81}{\beta}+\frac{5\beta}{4}}}-yX_1 t_{\text{(2.194)}}^{{6-\beta}} \\ +b^{{-\frac{1}{2}}}X_4 t_{\text{(2.734)}}^{{2-\frac{27}{2\beta}+\frac{9\beta}{8}}}+\ldots+\left[J_1\left(3+7X_2+4X_4\right)-2J_3\left(1+X_2\right)\right]t^{5}+\ldots
\end{multline}
up to order $t^5$, where the full expression, shown in~\S\ref{subsec:IB-plus-index}, has 154 terms. We see immediately that (\ref{eqn:IAplusIndex}) and (\ref{eqn:IBplusIndex}) do not match, hence I$_A^+$ and I$_B^+$ are not dual, as anticipated in~\cite{dualities1} and borne out by figure~\ref{fig:dP1phaseIII}. By contrast, computing the index for II$^-_-$ to the same order gives back all 180 terms of (\ref{eqn:IAplusIndex}), and computing the index for II$^-_+$ gives back all 154 terms of (\ref{eqn:IBplusIndex})! This is highly non-trivial evidence in favor of the dualities $\mathrm{I}_A^+ \longleftrightarrow \mathrm{II}^-_-$ and $\mathrm{I}_B^+ \longleftrightarrow \mathrm{II}^-_+$.

Likewise, the index for I$_A^-$ and $Q_{\rm D3}=8$ is
\begin{multline} \label{eqn:IAminusIndex}
1-(1+X_2) t^{2}-y X_1 t_{\text{(2.178)}}^{6-\beta}+by^{{-7}}t_{\text{(2.39)}}^{{-23+\frac{24}{\beta}+5\beta}}+by^{{-6}}X_1 t_{\text{(2.568)}}^{{-19+\frac{24}{\beta}+4\beta}} +2by^{{-5}}X_2 t_{\text{(2.746)}}^{{-15+\frac{24}{\beta}+3\beta}}  \\+\ldots+\left[9+10X_2+3X_4-2X_6+2J_2\left(3+8X_2+3X_4\right)-2J_4\left(1+X_2\right)\right] t^{6} + \ldots
\end{multline}
up to order $t^6$. The full result in~\S\ref{subsec:IA-minus-index} has 215 terms, all of which are reproduced in both I$_B^-$ and III$_-$! This is highly non-trivial evidence for the duality $\mathrm{I}_A^- \longleftrightarrow \mathrm{I}_B^-$ proposed in~\cite{dualities1}, now expanded into a triality $\mathrm{I}_A^- \longleftrightarrow \mathrm{I}_B^- \longleftrightarrow \mathrm{III}_-$.

Finally, the index for II$^+_-$ and $Q_{\rm D3}=8$ is\footnote{We omit the fugacity for the $\bZ_2$ discrete symmetry for simplicity.}
\begin{multline} \label{eqn:IIplusminusIndex}
1+b^{{-1}}y^{4} X_5 t_{\text{(1.077)}}^{{15-\frac{24}{\beta}-2\beta}}-\left(1+X_2\right)t^{2}+b^{{-1}}y^{4}J_1 X_5 t_{\text{(2.077)}}^{{16-\frac{24}{\beta}-2\beta}}+b^{{-2}}y^{8}\left(X_2+X_6+X_{10}\right) t_{\text{(2.153)}}^{{30-\frac{48}{\beta}-4\beta}}\\-yt_{\text{(2.178)}}^{{6-\beta}}X_1+\ldots+\left[J_1\left(5+14X_2+12X_4+7X_6+4X_8\right)-2J_3\left(1+X_2\right)\right]t^{5}+\ldots
\end{multline}
up to order $t^5$, where the full result in~\S\ref{subsec:II-plus-minus-index} has 216 terms, all of which are reproduced in III$_+$! This confirms the last remaining duality $\mathrm{II}^+_- \longleftrightarrow \mathrm{III}_+$ from figure~\ref{fig:dP1phaseIII}.

For completeness, we also give the first few terms of the index for II$^+_+$ and $Q_{\rm D3}=8$:
\begin{multline} \label{eqn:IIplusplusindex}
1+\frac{b^{3/2} t_{(0.830)}^{-\frac{21}{2}+\frac{36}{\beta 
}+\frac{\beta }{2}}}{y^{7/2}}+\frac{\sqrt{b} 
t_{(1.017)}^{-\frac{31}{2}+\frac{12}{\beta }+\frac{7 \beta 
}{2}}}{y^{9/2}}+\frac{y^4 X_5 t_{(1.077)}^{15-\frac{24}{\beta }-2 
\beta } }{b}+\frac{\sqrt{b} 
X_1 t_{(1.195)}^{-\frac{23}{2}+\frac{12}{\beta }+\frac{5 \beta }{2}} 
}{y^{7/2}}\\+\frac{\sqrt{b} 
X_2 t_{(1.373)}^{-\frac{15}{2}+\frac{12}{\beta }+\frac{3 \beta }{2}} 
}{y^{5/2}}+\frac{\sqrt{b} 
X_3 t_{(1.551)}^{-\frac{7}{2}+\frac{12}{\beta }+\frac{\beta }{2}} 
}{y^{3/2}}+\frac{b^3 t_{(1.660)}^{-21+\frac{72}{\beta }+\beta 
}}{y^7}
+\frac{\sqrt{b} X_4 t_{(1.729)}^{\frac{1}{2}+\frac{12}{\beta 
}-\frac{\beta }{2}} }{\sqrt{y}} + \ldots
\end{multline}
Notice that (\ref{eqn:IIplusplusindex}) does not match (\ref{eqn:IAminusIndex}) or (\ref{eqn:IIplusminusIndex}), just as (\ref{eqn:IAplusIndex}) and (\ref{eqn:IBplusIndex}) do not match, even for the first few terms. Thus, the superconformal index is a highly sensitive test of S-dualities. The S-dual theories in figure~\ref{fig:dP1phaseIII} have indices which match hundreds of terms (and conjecturally exactly), whereas the $dP_1$ theories which are not S-dual have indices which disagree even in the first few terms!\footnote{To be precise, the leading baryons generally disagree. For larger values of $Q_{\rm D3}$, these baryons will appear with higher powers of $t$ and the leading $\mathcal{O}(b^0)$ terms will match, even between non-S-dual theories, as the index approaches a large $N$ limit.}

We have also matched the superconformal indices between the dual theories for other values of $Q_{\rm D3} \le 10$. Our results are summarized in table~\ref{tab:SCI-checks}.

\begin{table}
\centering
\begin{subfigure}{0.45\textwidth}
  \centering
  \begin{tabular}{c|cccc}
\multirow{2}{*}{$Q_{\rm D3}$} & \multicolumn{2}{c}{$\IA^+ = \mathrm{II}^-_-$} & \multicolumn{2}{c}{$\IB^+= \mathrm{II}^-_+$}  \\
 & Terms & Order & Terms & Order \\
\hline
5 & *275 & $\mathcal{O}(t^2)$ & *222 & $\mathcal{O}(t^2)$ \\
7 & 221 & $\mathcal{O}(t^4)$ & *187 & $\mathcal{O}(t^4)$ \\
9 & 180 & $\mathcal{O}(t^5)$ & 154 & $\mathcal{O}(t^5)$ \\
  \end{tabular}
\end{subfigure}
\hfill
\begin{subfigure}{0.45\textwidth}
  \begin{tabular}{c|cccc}
\multirow{2}{*}{$Q_{\rm D3}$} & \multicolumn{2}{c}{$\IA^-=\IB^-=\mathrm{III}_-$} & \multicolumn{2}{c}{$\mathrm{II}^+_- = \mathrm{III}_+$} \\
 & Terms & Order & Terms & Order \\
\hline
6 & 145 & $\mathcal{O}(t^4)$ & *135 & $\mathcal{O}(t^3)$ \\
8 & 215 & $\mathcal{O}(t^6)$ & 216 & $\mathcal{O}(t^5)$ \\
10 & 110 & $\mathcal{O}(t^6)$ & 1 & $\mathcal{O}(t^{5/4})$ \\
  \end{tabular}
\end{subfigure}

\caption{Summary of S-duality checks using the superconformal index. For $\mathrm{III}_+$ and $Q_{\rm D3}=10$, we are unable to compute any non-vanishing terms in the index beyond the leading $1$, but can show that several contributions cancel as required to match the S-dual theory. The starred theories have accidental symmetries, but the index still matches for the manifest flavor symmetries.}
\label{tab:SCI-checks}
\end{table}


\section{Conclusions} \label{sec:conclusions}


In this paper we have argued that in studying $\cN=1$ S-duality in Calabi-Yau orientifolds, as initiated in~\cite{dualities1,Bianchi:2013gka,dualities2}, one is forced to include strongly-coupled sectors. In toric examples, these sectors arise from more than two NS5 branes intersecting atop an O5 plane on a D5 brane stack, such as the quad CFTs described in~\S\ref{subsec:quadCFTs} and constructed in~\S\ref{sec:deconfinement}.

In the cases we have studied, these sectors have simple yet non-trivial properties. They are characterized by a large flavor symmetry group and a moduli space with mesonic and baryonic directions, where the mesonic directions relate CFTs with different ranks, and the baryonic directions relate the CFT to weakly-coupled chiral superfields describing brane recombinations which avoid the higher-multiplicity intersection. A remarkable property of these CFTs is the appearance of spinor representations of the $\SO(2(M+4))$ flavor symmetry in the baryonic spectrum for arbitrary $M$. This property is hidden in the deconfined gauge theory description, and indeed it is difficult to imagine how it could have been manifest, since the dimension of the spinor representation grows exponentially with $M$, in conflict with asymptotic freedom. This confirms the intrinsically strongly-coupled (non-Lagrangian) nature of the quad CFTs.

By coupling these quad CFTs to quiver gauge theories, as dictated by the brane tiling construction, we have obtained a complete picture of S-dualities in the $dP_1$ orientifold considered in this paper. Our description has been subjected to stringent consistency checks using one of the best instruments available in minimally supersymmetric theories, the superconformal index. We expect that similar ingredients can be used to understand S-duality in a broad class of toric orientifolds~\cite{toricII}, if not beyond.

The quad CFTs --- and similar strongly-coupled sectors arising from $2k$ NS5 branes intersecting atop an O5 plane with $k>2$ --- are interesting in their own right, and deserve further study. It would be instructive to determine the string-theory origin of the non-abelian enhancement to $\Sp(2M)$ or $\SO(2(M+4))$, and in particular the appearance of spinor representations in the latter case. Other descriptions of these CFTs, such as by symmetric tensor deconfinement, could be developed further. The superconformal index of these CFTs may admit a description which is independent of the deconfinement procedure, making their properties and symmetries manifest. It would also be interesting to determine whether these same CFTs play a role in orientifolds of non-toric Calabi-Yau singularities.

Recently, there has been some progress in describing certain S-dualities of chiral $\cN =1$ theories in terms of a 6d $(1,0)$ theory compactified on a punctured Riemann surface~\cite{Gaiotto:2015usa, Franco:2015jna, Hanany:2015pfa}. It remains to be seen if our S-dualities fall into this class, in which case they might form part of a yet-unknown unified picture of $\cN=1$ dualities.

Finally, while our discussion has focused on the brane tiling in the $g_s \to \infty$ limit, it should be possible to describe the new phases in the $g_s \to 0$ limit as well as in the mirror description with intersecting D6 branes. At present, this is an open problem.

\acknowledgments

We thank A.~Hanany for discussions, and T.~Wrase for
collaboration in related previous work
\cite{dualities1,dualities2}. BH is supported by the Fundamental Laws Initiative of the Harvard Center for the Fundamental Laws of Nature. I.~G.-E.\ thanks N.~Hasegawa for kind
encouragement and constant support.

\appendix

\section{Del Pezzo orientifold singularities} \label{app:delPezzo}

In this appendix,  we construct the ten del Pezzo singularities and
their orientifolds with the del Pezzo as an isolated fixed
plane.\footnote{Embeddings for the non-toric del Pezzo singularities
  are taken from~\cite{Malyshev:2007yb} and references therein.}

The del Pezzo singularities for $dP_k$, $k\ge 5$, are complete intersection singularities. Let $P_r^{(a_1,\ldots,a_p)}$ denote a generic quasi-homogeneous polynomial of weight $r$ in $p$ variables with weights $(a_1,\ldots,a_p)$. The $dP_6$, $dP_7$, and $dP_8$ singularities are the hypersurfaces $P_3^{(1,1,1,1)}=0$, $P_4^{(1,1,1,2)}=0$ and $P_6^{(1,1,2,3)}=0$ in $\bC^4$, respectively, whereas the $dP_5$ singularity is the complete intersection of two quadrics $P_2^{(1,1)} = R_2^{(1,1)} = 0$ in $\bC^5$. In each case, the involution $\sigma$ reflects the odd-weight coordinates and leaves the even-weight coordinates invariant.

The del Pezzo surfaces $dP_k$ for $k\le3$ and $\bF_0$ are toric, as are the corresponding singularities. These singularities are most easily described as the classical moduli space of a gauged linear sigma model (GLSM) with vanishing Fayet-Iliopoulos (FI) parameters:
\begin{align}
dP_0&: \begin{array}{c|ccc} & x & y & z \\ \hline \bZ_3 & \omega_3 & \omega_3 & \omega_3 \end{array}\,, &dP_1&:\begin{array}{c|cccc} & x & y & z & w \\ \hline U(1) & 2 & 2 & -1 & -3 \end{array}\,, &\bF_0&: \begin{array}{c|cccc} & y & z & w & u \\ \hline U(1) & 1 & -1 & -1 & 1 \\ \bZ_2 & + & - & - & + \end{array}\,, \\
dP_2&: \begin{array}{c|ccccc} & x & y & z & w & u \\ \hline U(1)_1 & 2 & 2 & -1 & -3 & 0 \\ U(1)_2 & 0 & 1 & -1 & -1 & 1 \end{array}\,, &dP_3&:\begin{array}{c|cccccc} & x & y & z & w & u & v \\ \hline U(1)_1 & 2 & 2 & -1 & -3 & 0 & 0 \\ U(1)_2 & 0 & 1 & -1 & -1 & 1 & 0 \\ U(1)_3 & 1 & 0 & -1 & -1 & 0 & 1 \end{array} \,,
\end{align}
where $\sigma$ reflects $x,y,z$ for $dP_k$ and takes $z \to i z$, $w \to i w$ for $\bF_0$. Note that the $dP_0$ and $\bF_0$ singularities are orbifolds of $\bC^3$ and the conifold, respectively.

By contrast, the $dP_4$ singularity is neither toric nor a complete intersection singularity. A hybrid approach is most convenient. Consider the affine variety defined by the GLSM
\be
\begin{array}{c|ccccc} & z_1 & z_2 & z_3 & u_1 & u_2 \\ \hline U(1) & 1 & 1 & 1 & -1 & -1 \end{array}\,,
\ee
with vanishing FI parameter. The $dP_4$ singularity is the hypersurface $P_2(z) u_1 + R_2(z) u_2 = 0$ in this variety for generic quadrics $P_2$ and $R_2$. The involution $\sigma$ takes $z_i \to - z_i$.

\section{Exceptional collections and brane charges}

\label{app:dP1-microscopics}

In this appendix we study the dynamics of D3 branes probing the $dP_1$
singularity from the point of view of exceptional collections. This
allows us to prove, without needing to resolve the singularity, that
the rank relations necessary for the duality to work are equivalent to
D3 charge conservation. In~\cite{dualities2} it was proposed that the microscopic process behind the duality is a somewhat
mysterious orientifold transition for (collapsed) O7 planes, and
this was verified for the $dP_0$ singularity. The exceptional
collection language will allow us to see that the same process is
compatible with the duality $\IA^- \longleftrightarrow \IB^-$ proposed in~\cite{dualities1} .
Somewhat
unsatisfactorily, the discussion below is limited to the
``classical'' phases of $dP_1$, since we lack a description in
terms of exceptional collections of the strongly-coupled phases which
are the main focus of our paper. It would be very interesting to
understand if a generalization of the orientifold transition could
generate the strongly coupled phases as well. 

\begin{figure}
  \centering
  \includegraphics{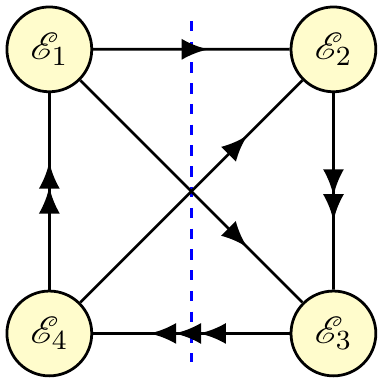}

  \caption{The map between sheaves and nodes in the quiver for the $dP_1$
    singularity. The dashed line indicates the orientifold involution
    studied in the text.}

  \label{fig:orientifolded-dP1-quiver}
\end{figure}

A standard basis for projective objects on $dP_1$ is given by:
\begin{align}
  \label{eq:dP1-projective}
  \cP = \{\cO, \cO(e), \cO(\ell), \cO(2\ell) \}\, ,
\end{align}
where $\ell$ is the hyperplane class of $dP_1$, and $e$ the class of
the exceptional $\bP^1$. The basis of fractional branes obtained by
mutation of~\eqref{eq:dP1-projective} has the following Chern
characters:
\be
  \label{eq:dP1-naive-basis}
    \ch(\cE_1) = 1 \;\;,\;\; 
    \ch(\cE_2) = e-\frac{1}{2}\ell^2 \;\;,\;\;
    \ch(\cE_3) = -2-2e+\ell+\frac{3}{2}\ell^2 \;\;,\;\;
    \ch(\cE_4) = 1 + e - \ell\, ,
\ee
and the resulting quiver is shown in
figure~\ref{fig:orientifolded-dP1-quiver}, which reproduces the $dP_1$
quiver studied in the text.

Now consider an O7 plane on the contracting $dP_1$. Clearly the
basis~\eqref{eq:dP1-naive-basis} cannot be invariant under the large
volume orientifold action \cite{dualities2}
\begin{align}
    i_*\cE[k] \longrightarrow i_*(\cE^\vee\otimes K_\cS\otimes \cL_{2B_2})[2-k] \, .
  \label{eq:orientifold-action-B}
\end{align}
For instance, $\cE_2$ is the only sheaf with rank 0,
and~\eqref{eq:orientifold-action-B} preserves rank. Instead, we will
find that the orientifold action on this basis is one of the
{\ae}rientifolds introduced in \cite{dualities2}.

Let us motivate the particular {\ae}rientifold we will study by
comparing with the description of the orientifolded $\bC^3/\bZ_3$
\cite{dualities2}. As we found in
\S\ref{sec:classical-phases-torsion}, one can go from the $dP_1$
quiver in figure~\ref{fig:orientifolded-dP1-quiver} to the quiver for
phase 1 of $\bC^3/\bZ_3$ by giving a vev to the field $Z$ between $\cE_1$
and $\cE_2$. 
Under the blow down map
$\pi\colon dP_1\to\bP^2$, 
the exceptional divisor $e$
contracts to a point, and the hyperplane $\ell$ in $dP_1$ becomes the
hyperplane in $\bP^2$. By looking to the Chern characters
in~\eqref{eq:dP1-naive-basis}, and the fractional basis for $\bP^2$
given in \cite{dualities2}, a very plausible description of the
Higgsing in terms of sheaves is given by
\be
  \pi_*(\cE_1+\cE_2)=\cO \qquad;\qquad \pi_*\cE_3 =
\Omega(1)[1]\qquad ; \qquad \pi_*\cE_4 = \cO(-1)[2]\, .
\ee
From this point of view, it is no surprise that the
collection~\eqref{eq:dP1-naive-basis} is not invariant under the
ordinary orientifold action~\eqref{eq:orientifold-action-B}, since in the $\bP^2$ case we would need to
compose~\eqref{eq:orientifold-action-B} with an auto-equivalence of
the category to be able to exchange $\pi_*\cE_3$ and $\pi_*\cE_4$,
leaving $\pi_*(\cE_1+\cE_2)$ invariant. Given this correspondence, and
the form of the quantum monodromy in the $\bC^3/\bZ_3$ case, we can
make a natural suggestion for the auto-equivalence in $dP_1$: 
\be
  \cM_q = \mathfrak{L}^{-1}\cdot \mathfrak{M}_{\cO(-\ell)}\cdot \mathfrak{L}\, ,
\ee
where $\mathfrak{L}$ is the operation of tensoring a sheaf with
$\cO(-\ell)$, and $\mathfrak{M}_{\cO(-\ell)}$ is the monodromy around
the locus where $\cO(-\ell)$ becomes massless, which we assume to exist, in analogy with
the $\bC^3/\bZ_3$ case. Acting on the Chern characters,
we have:
\begin{align}
  \ch(\mathfrak{M}_{\cO(-\ell)}\cF) = \ch(\cF) -
  \dsz{\cF}{\cO(-\ell)}\ch(\cO(-\ell))\, .
\end{align}
After acting with $\cM_q$, we have a new basis of branes given by:
\begin{equation}
  \label{eq:dP1-orientifold-basis}
    \cE_1  = \cO[2] \;\;,\;\;
    \cE_2 = \cO(e)[0] \;\;,\;\;
    \cE_3 = \cO(2e-\ell)[1] \;\;,\;\;
    \cE_4 = \cO(\ell-e)[1]\, .
\end{equation}
It is now straightforward to verify that this basis is
invariant under the orientifold
action~\eqref{eq:orientifold-action-B} with $\cL_{2B}=\cO(3\ell)$,
giving the involution on the quiver shown in 
figure~\ref{fig:orientifolded-dP1-quiver}. The corresponding charge
vectors are given by:
\begin{equation}
  \label{eq:dP1-brane-charges}
  \begin{aligned}
    e^{-B}\Gamma(\cE_1) & = 1 - \frac{1}{2}e + \frac{1}{24}\ell^2 \,,\\
    e^{-B}\Gamma(\cE_2) & = 1 + \frac{1}{2}e + \frac{1}{24}\ell^2 \,,
    \end{aligned}
    \qquad 
    \begin{aligned}
    e^{-B}\Gamma(\cE_3) & = -1 - \frac{3}{2}e + \ell +
    \frac{11}{24}\ell^2 \,,\\
    e^{-B}\Gamma(\cE_4) & = -1 + \frac{3}{2}e - \ell +
    \frac{11}{24}\ell^2\, .
  \end{aligned}
\end{equation}
After applying the orientifold action~\eqref{eq:orientifold-action-B}
we could apply $\cM_q$ in order to obtain the action on the original
basis of branes, but for simplicity we will stay with the
basis~\eqref{eq:dP1-orientifold-basis} for the remainder of this
section.

The spectrum for the quiver can now be computed using the same
techniques as in \cite{dualities2}. We first consider an
O7$^+$ plane wrapping the contracting $dP_1$ surface,
which requires a configuration of the form
$\mathrm{O7}^++4(\cE_3+\cE_4)+(N-4)\,\mathrm{D3s}$ in order to
cancel tadpoles, where $N-4$ counts D3 branes in
covering space conventions. The resulting gauge group is
$\SU(N-4)\times \SU(N)$, and the matter content of the theory is given in the following table:
\begin{align}
  \label{eq:dP1+-quiver}
  \begin{array}{cc|cc}
    SU(N-4) & SU(N) & \SU(2)\\
    \hline
    \fund & \ov\fund & \fund\\
    \fund & \ov\fund & {\bf 1}\\
    \ov\symm & {\bf 1} & {\bf 1}\\
    {\bf 1} & \asymm & \fund\\
    {\bf 1} & \asymm & {\bf 1}
  \end{array}
\end{align}
We identify this configuration as phase $\IA$. 
To understand the origin of the $\SU(2)$ flavor symmetry, we note that
$dP_1$ is the moduli space of the gauged linear
sigma model
\begin{align}
  \begin{array}{c|cccc}
    & x_0 & x_1 & x_2 & x_3\\
    \hline
    \bC^* & -1 & 1 & 0 & 1\\
    \bC^* & 1 & 0 & 1 & 0
    \end{array}
\end{align}
Notice that $x_1=0$ and $x_3=0$ are linearly equivalent divisors, and
thus there is a $\SU(2)$ symmetry rotating the corresponding
sections. As an example, the doublet between $\cE_1$ and $\cE_4$ comes
from:
\begin{equation}
    \Ext_X^1(i_*\cE_1,\cE_4) = \Ext_X^0(i_*\cO,i_*\cO(\ell-e))
    = \Ext_\cS^0(\cO,\cO(\ell-e)) = \{x_1,x_3\} = \bC^2\, ,
\end{equation}
where $\cS=dP_1$, and $\{x_1,x_3\}$ denotes the space spanned by the
$x_1,x_3$ sections. The ${\bf 3}={\bf 2}+{\bf 1}$ split in the
bifundamentals between $\cE_3$ and $\cE_4$ can be understood in a
similar way:
\begin{align}
  \begin{split}
    \Ext_X^1(i_*\cE_4, i_*\cE_3) & = \Ext_\cS^1(\cO(\ell-e),
    \cO(2e-\ell))\\
    & = \left\{\frac{1}{x_1x_3}x_0, \frac{1}{x_1x_3} x_2 \frac{1}{x_1},
    \frac{1}{x_1x_3} x_2 \frac{1}{x_3}\right\} = \bC\otimes\bC^2\, ,
  \end{split}
\end{align}
where we have used the ``rationom'' counting procedure of
\cite{Blumenhagen:2010pv,Jow-cohomology,Rahn:2010fm} for representing
the generators of $H^1(dP_1, \cO(3e-2\ell))$.


Applying S-duality to $\IA^-$, we expected a transition in the orientifold type of the form
\begin{align}
  O7^+ + 8(\cE_3+\cE_4) \longleftrightarrow O7^- + n\, D3s\, .
\end{align}
Charge conservation implies $n=6$, and we obtain a dual theory given
by:
\begin{align}
  \label{eq:dP1--quiver}
  \begin{array}{cc|cc}
    SU(N+2) & SU(N-2) & \SU(2)\\
    \hline
    \fund & \ov\fund & \fund\\
    \ov\fund & \ov\fund & {\bf 1}\\
    \ov\asymm & {\bf 1} & {\bf 1}\\
    {\bf 1} & \symm & \fund\\
    {\bf 1} & \symm & {\bf 1}
  \end{array}
\end{align}
which we identify as phase $\IB$ with $\tN=N-2$, agreeing with our
discussion in \cite{dualities1} and \S\ref{sec:dP1-tiling}.

\section{Results for the superconformal index of \alt{$dP_1$}{dP1}}
\label{app:SCI}

In this appendix, we present a number of lengthy formulas for the superconformal index of the $dP_1$ theories expanded in powers of $t$. Our conventions are explained in~\S\ref{subsec:SCImatching}. The code that we used for our computations, based around the computer algebra package \texttt{LiE}~\cite{LiE}, is attached to the {\tt arXiv} submission.

\subsection{\alt{$\II^+_- = \III_+$}{II+-=III+}, \alt{$Q_{\rm D3}=8$}{QD3=8}} \label{subsec:II-plus-minus-index}

 \begin{dmath*}[style={\small},indentstep={0pt},spread={0pt}]
 1+b^{{-1}}y^{4}t_{\text{(1.077)}}^{{15-\frac{24}{\beta}-2\beta}}X_5+t^{2}\left(-1-X_2\right)+b^{{-1}}y^{4}J_1t_{\text{(2.077)}}^{{16-\frac{24}{\beta}-2\beta}}X_5+b^{{-2}}y^{8}t_{\text{(2.153)}}^{{30-\frac{48}{\beta}-4\beta}}\left(X_2+X_6+X_{10}\right)-yt_{\text{(2.178)}}^{{6-\beta}}X_1+by^{{-7}}t_{\text{(2.39)}}^{{-23+\frac{24}{\beta}+5\beta}}+by^{{-6}}t_{\text{(2.568)}}^{{-19+\frac{24}{\beta}+4\beta}}X_1+2by^{{-5}}t_{\text{(2.746)}}^{{-15+\frac{24}{\beta}+3\beta}}X_2-y^{{-1}}J_1t_{\text{(2.822)}}^{{-1+\beta}}X_1-b^{{-1}}y^{3}t_{\text{(2.899)}}^{{13-\frac{24}{\beta}-\beta}}X_4+by^{{-4}}t_{\text{(2.923)}}^{{-11+\frac{24}{\beta}+2\beta}}\left(X_1+2X_3\right)-2t^{3}J_1\left(1+X_2\right)+b^{{-1}}y^{4}t_{\text{(3.077)}}^{{17-\frac{24}{\beta}-2\beta}}\left(-X_3+\left(-2+J_2\right)X_5-X_7\right)+by^{{-3}}t_{\text{(3.101)}}^{{-7+\frac{24}{\beta}+\beta}}\left(2+X_2+2X_4\right)+b^{{-2}}y^{8}J_1t_{\text{(3.153)}}^{{31-\frac{48}{\beta}-4\beta}}\left(1+X_2+X_4+X_6+X_8+X_{10}\right)-yJ_1t_{\text{(3.178)}}^{{7-\beta}}\left(X_1+X_3\right)+b^{{-3}}y^{{12}}t_{\text{(3.23)}}^{{45-\frac{72}{\beta}-6\beta}}\left(X_3+X_5+X_7+X_9+X_{11}+X_{15}\right)-b^{{-1}}y^{5}t_{\text{(3.254)}}^{{21-\frac{24}{\beta}-3\beta}}\left(X_2+X_4+X_6\right)+by^{{-2}}t_{\text{(3.279)}}^{{-3+\frac{24}{\beta}}}\left(X_1+X_3+X_5\right)+b^{3}y^{{-9}}t_{\text{(3.304)}}^{{-27+\frac{72}{\beta}+3\beta}}+by^{{-7}}J_1t_{\text{(3.39)}}^{{-22+\frac{24}{\beta}+5\beta}}+by^{{-1}}t_{\text{(3.457)}}^{{1+\frac{24}{\beta}-\beta}}\left(X_2+X_6\right)+y^{{-3}}t_{\text{(3.466)}}^{{-8+3\beta}}\left(X_1+X_3+X_5\right)+by^{{-6}}J_1t_{\text{(3.568)}}^{{-18+\frac{24}{\beta}+4\beta}}X_1+y^{{-2}}t_{\text{(3.644)}}^{{-4+2\beta}}\left(1+4X_2+2X_4+X_6\right)+2by^{{-5}}J_1t_{\text{(3.746)}}^{{-14+\frac{24}{\beta}+3\beta}}X_2-y^{{-1}}t_{\text{(3.822)}}^{{\beta}}\left(\left(-5+J_2\right)X_1-7X_3-2\left(2X_5+X_7\right)\right)-b^{{-1}}y^{3}J_1t_{\text{(3.899)}}^{{14-\frac{24}{\beta}-\beta}}\left(X_2+2X_4+X_6\right)+by^{{-4}}J_1t_{\text{(3.923)}}^{{-10+\frac{24}{\beta}+2\beta}}\left(X_1+2X_3\right)-b^{{-2}}y^{7}t_{\text{(3.975)}}^{{28-\frac{48}{\beta}-3\beta}}\left(X_1+X_3+X_5+X_7+X_9\right)+t^{4}\left(1+8X_2-2J_2\left(1+X_2\right)+8X_4+4X_6+2X_8\right)-b^{{-1}}y^{4}t_{\text{(4.077)}}^{{18-\frac{24}{\beta}-2\beta}}\left(-J_3X_5+J_1\left(X_1+3\left(X_3+2X_5+X_7\right)\right)\right)+by^{{-3}}J_1t_{\text{(4.101)}}^{{-6+\frac{24}{\beta}+\beta}}\left(2+X_2+2X_4\right)+b^{{-2}}y^{8}t_{\text{(4.153)}}^{{32-\frac{48}{\beta}-4\beta}}\left(-2X_2-X_4-2X_6-X_8-2X_{10}+J_2\left(1+2X_2+X_4+2X_6+X_8+2X_{10}\right)-X_{12}\right)-yt_{\text{(4.178)}}^{{8-\beta}}\left(\left(-6+J_2\right)X_1+\left(-9+J_2\right)X_3-2\left(4X_5+2X_7+X_9\right)\right)+b^{{-3}}y^{{12}}J_1t_{\text{(4.23)}}^{{46-\frac{72}{\beta}-6\beta}}\left(X_1+2X_3+3X_5+3X_7+2X_9+2X_{11}+X_{13}+X_{15}\right)-b^{{-1}}y^{5}J_1t_{\text{(4.254)}}^{{22-\frac{24}{\beta}-3\beta}}\left(1+2X_2+3X_4+3X_6+X_8\right)+by^{{-2}}J_1t_{\text{(4.279)}}^{{-2+\frac{24}{\beta}}}\left(X_1+X_3+X_5\right)+b^{3}y^{{-9}}J_1t_{\text{(4.304)}}^{{-26+\frac{72}{\beta}+3\beta}}+b^{{-4}}y^{{16}}t_{\text{(4.306)}}^{{60-\frac{96}{\beta}-8\beta}}\left(1+2X_4+X_6+2X_8+X_{10}+2X_{12}+X_{14}+X_{16}+X_{20}\right)-b^{{-2}}y^{9}t_{\text{(4.331)}}^{{36-\frac{48}{\beta}-5\beta}}\left(X_1+2X_3+2X_5+2X_7+X_9+X_{11}\right)+y^{2}t_{\text{(4.356)}}^{{12-2\beta}}\left(2+4X_2+4X_4+5X_6+2X_8+X_{10}\right)+by^{{-7}}t_{\text{(4.39)}}^{{-21+\frac{24}{\beta}+5\beta}}\left(-1+J_2-2X_2\right)+by^{{-1}}J_1t_{\text{(4.457)}}^{{2+\frac{24}{\beta}-\beta}}\left(X_2+X_6\right)+y^{{-3}}J_1t_{\text{(4.466)}}^{{-7+3\beta}}\left(X_1+X_3+2X_5\right)+y^{3}t_{\text{(4.534)}}^{{16-3\beta}}\left(X_1+2X_3+2X_5+2X_7+X_9+X_{11}\right)+b^{{-1}}yt_{\text{(4.543)}}^{{7-\frac{24}{\beta}+\beta}}\left(3X_2+2X_4+3X_6+X_8+X_{10}\right)+by^{{-6}}t_{\text{(4.568)}}^{{-17+\frac{24}{\beta}+4\beta}}\left(\left(-3+J_2\right)X_1-3X_3\right)+y^{{-2}}J_1t_{\text{(4.644)}}^{{-3+2\beta}}\left(2+3X_2+3X_4+2X_6\right)+b^{{-1}}y^{2}t_{\text{(4.721)}}^{{11-\frac{24}{\beta}}}\left(5X_1+8X_3+7X_5+6X_7+2X_9+X_{11}\right)+by^{{-5}}t_{\text{(4.746)}}^{{-13+\frac{24}{\beta}+3\beta}}\left(-4+2\left(-3+J_2\right)X_2-5X_4\right)+b^{2}y^{{-14}}t_{\text{(4.779)}}^{{-46+\frac{48}{\beta}+10\beta}}+y^{{-1}}t_{\text{(4.822)}}^{{1+\beta}}\left(-J_3X_1+J_1\left(7X_1+10X_3+6X_5+4X_7\right)\right)-b^{{-1}}y^{3}t_{\text{(4.899)}}^{{15-\frac{24}{\beta}-\beta}}\left(-8+\left(-13+J_2\right)X_2+\left(-19+3J_2\right)X_4-14X_6+2J_2X_6-11X_8-4X_{10}-2X_{12}\right)+by^{{-4}}t_{\text{(4.923)}}^{{-9+\frac{24}{\beta}+2\beta}}\left(\left(-8+J_2\right)X_1+\left(-9+2J_2\right)X_3-5X_5\right)+b^{2}y^{{-13}}t_{\text{(4.957)}}^{{-42+\frac{48}{\beta}+9\beta}}X_1-b^{{-2}}y^{7}J_1t_{\text{(4.975)}}^{{29-\frac{48}{\beta}-3\beta}}\left(3X_1+4X_3+4X_5+4X_7+3X_9+X_{11}\right)+t^{5}\left(-2J_3\left(1+X_2\right)+J_1\left(5+14X_2+12X_4+7X_6+4X_8\right)\right)+\ldots
 \end{dmath*}

\subsection{\alt{$\IA^- = \IB^- = \mathrm{III}_{-}$}{IA-=IB-=III-}, \alt{$Q_{\rm D3}=8$}{QD3=8}} \label{subsec:IA-minus-index}

 \begin{dmath*}[style={\small},indentstep={0pt},spread={-1.8pt}]
 1+t^{2}\left(-1-X_2\right)-yt_{\text{(2.178)}}^{{6-\beta}}X_1+by^{{-7}}t_{\text{(2.39)}}^{{-23+\frac{24}{\beta}+5\beta}}+by^{{-6}}t_{\text{(2.568)}}^{{-19+\frac{24}{\beta}+4\beta}}X_1+2by^{{-5}}t_{\text{(2.746)}}^{{-15+\frac{24}{\beta}+3\beta}}X_2-y^{{-1}}J_1t_{\text{(2.822)}}^{{-1+\beta}}X_1+by^{{-4}}t_{\text{(2.923)}}^{{-11+\frac{24}{\beta}+2\beta}}\left(X_1+2X_3\right)-2t^{3}J_1\left(1+X_2\right)+by^{{-3}}t_{\text{(3.101)}}^{{-7+\frac{24}{\beta}+\beta}}\left(2+X_2+2X_4\right)-yJ_1t_{\text{(3.178)}}^{{7-\beta}}\left(X_1+X_3\right)+by^{{-2}}t_{\text{(3.279)}}^{{-3+\frac{24}{\beta}}}\left(X_1+X_3+X_5\right)+b^{3}y^{{-9}}t_{\text{(3.304)}}^{{-27+\frac{72}{\beta}+3\beta}}+by^{{-7}}J_1t_{\text{(3.39)}}^{{-22+\frac{24}{\beta}+5\beta}}+by^{{-1}}t_{\text{(3.457)}}^{{1+\frac{24}{\beta}-\beta}}\left(X_2+X_6\right)+by^{{-6}}J_1t_{\text{(3.568)}}^{{-18+\frac{24}{\beta}+4\beta}}X_1+y^{{-2}}t_{\text{(3.644)}}^{{-4+2\beta}}X_2+2by^{{-5}}J_1t_{\text{(3.746)}}^{{-14+\frac{24}{\beta}+3\beta}}X_2-y^{{-1}}t_{\text{(3.822)}}^{{\beta}}\left(J_2X_1-X_3\right)+by^{{-4}}J_1t_{\text{(3.923)}}^{{-10+\frac{24}{\beta}+2\beta}}\left(X_1+2X_3\right)+t^{4}\left(-2+X_2-2J_2\left(1+X_2\right)+X_4\right)+by^{{-3}}J_1t_{\text{(4.101)}}^{{-6+\frac{24}{\beta}+\beta}}\left(2+X_2+2X_4\right)-yt_{\text{(4.178)}}^{{8-\beta}}\left(\left(-1+J_2\right)X_1+\left(-1+J_2\right)X_3-X_5\right)+by^{{-2}}J_1t_{\text{(4.279)}}^{{-2+\frac{24}{\beta}}}\left(X_1+X_3+X_5\right)+b^{3}y^{{-9}}J_1t_{\text{(4.304)}}^{{-26+\frac{72}{\beta}+3\beta}}+y^{2}t_{\text{(4.356)}}^{{12-2\beta}}\left(1+X_6\right)+by^{{-7}}t_{\text{(4.39)}}^{{-21+\frac{24}{\beta}+5\beta}}\left(-1+J_2-2X_2\right)+by^{{-1}}J_1t_{\text{(4.457)}}^{{2+\frac{24}{\beta}-\beta}}\left(X_2+X_6\right)+b^{{-1}}yt_{\text{(4.543)}}^{{7-\frac{24}{\beta}+\beta}}\left(2X_2+X_4+X_6\right)+by^{{-6}}t_{\text{(4.568)}}^{{-17+\frac{24}{\beta}+4\beta}}\left(\left(-3+J_2\right)X_1-3X_3\right)+y^{{-2}}J_1t_{\text{(4.644)}}^{{-3+2\beta}}\left(1+X_2\right)+2b^{{-1}}y^{2}t_{\text{(4.721)}}^{{11-\frac{24}{\beta}}}\left(2X_1+3X_3+2X_5+X_7\right)+by^{{-5}}t_{\text{(4.746)}}^{{-13+\frac{24}{\beta}+3\beta}}\left(-4+2\left(-3+J_2\right)X_2-5X_4\right)+b^{2}y^{{-14}}t_{\text{(4.779)}}^{{-46+\frac{48}{\beta}+10\beta}}+y^{{-1}}t_{\text{(4.822)}}^{{1+\beta}}\left(-J_3X_1+3J_1\left(X_1+X_3\right)\right)+3b^{{-1}}y^{3}t_{\text{(4.899)}}^{{15-\frac{24}{\beta}-\beta}}\left(2+3X_2+4X_4+2X_6+X_8\right)+by^{{-4}}t_{\text{(4.923)}}^{{-9+\frac{24}{\beta}+2\beta}}\left(\left(-8+J_2\right)X_1+\left(-9+2J_2\right)X_3-5X_5\right)+b^{2}y^{{-13}}t_{\text{(4.957)}}^{{-42+\frac{48}{\beta}+9\beta}}X_1+t^{5}\left(-2J_3\left(1+X_2\right)+J_1\left(3+7X_2+4X_4\right)\right)+b^{{-1}}y^{4}t_{\text{(5.077)}}^{{19-\frac{24}{\beta}-2\beta}}\left(10X_1+14X_3+13X_5+7X_7+3X_9\right)+by^{{-3}}t_{\text{(5.101)}}^{{-5+\frac{24}{\beta}+\beta}}\left(-4-12X_2-9X_4+J_2\left(2+X_2+2X_4\right)-5X_6\right)-b^{3}y^{{-10}}t_{\text{(5.126)}}^{{-29+\frac{72}{\beta}+4\beta}}X_1+3b^{2}y^{{-12}}t_{\text{(5.135)}}^{{-38+\frac{48}{\beta}+8\beta}}X_2+yt_{\text{(5.178)}}^{{9-\beta}}\left(-J_3\left(X_1+X_3\right)+3J_1\left(2X_1+2X_3+X_5\right)\right)-by^{{-8}}J_1t_{\text{(5.212)}}^{{-24+\frac{24}{\beta}+6\beta}}X_1+b^{{-1}}y^{5}t_{\text{(5.254)}}^{{23-\frac{24}{\beta}-3\beta}}\left(11X_2+2\left(1+5X_4+5X_6+2X_8+X_{10}\right)\right)+by^{{-2}}t_{\text{(5.279)}}^{{-1+\frac{24}{\beta}}}\left(\left(-7+J_2\right)X_1+\left(-10+J_2\right)X_3-7X_5+J_2X_5-3X_7\right)+b^{3}y^{{-9}}t_{\text{(5.304)}}^{{-25+\frac{72}{\beta}+3\beta}}\left(-1+J_2-X_2\right)+b^{2}y^{{-11}}t_{\text{(5.313)}}^{{-34+\frac{48}{\beta}+7\beta}}\left(3X_1+4X_3\right)+y^{2}J_1t_{\text{(5.356)}}^{{13-2\beta}}\left(1+2X_2+2X_4+X_6\right)-by^{{-7}}t_{\text{(5.39)}}^{{-20+\frac{24}{\beta}+5\beta}}\left(-J_3+J_1\left(4+5X_2\right)\right)+b^{{-1}}y^{6}t_{\text{(5.432)}}^{{27-\frac{24}{\beta}-4\beta}}\left(3X_1+5X_3+5X_5+4X_7+2X_9+X_{11}\right)+by^{{-1}}t_{\text{(5.457)}}^{{3+\frac{24}{\beta}-\beta}}\left(-3+\left(-5+J_2\right)X_2-6X_4-4X_6+J_2X_6-2X_8\right)-y^{{-3}}t_{\text{(5.466)}}^{{-6+3\beta}}X_1-b^{3}y^{{-8}}t_{\text{(5.482)}}^{{-21+\frac{72}{\beta}+2\beta}}X_1+b^{2}y^{{-10}}t_{\text{(5.491)}}^{{-30+\frac{48}{\beta}+6\beta}}\left(7+5X_2+7X_4\right)+b^{{-1}}yJ_1t_{\text{(5.543)}}^{{8-\frac{24}{\beta}+\beta}}\left(2X_2+X_4+X_6\right)-by^{{-6}}t_{\text{(5.568)}}^{{-16+\frac{24}{\beta}+4\beta}}\left(-J_3X_1+2J_1\left(5X_1+4X_3\right)\right)+b^{{-1}}y^{7}t_{\text{(5.61)}}^{{31-\frac{24}{\beta}-5\beta}}\left(1+X_4+X_8\right)-bt_{\text{(5.635)}}^{{7+\frac{24}{\beta}-2\beta}}\left(X_1+2X_3+2X_5+X_7\right)+y^{{-2}}t_{\text{(5.644)}}^{{-2+2\beta}}\left(-X_2+J_2\left(2+X_2\right)-2X_4\right)+b^{2}y^{{-9}}t_{\text{(5.669)}}^{{-26+\frac{48}{\beta}+5\beta}}\left(11X_1+11X_3+7X_5\right)+b^{4}y^{{-16}}t_{\text{(5.694)}}^{{-50+\frac{96}{\beta}+8\beta}}+2b^{{-1}}y^{2}J_1t_{\text{(5.721)}}^{{12-\frac{24}{\beta}}}\left(2X_1+3X_3+2X_5+X_7\right)-2by^{{-5}}t_{\text{(5.746)}}^{{-12+\frac{24}{\beta}+3\beta}}\left(-J_3X_2+J_1\left(5+9X_2+6X_4\right)\right)+b^{2}y^{{-14}}J_1t_{\text{(5.779)}}^{{-45+\frac{48}{\beta}+10\beta}}+y^{{-1}}t_{\text{(5.822)}}^{{2+\beta}}\left(\left(4+7J_2-J_4\right)X_1+5J_2X_3-2X_5\right)+b^{2}y^{{-8}}t_{\text{(5.847)}}^{{-22+\frac{48}{\beta}+4\beta}}\left(4+23X_2+12X_4+9X_6\right)+b^{4}y^{{-15}}t_{\text{(5.872)}}^{{-46+\frac{96}{\beta}+7\beta}}X_1+b^{{-1}}y^{3}J_1t_{\text{(5.899)}}^{{16-\frac{24}{\beta}-\beta}}\left(6+8X_2+11X_4+6X_6+3X_8\right)-by^{{-4}}t_{\text{(5.923)}}^{{-8+\frac{24}{\beta}+2\beta}}\left(-J_3\left(X_1+2X_3\right)+J_1\left(20X_1+24X_3+13X_5\right)\right)-b^{3}y^{{-11}}J_1t_{\text{(5.948)}}^{{-32+\frac{72}{\beta}+5\beta}}+2b^{2}y^{{-13}}J_1t_{\text{(5.957)}}^{{-41+\frac{48}{\beta}+9\beta}}X_1+t^{6}\left(9+10X_2-2J_4\left(1+X_2\right)+3X_4+2J_2\left(3+8X_2+3X_4\right)-2X_6\right)+\ldots
 \end{dmath*}

\subsection{\alt{$\IA^+ = \II_-^-$}{IA+=II--}, \alt{$Q_{\rm D3}=9$}{QD3=9}} \label{subsec:IA-plus-index}

 \begin{dmath*}[style={\small},indentstep={0pt},spread={-1pt}]
 1+b^{{\frac{1}{2}}}y^{{-5}}t_{\text{(1.294)}}^{{-17+\frac{27}{2\beta}+\frac{31\beta}{8}}}+b^{{\frac{1}{2}}}y^{{-4}}t_{\text{(1.488)}}^{{-13+\frac{27}{2\beta}+\frac{23\beta}{8}}}X_1+b^{{\frac{1}{2}}}y^{{-3}}t_{\text{(1.683)}}^{{-9+\frac{27}{2\beta}+\frac{15\beta}{8}}}X_2+b^{{\frac{1}{2}}}y^{{-2}}t_{\text{(1.877)}}^{{-5+\frac{27}{2\beta}+\frac{7\beta}{8}}}X_3+t^{2}\left(-1-X_2\right)+b^{{\frac{1}{2}}}y^{{-1}}t_{\text{(2.072)}}^{{-1+\frac{27}{2\beta}-\frac{\beta}{8}}}X_4-yt_{\text{(2.194)}}^{{6-\beta}}X_1+b^{{\frac{1}{2}}}t_{\text{(2.266)}}^{{3+\frac{27}{2\beta}-\frac{9\beta}{8}}}X_5+b^{{\frac{1}{2}}}y^{{-5}}J_1t_{\text{(2.294)}}^{{-16+\frac{27}{2\beta}+\frac{31\beta}{8}}}+b^{{\frac{1}{2}}}y^{{-4}}J_1t_{\text{(2.488)}}^{{-12+\frac{27}{2\beta}+\frac{23\beta}{8}}}X_1+by^{{-10}}t_{\text{(2.588)}}^{{-34+\frac{27}{\beta}+\frac{31\beta}{4}}}+b^{{\frac{1}{2}}}y^{{-3}}J_1t_{\text{(2.683)}}^{{-8+\frac{27}{2\beta}+\frac{15\beta}{8}}}X_2+by^{{-9}}t_{\text{(2.783)}}^{{-30+\frac{27}{\beta}+\frac{27\beta}{4}}}X_1-y^{{-1}}J_1t_{\text{(2.806)}}^{{-1+\beta}}X_1+b^{{\frac{1}{2}}}y^{{-2}}J_1t_{\text{(2.877)}}^{{-4+\frac{27}{2\beta}+\frac{7\beta}{8}}}X_3+2by^{{-8}}t_{\text{(2.977)}}^{{-26+\frac{27}{\beta}+\frac{23\beta}{4}}}X_2-2t^{3}J_1\left(1+X_2\right)+b^{{\frac{1}{2}}}y^{{-1}}J_1t_{\text{(3.072)}}^{{\frac{27}{2\beta}-\frac{\beta}{8}}}X_4+by^{{-7}}t_{\text{(3.171)}}^{{-22+\frac{27}{\beta}+\frac{19\beta}{4}}}\left(X_1+2X_3\right)-yJ_1t_{\text{(3.194)}}^{{7-\beta}}\left(X_1+X_3\right)+b^{{\frac{1}{2}}}J_1t_{\text{(3.266)}}^{{4+\frac{27}{2\beta}-\frac{9\beta}{8}}}X_5+b^{{\frac{1}{2}}}y^{{-5}}t_{\text{(3.294)}}^{{-15+\frac{27}{2\beta}+\frac{31\beta}{8}}}\left(-1+J_2-X_2\right)+by^{{-6}}t_{\text{(3.366)}}^{{-18+\frac{27}{\beta}+\frac{15\beta}{4}}}\left(2+X_2+3X_4\right)+b^{{\frac{1}{2}}}y^{{-4}}t_{\text{(3.488)}}^{{-11+\frac{27}{2\beta}+\frac{23\beta}{8}}}\left(\left(-2+J_2\right)X_1-2X_3\right)+by^{{-5}}t_{\text{(3.56)}}^{{-14+\frac{27}{\beta}+\frac{11\beta}{4}}}\left(2X_1+2X_3+3X_5\right)+by^{{-10}}J_1t_{\text{(3.588)}}^{{-33+\frac{27}{\beta}+\frac{31\beta}{4}}}+y^{{-2}}t_{\text{(3.611)}}^{{-4+2\beta}}X_2+b^{3}y^{{-10}}t_{\text{(3.653)}}^{{-30+\frac{81}{\beta}+\frac{13\beta}{4}}}+b^{{\frac{1}{2}}}y^{{-3}}t_{\text{(3.683)}}^{{-7+\frac{27}{2\beta}+\frac{15\beta}{8}}}\left(-1+\left(-3+J_2\right)X_2-2X_4\right)+by^{{-4}}t_{\text{(3.755)}}^{{-10+\frac{27}{\beta}+\frac{7\beta}{4}}}\left(4X_2+2X_4+3X_6\right)+2by^{{-9}}J_1t_{\text{(3.783)}}^{{-29+\frac{27}{\beta}+\frac{27\beta}{4}}}X_1-y^{{-1}}t_{\text{(3.806)}}^{{\beta}}\left(J_2X_1-X_3\right)+b^{{\frac{1}{2}}}y^{{-2}}t_{\text{(3.877)}}^{{-3+\frac{27}{2\beta}+\frac{7\beta}{8}}}\left(-2X_1+\left(-3+J_2\right)X_3-2X_5\right)+b^{{\frac{3}{2}}}y^{{-15}}t_{\text{(3.882)}}^{{-51+\frac{81}{2\beta}+\frac{93\beta}{8}}}+by^{{-3}}t_{\text{(3.949)}}^{{-6+\frac{27}{\beta}+\frac{3\beta}{4}}}\left(2X_1+3X_3+2\left(X_5+X_7\right)\right)+by^{{-8}}J_1t_{\text{(3.977)}}^{{-25+\frac{27}{\beta}+\frac{23\beta}{4}}}\left(1+3X_2\right)+t^{4}\left(-2+X_2-2J_2\left(1+X_2\right)+X_4\right)+b^{{\frac{1}{2}}}y^{{-1}}t_{\text{(4.072)}}^{{1+\frac{27}{2\beta}-\frac{\beta}{8}}}\left(-1-2X_2+\left(-3+J_2\right)X_4-2X_6\right)+b^{{\frac{3}{2}}}y^{{-14}}t_{\text{(4.077)}}^{{-47+\frac{81}{2\beta}+\frac{85\beta}{8}}}X_1-b^{{\frac{1}{2}}}y^{{-6}}J_1t_{\text{(4.1)}}^{{-18+\frac{27}{2\beta}+\frac{39\beta}{8}}}X_1+by^{{-2}}t_{\text{(4.143)}}^{{-2+\frac{27}{\beta}-\frac{\beta}{4}}}\left(2+X_2+3X_4+X_6+2X_8\right)+2by^{{-7}}J_1t_{\text{(4.171)}}^{{-21+\frac{27}{\beta}+\frac{19\beta}{4}}}\left(X_1+2X_3\right)+b^{{-\frac{3}{2}}}y^{5}t_{\text{(4.174)}}^{{21-\frac{81}{2\beta}-\frac{13\beta}{8}}}\left(X_2+X_6\right)-yt_{\text{(4.194)}}^{{8-\beta}}\left(\left(-1+J_2\right)X_1+\left(-1+J_2\right)X_3-X_5\right)-b^{{\frac{1}{2}}}t_{\text{(4.266)}}^{{5+\frac{27}{2\beta}-\frac{9\beta}{8}}}\left(X_1+2X_3+3X_5-J_2X_5+X_7\right)+2b^{{\frac{3}{2}}}y^{{-13}}t_{\text{(4.271)}}^{{-43+\frac{81}{2\beta}+\frac{77\beta}{8}}}X_2-b^{{\frac{1}{2}}}y^{{-5}}t_{\text{(4.294)}}^{{-14+\frac{27}{2\beta}+\frac{31\beta}{8}}}\left(-J_3+4J_1\left(1+X_2\right)\right)+by^{{-1}}t_{\text{(4.338)}}^{{2+\frac{27}{\beta}-\frac{5\beta}{4}}}\left(X_1+X_3+X_5+X_7+X_9\right)+by^{{-6}}J_1t_{\text{(4.366)}}^{{-17+\frac{27}{\beta}+\frac{15\beta}{4}}}\left(2+3X_2+5X_4\right)+b^{{-\frac{3}{2}}}y^{6}t_{\text{(4.368)}}^{{25-\frac{81}{2\beta}-\frac{21\beta}{8}}}\left(X_1+2X_3+X_5+X_7\right)+y^{2}t_{\text{(4.389)}}^{{12-2\beta}}\left(1+X_6\right)-b^{{\frac{1}{2}}}yt_{\text{(4.461)}}^{{9+\frac{27}{2\beta}-\frac{17\beta}{8}}}\left(X_2+X_4+X_6\right)+b^{{\frac{3}{2}}}y^{{-12}}t_{\text{(4.465)}}^{{-39+\frac{81}{2\beta}+\frac{69\beta}{8}}}\left(X_1+3X_3\right)-b^{{\frac{1}{2}}}y^{{-4}}t_{\text{(4.488)}}^{{-10+\frac{27}{2\beta}+\frac{23\beta}{8}}}\left(-J_3X_1+J_1\left(8X_1+6X_3\right)\right)+bt_{\text{(4.532)}}^{{6+\frac{27}{\beta}-\frac{9\beta}{4}}}\left(X_2+X_6+X_{10}\right)+by^{{-5}}J_1t_{\text{(4.56)}}^{{-13+\frac{27}{\beta}+\frac{11\beta}{4}}}\left(3X_1+4X_3+6X_5\right)+b^{{-\frac{3}{2}}}y^{7}t_{\text{(4.563)}}^{{29-\frac{81}{2\beta}-\frac{29\beta}{8}}}\left(1+X_2+2X_4+X_6+X_8\right)+by^{{-10}}t_{\text{(4.588)}}^{{-32+\frac{27}{\beta}+\frac{31\beta}{4}}}\left(-1+2J_2-X_2\right)+y^{{-2}}J_1t_{\text{(4.611)}}^{{-3+2\beta}}\left(1+X_2\right)+b^{3}y^{{-10}}J_1t_{\text{(4.653)}}^{{-29+\frac{81}{\beta}+\frac{13\beta}{4}}}+b^{{\frac{3}{2}}}y^{{-11}}t_{\text{(4.66)}}^{{-35+\frac{81}{2\beta}+\frac{61\beta}{8}}}\left(3+2X_2+4X_4\right)-b^{{\frac{1}{2}}}y^{{-3}}t_{\text{(4.683)}}^{{-6+\frac{27}{2\beta}+\frac{15\beta}{8}}}\left(-J_3X_2+J_1\left(4+10X_2+7X_4\right)\right)+by^{{-4}}J_1t_{\text{(4.755)}}^{{-9+\frac{27}{\beta}+\frac{7\beta}{4}}}\left(1+5X_2+5X_4+5X_6\right)+b^{{-\frac{3}{2}}}y^{8}t_{\text{(4.757)}}^{{33-\frac{81}{2\beta}-\frac{37\beta}{8}}}\left(X_3+X_5+X_9\right)+by^{{-9}}t_{\text{(4.783)}}^{{-28+\frac{27}{\beta}+\frac{27\beta}{4}}}\left(\left(-1+3J_2\right)X_1-2X_3\right)+y^{{-1}}t_{\text{(4.806)}}^{{1+\beta}}\left(-J_3X_1+3J_1\left(X_1+X_3\right)\right)+b^{{\frac{3}{2}}}y^{{-10}}t_{\text{(4.854)}}^{{-31+\frac{81}{2\beta}+\frac{53\beta}{8}}}\left(5X_1+4X_3+5X_5\right)-b^{{\frac{1}{2}}}y^{{-2}}t_{\text{(4.877)}}^{{-2+\frac{27}{2\beta}+\frac{7\beta}{8}}}\left(-J_3X_3+J_1\left(6X_1+11X_3+7X_5\right)\right)+b^{{\frac{3}{2}}}y^{{-15}}J_1t_{\text{(4.882)}}^{{-50+\frac{81}{2\beta}+\frac{93\beta}{8}}}+b^{{\frac{1}{2}}}y^{{-7}}t_{\text{(4.905)}}^{{-21+\frac{27}{2\beta}+\frac{47\beta}{8}}}X_2+b^{{\frac{7}{2}}}y^{{-15}}t_{\text{(4.947)}}^{{-47+\frac{189}{2\beta}+\frac{57\beta}{8}}}+by^{{-3}}J_1t_{\text{(4.949)}}^{{-5+\frac{27}{\beta}+\frac{3\beta}{4}}}\left(3X_1+5X_3+4\left(X_5+X_7\right)\right)+by^{{-8}}t_{\text{(4.977)}}^{{-24+\frac{27}{\beta}+\frac{23\beta}{4}}}\left(-1-4X_2+J_2\left(1+5X_2\right)-4X_4\right)+t^{5}\left(-2J_3\left(1+X_2\right)+J_1\left(3+7X_2+4X_4\right)\right)+\ldots
 \end{dmath*}

\subsection{\alt{$\IB^+ = \II^-_+$}{IB+=II-+}, \alt{$Q_{\rm D3}=9$}{QD3=9}} \label{subsec:IB-plus-index}
\begin{dmath*}[style={\small},indentstep={0pt},spread={0pt}]
1+b^{{\frac{3}{2}}}y^{{-4}}t_{\text{(1.021)}}^{{-12+\frac{81}{2\beta}+\frac{5\beta}{8}}}+t^{2}\left(-1-X_2\right)+b^{{\frac{3}{2}}}y^{{-4}}J_1t_{\text{(2.021)}}^{{-11+\frac{81}{2\beta}+\frac{5\beta}{8}}}+b^{3}y^{{-8}}t_{\text{(2.041)}}^{{-24+\frac{81}{\beta}+\frac{5\beta}{4}}}-yt_{\text{(2.194)}}^{{6-\beta}}X_1+b^{{-\frac{1}{2}}}t_{\text{(2.734)}}^{{2-\frac{27}{2\beta}+\frac{9\beta}{8}}}X_4-y^{{-1}}J_1t_{\text{(2.806)}}^{{-1+\beta}}X_1-b^{{\frac{3}{2}}}y^{{-5}}t_{\text{(2.826)}}^{{-14+\frac{81}{2\beta}+\frac{13\beta}{8}}}X_1+b^{{-\frac{1}{2}}}yt_{\text{(2.928)}}^{{6-\frac{27}{2\beta}+\frac{\beta}{8}}}\left(X_1+X_3+X_5\right)+by^{{-8}}t_{\text{(2.977)}}^{{-26+\frac{27}{\beta}+\frac{23\beta}{4}}}-2t^{3}J_1\left(1+X_2\right)+b^{{\frac{3}{2}}}y^{{-4}}t_{\text{(3.021)}}^{{-10+\frac{81}{2\beta}+\frac{5\beta}{8}}}\left(-1+J_2-X_2\right)+b^{3}y^{{-8}}J_1t_{\text{(3.041)}}^{{-23+\frac{81}{\beta}+\frac{5\beta}{4}}}+b^{{\frac{9}{2}}}y^{{-12}}t_{\text{(3.062)}}^{{-36+\frac{243}{2\beta}+\frac{15\beta}{8}}}+b^{{-\frac{1}{2}}}y^{2}t_{\text{(3.123)}}^{{10-\frac{27}{2\beta}-\frac{7\beta}{8}}}\left(2X_2+X_4+X_6\right)+by^{{-7}}t_{\text{(3.171)}}^{{-22+\frac{27}{\beta}+\frac{19\beta}{4}}}X_1-yJ_1t_{\text{(3.194)}}^{{7-\beta}}\left(X_1+X_3\right)-b^{{\frac{3}{2}}}y^{{-3}}t_{\text{(3.215)}}^{{-6+\frac{81}{2\beta}-\frac{3\beta}{8}}}X_1+b^{{-\frac{1}{2}}}y^{3}t_{\text{(3.317)}}^{{14-\frac{27}{2\beta}-\frac{15\beta}{8}}}\left(X_1+X_3+X_5+X_7\right)+2by^{{-6}}t_{\text{(3.366)}}^{{-18+\frac{27}{\beta}+\frac{15\beta}{4}}}X_2+b^{{-\frac{1}{2}}}y^{4}t_{\text{(3.512)}}^{{18-\frac{27}{2\beta}-\frac{23\beta}{8}}}\left(1+X_4+X_8\right)+by^{{-5}}t_{\text{(3.56)}}^{{-14+\frac{27}{\beta}+\frac{11\beta}{4}}}\left(X_1+2X_3\right)+y^{{-2}}t_{\text{(3.611)}}^{{-4+2\beta}}X_2-b^{{\frac{3}{2}}}y^{{-6}}J_1t_{\text{(3.632)}}^{{-17+\frac{81}{2\beta}+\frac{21\beta}{8}}}+b^{{-\frac{1}{2}}}J_1t_{\text{(3.734)}}^{{3-\frac{27}{2\beta}+\frac{9\beta}{8}}}X_4+by^{{-4}}t_{\text{(3.755)}}^{{-10+\frac{27}{\beta}+\frac{7\beta}{4}}}\left(2+X_2+3X_4\right)-y^{{-1}}t_{\text{(3.806)}}^{{\beta}}\left(J_2X_1-X_3\right)-2b^{{\frac{3}{2}}}y^{{-5}}J_1t_{\text{(3.826)}}^{{-13+\frac{81}{2\beta}+\frac{13\beta}{8}}}X_1-b^{3}y^{{-9}}t_{\text{(3.847)}}^{{-26+\frac{81}{\beta}+\frac{9\beta}{4}}}X_1+b^{{-\frac{1}{2}}}yJ_1t_{\text{(3.928)}}^{{7-\frac{27}{2\beta}+\frac{\beta}{8}}}\left(X_1+X_3+X_5\right)+2by^{{-3}}t_{\text{(3.949)}}^{{-6+\frac{27}{\beta}+\frac{3\beta}{4}}}\left(X_1+X_3+X_5\right)+by^{{-8}}J_1t_{\text{(3.977)}}^{{-25+\frac{27}{\beta}+\frac{23\beta}{4}}}+b^{{\frac{5}{2}}}y^{{-12}}t_{\text{(3.998)}}^{{-38+\frac{135}{2\beta}+\frac{51\beta}{8}}}+t^{4}\left(-2+X_2-2J_2\left(1+X_2\right)+X_4\right)-b^{{\frac{3}{2}}}y^{{-4}}t_{\text{(4.021)}}^{{-9+\frac{81}{2\beta}+\frac{5\beta}{8}}}\left(-J_3+3J_1\left(1+X_2\right)\right)+b^{3}y^{{-8}}t_{\text{(4.041)}}^{{-22+\frac{81}{\beta}+\frac{5\beta}{4}}}\left(-1+2J_2-X_2\right)+b^{{\frac{9}{2}}}y^{{-12}}J_1t_{\text{(4.062)}}^{{-35+\frac{243}{2\beta}+\frac{15\beta}{8}}}+b^{6}y^{{-16}}t_{\text{(4.083)}}^{{-48+\frac{162}{\beta}+\frac{5\beta}{2}}}+b^{{-\frac{1}{2}}}y^{2}J_1t_{\text{(4.123)}}^{{11-\frac{27}{2\beta}-\frac{7\beta}{8}}}\left(2X_2+X_4+X_6\right)+by^{{-2}}t_{\text{(4.143)}}^{{-2+\frac{27}{\beta}-\frac{\beta}{4}}}\left(3X_2+X_4+2X_6\right)+by^{{-7}}J_1t_{\text{(4.171)}}^{{-21+\frac{27}{\beta}+\frac{19\beta}{4}}}X_1+b^{{\frac{5}{2}}}y^{{-11}}t_{\text{(4.192)}}^{{-34+\frac{135}{2\beta}+\frac{43\beta}{8}}}X_1-yt_{\text{(4.194)}}^{{8-\beta}}\left(\left(-1+J_2\right)X_1+\left(-1+J_2\right)X_3-X_5\right)-b^{{\frac{3}{2}}}y^{{-3}}J_1t_{\text{(4.215)}}^{{-5+\frac{81}{2\beta}-\frac{3\beta}{8}}}\left(2X_1+X_3\right)-b^{3}y^{{-7}}t_{\text{(4.236)}}^{{-18+\frac{81}{\beta}+\frac{\beta}{4}}}X_1+b^{{-\frac{1}{2}}}y^{3}J_1t_{\text{(4.317)}}^{{15-\frac{27}{2\beta}-\frac{15\beta}{8}}}\left(X_1+X_3+X_5+X_7\right)+by^{{-1}}t_{\text{(4.338)}}^{{2+\frac{27}{\beta}-\frac{5\beta}{4}}}\left(X_1+X_3+X_5+X_7\right)+2by^{{-6}}J_1t_{\text{(4.366)}}^{{-17+\frac{27}{\beta}+\frac{15\beta}{4}}}X_2+2b^{{\frac{5}{2}}}y^{{-10}}t_{\text{(4.387)}}^{{-30+\frac{135}{2\beta}+\frac{35\beta}{8}}}X_2+y^{2}t_{\text{(4.389)}}^{{12-2\beta}}\left(1+X_6\right)+b^{{\frac{3}{2}}}y^{{-7}}t_{\text{(4.437)}}^{{-20+\frac{81}{2\beta}+\frac{29\beta}{8}}}X_1+b^{{-\frac{1}{2}}}y^{4}J_1t_{\text{(4.512)}}^{{19-\frac{27}{2\beta}-\frac{23\beta}{8}}}\left(1+X_4+X_8\right)+bt_{\text{(4.532)}}^{{6+\frac{27}{\beta}-\frac{9\beta}{4}}}\left(1+X_4+X_8\right)-b^{{-\frac{1}{2}}}y^{{-1}}t_{\text{(4.539)}}^{{-\frac{27}{2\beta}+\frac{17\beta}{8}}}X_3+by^{{-5}}J_1t_{\text{(4.56)}}^{{-13+\frac{27}{\beta}+\frac{11\beta}{4}}}\left(X_1+2X_3\right)+b^{{\frac{5}{2}}}y^{{-9}}t_{\text{(4.581)}}^{{-26+\frac{135}{2\beta}+\frac{27\beta}{8}}}\left(X_1+2X_3\right)+y^{{-2}}J_1t_{\text{(4.611)}}^{{-3+2\beta}}\left(1+X_2\right)-b^{{\frac{3}{2}}}y^{{-6}}t_{\text{(4.632)}}^{{-16+\frac{81}{2\beta}+\frac{21\beta}{8}}}\left(J_2-X_2\right)-b^{3}y^{{-10}}J_1t_{\text{(4.653)}}^{{-29+\frac{81}{\beta}+\frac{13\beta}{4}}}-b^{{-\frac{1}{2}}}t_{\text{(4.734)}}^{{4-\frac{27}{2\beta}+\frac{9\beta}{8}}}\left(1+2X_2-\left(-3+J_2\right)X_4+X_6\right)+by^{{-4}}J_1t_{\text{(4.755)}}^{{-9+\frac{27}{\beta}+\frac{7\beta}{4}}}\left(2+X_2+4X_4\right)+b^{{\frac{5}{2}}}y^{{-8}}t_{\text{(4.775)}}^{{-22+\frac{135}{2\beta}+\frac{19\beta}{8}}}\left(2+X_2+3X_4\right)+y^{{-1}}t_{\text{(4.806)}}^{{1+\beta}}\left(-J_3X_1+3J_1\left(X_1+X_3\right)\right)-b^{{\frac{3}{2}}}y^{{-5}}t_{\text{(4.826)}}^{{-12+\frac{81}{2\beta}+\frac{13\beta}{8}}}\left(3J_2X_1-2X_3\right)-3b^{3}y^{{-9}}J_1t_{\text{(4.847)}}^{{-25+\frac{81}{\beta}+\frac{9\beta}{4}}}X_1-b^{{\frac{9}{2}}}y^{{-13}}t_{\text{(4.868)}}^{{-38+\frac{243}{2\beta}+\frac{23\beta}{8}}}X_1+b^{{-\frac{1}{2}}}yt_{\text{(4.928)}}^{{8-\frac{27}{2\beta}+\frac{\beta}{8}}}\left(\left(-4+J_2\right)X_1+\left(-6+J_2\right)X_3-5X_5+J_2X_5-2X_7\right)+3by^{{-3}}J_1t_{\text{(4.949)}}^{{-5+\frac{27}{\beta}+\frac{3\beta}{4}}}\left(X_1+X_3+X_5\right)+2b^{{\frac{5}{2}}}y^{{-7}}t_{\text{(4.97)}}^{{-18+\frac{135}{2\beta}+\frac{11\beta}{8}}}\left(X_1+X_3+X_5\right)+by^{{-8}}t_{\text{(4.977)}}^{{-24+\frac{27}{\beta}+\frac{23\beta}{4}}}\left(-1+J_2-2X_2\right)+2b^{{\frac{5}{2}}}y^{{-12}}J_1t_{\text{(4.998)}}^{{-37+\frac{135}{2\beta}+\frac{51\beta}{8}}}+t^{5}\left(-2J_3\left(1+X_2\right)+J_1\left(3+7X_2+4X_4\right)\right)+\ldots
\end{dmath*}

\bibliographystyle{JHEP}
\bibliography{refs}

\providecommand{\href}[2]{#2}\begingroup\raggedright\begin{thebibliography}{10}

\bibitem{Witten:1998xy}
E.~Witten, {\it {Baryons and branes in anti-de Sitter space}},  {\em JHEP} {\bf
  9807} (1998) 006, [\href{http://arxiv.org/abs/hep-th/9805112}{{\tt
  hep-th/9805112}}].

\bibitem{Leigh:1995ep}
R.~G. Leigh and M.~J. Strassler, {\it {Exactly marginal operators and duality
  in four-dimensional N=1 supersymmetric gauge theory}},  {\em Nucl.Phys.} {\bf
  B447} (1995) 95--136, [\href{http://arxiv.org/abs/hep-th/9503121}{{\tt
  hep-th/9503121}}].

\bibitem{Argyres:1999xu}
P.~C. Argyres, K.~A. Intriligator, R.~G. Leigh, and M.~J. Strassler, {\it {On
  inherited duality in {$\mathcal{N}=1$} {$d = 4$} supersymmetric gauge
  theories}},  {\em JHEP} {\bf 0004} (2000) 029,
  [\href{http://arxiv.org/abs/hep-th/9910250}{{\tt hep-th/9910250}}].

\bibitem{Benini:2009mz}
F.~Benini, Y.~Tachikawa, and B.~Wecht, {\it {Sicilian gauge theories and N=1
  dualities}},  {\em JHEP} {\bf 1001} (2010) 088,
  [\href{http://arxiv.org/abs/0909.1327}{{\tt arXiv:0909.1327}}].

\bibitem{Seiberg:1994pq}
N.~Seiberg, {\it {Electric-magnetic duality in supersymmetric non-Abelian gauge
  theories}},  {\em Nucl.Phys.} {\bf B435} (1995) 129--146,
  [\href{http://arxiv.org/abs/hep-th/9411149}{{\tt hep-th/9411149}}].

\bibitem{Seiberg:1995ac}
N.~Seiberg, {\it {The Power of duality: Exact results in 4D SUSY field
  theory}},  {\em Int.J.Mod.Phys.} {\bf A16} (2001) 4365--4376,
  [\href{http://arxiv.org/abs/hep-th/9506077}{{\tt hep-th/9506077}}].

\bibitem{dualities1}
I.~Garc\'ia-Etxebarria, B.~Heidenreich, and T.~Wrase, {\it {New $\mathcal{N}=1$
  dualities from orientifold transitions -- Part I: Field Theory}},  {\em JHEP}
  {\bf 1310} (2013) 007, [\href{http://arxiv.org/abs/1210.7799}{{\tt
  arXiv:1210.7799}}].

\bibitem{Bianchi:2013gka}
M.~Bianchi, G.~Inverso, J.~F. Morales, and D.~R. Pacifici, {\it {Unoriented
  Quivers with Flavour}},  {\em JHEP} {\bf 1401} (2014) 128,
  [\href{http://arxiv.org/abs/1307.0466}{{\tt arXiv:1307.0466}}].

\bibitem{dualities2}
I.~Garc\'ia-Etxebarria, B.~Heidenreich, and T.~Wrase, {\it {New $\mathcal{N}=1$
  dualities from orientifold transitions -- Part II: String Theory}},  {\em
  JHEP} {\bf 1310} (2013) 006, [\href{http://arxiv.org/abs/1307.1701}{{\tt
  arXiv:1307.1701}}].

\bibitem{Uranga:1999ib}
A.~M. Uranga, {\it {Comments on nonsupersymmetric orientifolds at strong
  coupling}},  {\em JHEP} {\bf 0002} (2000) 041,
  [\href{http://arxiv.org/abs/hep-th/9912145}{{\tt hep-th/9912145}}].

\bibitem{Sugimoto:2012rt}
S.~Sugimoto, {\it {Confinement and Dynamical Symmetry Breaking in non-SUSY
  Gauge Theory from S-duality in String Theory}},  {\em Prog.Theor.Phys.} {\bf
  128} (2012) 1175--1209, [\href{http://arxiv.org/abs/1207.2203}{{\tt
  arXiv:1207.2203}}].

\bibitem{Hook:2013vza}
A.~Hook and G.~Torroba, {\it {S-duality of nonsupersymmetric gauge theories}},
  {\em Phys.Rev.} {\bf D89} (2014), no.~2 025006,
  [\href{http://arxiv.org/abs/1309.5948}{{\tt arXiv:1309.5948}}].

\bibitem{Gaiotto:2015usa}
D.~Gaiotto and S.~S. Razamat, {\it {$N=1$ theories of class $S_k$}},
  \href{http://arxiv.org/abs/1503.05159}{{\tt arXiv:1503.05159}}.

\bibitem{Franco:2015jna}
S.~Franco, H.~Hayashi, and A.~Uranga, {\it {Charting Class ${\cal S}_k$
  Territory}},  \href{http://arxiv.org/abs/1504.05988}{{\tt arXiv:1504.05988}}.

\bibitem{Hanany:2015pfa}
A.~Hanany and K.~Maruyoshi, {\it {Chiral theories of class S}},
  \href{http://arxiv.org/abs/1505.05053}{{\tt arXiv:1505.05053}}.

\bibitem{Franco:2010jv}
S.~Franco and G.~Torroba, {\it {Gauge theories from D7-branes over vanishing
  4-cycles}},  {\em JHEP} {\bf 1101} (2011) 017,
  [\href{http://arxiv.org/abs/1010.4029}{{\tt arXiv:1010.4029}}].

\bibitem{Berkooz:1995km}
M.~Berkooz, {\it {The Dual of supersymmetric SU(2k) with an antisymmetric
  tensor and composite dualities}},  {\em Nucl.Phys.} {\bf B452} (1995)
  513--525, [\href{http://arxiv.org/abs/hep-th/9505067}{{\tt hep-th/9505067}}].

\bibitem{Pouliot:1995me}
P.~Pouliot, {\it {Duality in SUSY SU(N) with an antisymmetric tensor}},  {\em
  Phys.Lett.} {\bf B367} (1996) 151--156,
  [\href{http://arxiv.org/abs/hep-th/9510148}{{\tt hep-th/9510148}}].

\bibitem{Witten:1993yc}
E.~Witten, {\it {Phases of N=2 theories in two-dimensions}},  {\em Nucl.Phys.}
  {\bf B403} (1993) 159--222, [\href{http://arxiv.org/abs/hep-th/9301042}{{\tt
  hep-th/9301042}}].

\bibitem{GriffithsHarris}
P.~A. Griffiths and J.~Harris, {\em Principles of algebraic geometry}.
\newblock Wiley New York, 1978.

\bibitem{Malyshev:2007yb}
D.~Malyshev, {\it {Del Pezzo singularities and SUSY breaking}},  {\em Adv.High
  Energy Phys.} {\bf 2011} (2011) 630892,
  [\href{http://arxiv.org/abs/0705.3281}{{\tt arXiv:0705.3281}}].

\bibitem{Hanany:2005ve}
A.~Hanany and K.~D. Kennaway, {\it {Dimer models and toric diagrams}},
  \href{http://arxiv.org/abs/hep-th/0503149}{{\tt hep-th/0503149}}.

\bibitem{Franco:2005rj}
S.~Franco, A.~Hanany, K.~D. Kennaway, D.~Vegh, and B.~Wecht, {\it {Brane dimers
  and quiver gauge theories}},  {\em JHEP} {\bf 0601} (2006) 096,
  [\href{http://arxiv.org/abs/hep-th/0504110}{{\tt hep-th/0504110}}].

\bibitem{Franco:2007ii}
S.~Franco, A.~Hanany, D.~Krefl, J.~Park, A.~M. Uranga, et~al., {\it {Dimers and
  orientifolds}},  {\em JHEP} {\bf 0709} (2007) 075,
  [\href{http://arxiv.org/abs/0707.0298}{{\tt arXiv:0707.0298}}].

\bibitem{Kennaway:2007tq}
K.~D. Kennaway, {\it {Brane Tilings}},  {\em Int.J.Mod.Phys.} {\bf A22} (2007)
  2977--3038, [\href{http://arxiv.org/abs/0706.1660}{{\tt arXiv:0706.1660}}].

\bibitem{Yamazaki:2008bt}
M.~Yamazaki, {\it {Brane Tilings and Their Applications}},  {\em Fortsch.Phys.}
  {\bf 56} (2008) 555--686, [\href{http://arxiv.org/abs/0803.4474}{{\tt
  arXiv:0803.4474}}].

\bibitem{Ishii:2007}
A.~Ishii and K.~Ueda, {\it {On moduli spaces of quiver representations
  associated with dimer models}},  {\em RIMS {K\^{o}ky\^{u}roku} Bessatsu} {\bf
  B9} (2008) 127--141, [\href{http://arxiv.org/abs/0710.1898}{{\tt
  arXiv:0710.1898}}].

\bibitem{Hanany:2005ss}
A.~Hanany and D.~Vegh, {\it {Quivers, tilings, branes and rhombi}},  {\em JHEP}
  {\bf 0710} (2007) 029, [\href{http://arxiv.org/abs/hep-th/0511063}{{\tt
  hep-th/0511063}}].

\bibitem{Feng:2000mi}
B.~Feng, A.~Hanany, and Y.-H. He, {\it {D-brane gauge theories from toric
  singularities and toric duality}},  {\em Nucl.Phys.} {\bf B595} (2001)
  165--200, [\href{http://arxiv.org/abs/hep-th/0003085}{{\tt hep-th/0003085}}].

\bibitem{Beasley:2001zp}
C.~E. Beasley and M.~R. Plesser, {\it {Toric duality is Seiberg duality}},
  {\em JHEP} {\bf 0112} (2001) 001,
  [\href{http://arxiv.org/abs/hep-th/0109053}{{\tt hep-th/0109053}}].

\bibitem{Imamura:2008fd}
Y.~Imamura, K.~Kimura, and M.~Yamazaki, {\it {Anomalies and O-plane charges in
  orientifolded brane tilings}},  {\em JHEP} {\bf 0803} (2008) 058,
  [\href{http://arxiv.org/abs/0801.3528}{{\tt arXiv:0801.3528}}].

\bibitem{Angelantonj:1996uy}
C.~Angelantonj, M.~Bianchi, G.~Pradisi, A.~Sagnotti, and Y.~Stanev, {\it
  {Chiral asymmetry in four-dimensional open string vacua}},  {\em Phys.Lett.}
  {\bf B385} (1996) 96--102, [\href{http://arxiv.org/abs/hep-th/9606169}{{\tt
  hep-th/9606169}}].

\bibitem{Lykken:1997ub}
J.~D. Lykken, E.~Poppitz, and S.~P. Trivedi, {\it {M(ore) on chiral gauge
  theories from D-branes}},  {\em Nucl.Phys.} {\bf B520} (1998) 51--74,
  [\href{http://arxiv.org/abs/hep-th/9712193}{{\tt hep-th/9712193}}].

\bibitem{Kakushadze:1998tr}
Z.~Kakushadze, {\it {Gauge theories from orientifolds and large N limit}},
  {\em Nucl.Phys.} {\bf B529} (1998) 157--179,
  [\href{http://arxiv.org/abs/hep-th/9803214}{{\tt hep-th/9803214}}].

\bibitem{Wijnholt:2002qz}
M.~Wijnholt, {\it {Large volume perspective on branes at singularities}},  {\em
  Adv.Theor.Math.Phys.} {\bf 7} (2004) 1117--1153,
  [\href{http://arxiv.org/abs/hep-th/0212021}{{\tt hep-th/0212021}}].

\bibitem{Wijnholt:2007vn}
M.~Wijnholt, {\it {Geometry of Particle Physics}},  {\em Adv.Theor.Math.Phys.}
  {\bf 13} (2009) [\href{http://arxiv.org/abs/hep-th/0703047}{{\tt
  hep-th/0703047}}].

\bibitem{BJHorientifoldnote}
B.~Heidenreich.
\newblock Unpublished notes.

\bibitem{Evslin:2006cj}
J.~Evslin, {\it {What does(n't) K-theory classify?}},
  \href{http://arxiv.org/abs/hep-th/0610328}{{\tt hep-th/0610328}}.

\bibitem{Hanany:2000fq}
A.~Hanany and B.~Kol, {\it {On orientifolds, discrete torsion, branes and M
  theory}},  {\em JHEP} {\bf 0006} (2000) 013,
  [\href{http://arxiv.org/abs/hep-th/0003025}{{\tt hep-th/0003025}}].

\bibitem{AT}
A.~Hatcher, {\em Algebraic Topology}.
\newblock Cambridge University Press, 2002.
\newblock \url{http://www.math.cornell.edu/~hatcher/AT/ATpage.html}.

\bibitem{Liu:1998dra}
C.-H. Liu, {\it {On the isolated singularity of a seven space obtained by
  rolling Calabi-Yau threefolds through extremal transitions}},
  \href{http://arxiv.org/abs/hep-th/9801175}{{\tt hep-th/9801175}}.

\bibitem{kenyonintro}
R.~{Kenyon}, {\it {An introduction to the dimer model}},  {\em ArXiv
  Mathematics e-prints} (Oct., 2003)
  [\href{http://arxiv.org/abs/math/0310326}{{\tt math/0310326}}].

\bibitem{Feng:2005gw}
B.~Feng, Y.-H. He, K.~D. Kennaway, and C.~Vafa, {\it {Dimer models from mirror
  symmetry and quivering amoebae}},  {\em Adv.Theor.Math.Phys.} {\bf 12} (2008)
  489--545, [\href{http://arxiv.org/abs/hep-th/0511287}{{\tt hep-th/0511287}}].

\bibitem{GarciaEtxebarria:2006aq}
I.~Garcia-Etxebarria, F.~Saad, and A.~M. Uranga, {\it {Quiver gauge theories at
  resolved and deformed singularities using dimers}},  {\em JHEP} {\bf 0606}
  (2006) 055, [\href{http://arxiv.org/abs/hep-th/0603108}{{\tt
  hep-th/0603108}}].

\bibitem{toricII}
I.~Garc{\'i}a-Etxebarria and B.~Heidenreich.
\newblock To appear.

\bibitem{Evans:1997hk}
N.~J. Evans, C.~V. Johnson, and A.~D. Shapere, {\it {Orientifolds, branes, and
  duality of 4-D gauge theories}},  {\em Nucl.Phys.} {\bf B505} (1997)
  251--271, [\href{http://arxiv.org/abs/hep-th/9703210}{{\tt hep-th/9703210}}].

\bibitem{Hanany:1996ie}
A.~Hanany and E.~Witten, {\it {Type IIB superstrings, BPS monopoles, and
  three-dimensional gauge dynamics}},  {\em Nucl.Phys.} {\bf B492} (1997)
  152--190, [\href{http://arxiv.org/abs/hep-th/9611230}{{\tt hep-th/9611230}}].

\bibitem{Giveon:1998sr}
A.~Giveon and D.~Kutasov, {\it {Brane dynamics and gauge theory}},  {\em
  Rev.Mod.Phys.} {\bf 71} (1999) 983--1084,
  [\href{http://arxiv.org/abs/hep-th/9802067}{{\tt hep-th/9802067}}].

\bibitem{Imamura:2006ie}
Y.~Imamura, {\it {Global symmetries and 't Hooft anomalies in brane tilings}},
  {\em JHEP} {\bf 0612} (2006) 041,
  [\href{http://arxiv.org/abs/hep-th/0609163}{{\tt hep-th/0609163}}].

\bibitem{Intriligator:1995ne}
K.~A. Intriligator and P.~Pouliot, {\it {Exact superpotentials, quantum vacua
  and duality in supersymmetric SP(N(c)) gauge theories}},  {\em Phys.Lett.}
  {\bf B353} (1995) 471--476, [\href{http://arxiv.org/abs/hep-th/9505006}{{\tt
  hep-th/9505006}}].

\bibitem{Broomhead:2008an}
N.~Broomhead, {\it {Dimer models and Calabi-Yau algebras}},
  \href{http://arxiv.org/abs/0901.4662}{{\tt arXiv:0901.4662}}.

\bibitem{Hanany:2006nm}
A.~Hanany, C.~P. Herzog, and D.~Vegh, {\it {Brane tilings and exceptional
  collections}},  {\em JHEP} {\bf 0607} (2006) 001,
  [\href{http://arxiv.org/abs/hep-th/0602041}{{\tt hep-th/0602041}}].

\bibitem{Davey:2009bp}
J.~Davey, A.~Hanany, and J.~Pasukonis, {\it {On the Classification of Brane
  Tilings}},  {\em JHEP} {\bf 1001} (2010) 078,
  [\href{http://arxiv.org/abs/0909.2868}{{\tt arXiv:0909.2868}}].

\bibitem{Davey:2011sw}
J.~P. Davey, {\it {Brane Tilings, M2-branes and Orbifolds}},
  \href{http://arxiv.org/abs/1110.6658}{{\tt arXiv:1110.6658}}.

\bibitem{Franco:2006es}
S.~Franco and A.~M.~. Uranga, {\it {Dynamical SUSY breaking at meta-stable
  minima from D-branes at obstructed geometries}},  {\em JHEP} {\bf 0606}
  (2006) 031, [\href{http://arxiv.org/abs/hep-th/0604136}{{\tt
  hep-th/0604136}}].

\bibitem{Green:2010da}
D.~Green, Z.~Komargodski, N.~Seiberg, Y.~Tachikawa, and B.~Wecht, {\it {Exactly
  Marginal Deformations and Global Symmetries}},  {\em JHEP} {\bf 1006} (2010)
  106, [\href{http://arxiv.org/abs/1005.3546}{{\tt arXiv:1005.3546}}].

\bibitem{Intriligator:2003jj}
K.~A. Intriligator and B.~Wecht, {\it {The Exact superconformal R symmetry
  maximizes a}},  {\em Nucl.Phys.} {\bf B667} (2003) 183--200,
  [\href{http://arxiv.org/abs/hep-th/0304128}{{\tt hep-th/0304128}}].

\bibitem{Imamura:2007dc}
Y.~Imamura, H.~Isono, K.~Kimura, and M.~Yamazaki, {\it {Exactly marginal
  deformations of quiver gauge theories as seen from brane tilings}},  {\em
  Prog.Theor.Phys.} {\bf 117} (2007) 923--955,
  [\href{http://arxiv.org/abs/hep-th/0702049}{{\tt hep-th/0702049}}].

\bibitem{Dolan:2008qi}
F.~Dolan and H.~Osborn, {\it {Applications of the Superconformal Index for
  Protected Operators and q-Hypergeometric Identities to N=1 Dual Theories}},
  {\em Nucl.Phys.} {\bf B818} (2009) 137--178,
  [\href{http://arxiv.org/abs/0801.4947}{{\tt arXiv:0801.4947}}].

\bibitem{Luty:1996cg}
M.~A. Luty, M.~Schmaltz, and J.~Terning, {\it {A Sequence of duals for Sp(2N)
  supersymmetric gauge theories with adjoint matter}},  {\em Phys.Rev.} {\bf
  D54} (1996) 7815--7824, [\href{http://arxiv.org/abs/hep-th/9603034}{{\tt
  hep-th/9603034}}].

\bibitem{Sakai:1997xs}
T.~Sakai, {\it {Duality in supersymmetric SU(N) gauge theory with a symmetric
  tensor}},  {\em Mod.Phys.Lett.} {\bf A12} (1997) 1025--1034,
  [\href{http://arxiv.org/abs/hep-th/9701155}{{\tt hep-th/9701155}}].

\bibitem{Intriligator:1995id}
K.~A. Intriligator and N.~Seiberg, {\it {Duality, monopoles, dyons, confinement
  and oblique confinement in supersymmetric SO(N(c)) gauge theories}},  {\em
  Nucl.Phys.} {\bf B444} (1995) 125--160,
  [\href{http://arxiv.org/abs/hep-th/9503179}{{\tt hep-th/9503179}}].

\bibitem{Csaki:1996zb}
C.~Csaki, M.~Schmaltz, and W.~Skiba, {\it {Confinement in N=1 SUSY gauge
  theories and model building tools}},  {\em Phys.Rev.} {\bf D55} (1997)
  7840--7858, [\href{http://arxiv.org/abs/hep-th/9612207}{{\tt
  hep-th/9612207}}].

\bibitem{Romelsberger:2005eg}
C.~Romelsberger, {\it {Counting chiral primaries in N = 1, d=4 superconformal
  field theories}},  {\em Nucl.Phys.} {\bf B747} (2006) 329--353,
  [\href{http://arxiv.org/abs/hep-th/0510060}{{\tt hep-th/0510060}}].

\bibitem{Kinney:2005ej}
J.~Kinney, J.~M. Maldacena, S.~Minwalla, and S.~Raju, {\it {An Index for 4
  dimensional super conformal theories}},  {\em Commun.Math.Phys.} {\bf 275}
  (2007) 209--254, [\href{http://arxiv.org/abs/hep-th/0510251}{{\tt
  hep-th/0510251}}].

\bibitem{Romelsberger:2007ec}
C.~Romelsberger, {\it {Calculating the Superconformal Index and Seiberg
  Duality}},  \href{http://arxiv.org/abs/0707.3702}{{\tt arXiv:0707.3702}}.

\bibitem{Spiridonov:2009za}
V.~Spiridonov and G.~Vartanov, {\it {Elliptic Hypergeometry of Supersymmetric
  Dualities}},  {\em Commun.Math.Phys.} {\bf 304} (2011) 797--874,
  [\href{http://arxiv.org/abs/0910.5944}{{\tt arXiv:0910.5944}}].

\bibitem{Spiridonov:2010qv}
V.~Spiridonov and G.~Vartanov, {\it {Superconformal indices of ${\mathcal N}=4$
  SYM field theories}},  {\em Lett.Math.Phys.} {\bf 100} (2012) 97--118,
  [\href{http://arxiv.org/abs/1005.4196}{{\tt arXiv:1005.4196}}].

\bibitem{Blumenhagen:2010pv}
R.~Blumenhagen, B.~Jurke, T.~Rahn, and H.~Roschy, {\it {Cohomology of Line
  Bundles: A Computational Algorithm}},  {\em J.Math.Phys.} {\bf 51} (2010)
  103525, [\href{http://arxiv.org/abs/1003.5217}{{\tt arXiv:1003.5217}}].

\bibitem{Jow-cohomology}
S.-Y. {Jow}, {\it {Cohomology of toric line bundles via simplicial Alexander
  duality}},  {\em Journal of Mathematical Physics} {\bf 52} (Mar., 2011)
  033506, [\href{http://arxiv.org/abs/1006.0780}{{\tt arXiv:1006.0780}}].

\bibitem{Rahn:2010fm}
T.~Rahn and H.~Roschy, {\it {Cohomology of Line Bundles: Proof of the
  Algorithm}},  {\em J.Math.Phys.} {\bf 51} (2010) 103520,
  [\href{http://arxiv.org/abs/1006.2392}{{\tt arXiv:1006.2392}}].

\bibitem{LiE}
M.~A.~A. van Leeuwen, A.~M. Cohen, and B.~Lisser, {\it {LiE, A Package for Lie
  Group Computations}},  {\em Computer Algebra} (1992).
  \url{http://www-math.univ-poitiers.fr/~maavl/LiE}.

\end{thebibliography}\endgroup

\end{document}